  \documentclass[final,11pt]{article}
  \usepackage[letterpaper]{geometry}
  \usepackage{fancyhdr}
  \usepackage{ifpdf}
  \usepackage{amsfonts,amsmath, amssymb, amsthm, bbm}
  \usepackage{graphicx}
  \usepackage{slashed,enumerate}
  \usepackage{refcount,ifthen}
  \usepackage[all]{xy}

  \newcommand{\keywords}[1]{\mbox{}\\\noindent\emph{Keywords:} #1\\}

  \fancypagestyle{myheadsfoots}{
    \fancyhead[L]{\tiny\textit{\leftmark}}
    \fancyhead[R]{\tiny\textit{\rightmark}}
    \fancyfoot[C]{\small\thepage}
  }

\ifpdf
  \usepackage[pdftitle={TIME EVOLUTION OF THE EXTERNAL FIELD PROBLEM IN QED},
              pdfauthor={D.-A. Deckert, D. D\"urr, F. Merkl, M. Schottenloher},
              pdfpagemode=UseOutlines,
              pdfstartview=FitH,
              bookmarks=true,
              bookmarksopen=true,
              bookmarksnumbered=true,
              bookmarkstype=toc,
              colorlinks=true,
              linkcolor=blue,
              citecolor=blue,
              urlcolor=blue]{hyperref}
\fi

\numberwithin{equation}{section}
\newtheorem{our_theorem}{Theorem}[section]
\newtheorem*{our_mainresult}{Main Results}
\newtheorem{our_lemma}[our_theorem]{Lemma}
\newtheorem{our_corollary}[our_theorem]{Corollary}
\newtheorem{our_proposition}[our_theorem]{Proposition}
\newtheorem{our_definition}[our_theorem]{Definition}
\newtheorem{our_construction}[our_theorem]{Construction}

\usepackage[refpage,noprefix]{nomencl}

\newcommand{\labelref}[1]{ %
  \expandafter\ifx\csname xrf@#1\endcsname\relax %
    ?? %
  \else %
    {\csname xrf@#1\endcsname} %
  \fi %
}
\newcommand{\labelname}[2]{%
  \expandafter\edef\csname xrf@#1\endcsname{#2}\relax %
  \immediate\write\xrf{ %
    \string\expandafter\string\edef %
    \string\csname\space xrf@#1\string\endcsname{#2}\string\relax %
  } %
}

\newcounter{constNum}
\newcommand{\xconstl}[1]{ %
  {C_{\arabic{constNum}}} %
  \labelname{#1}{\arabic{constNum}} %
  \stepcounter{constNum} %
}
\newcommand{\xconstr}[1]{{C_{\labelref{#1}}}}

\newcounter{termNum}
\newcommand{\ppref}[1]                                  
{%
  \ifthenelse{\getpagerefnumber{#1}=\thepage}%
  {}{{\tiny{p.\pageref{#1}}}}%
}
\newcommand{\pref}[1]{\ref{#1}\ppref{#1}}               

\newcommand{\nomsep}{|}
\newcommand{\nomdsep}{\|}
\def\R{{\mathbb R}}  
\def\N{{\mathbb N}}  
\def\Z{{\mathbb Z}}  
\def\C{{\mathbb C}}  

\newcommand{\onorm}[1]{\|#1_\odd\|_{\hs}}
\newcommand{\pol}{\mathtt{Pol}}
\newcommand{\seas}{\mathtt{Seas}}
\newcommand{\iseas}{\mathtt{Seas}^{\perp}}
\newcommand{\ocean}{\mathtt{Ocean}}
\newcommand{\glm}{{\mathrm{GL}_-}}
\newcommand{\slg}{\mathrm{SL}}
\newcommand{\trcl}{{\mathrm{I_1}}}
\newcommand{\hs}{{\mathrm{I_2}}}
\newcommand{\cu}{{\mathrm U}}
\newcommand{\ur}{\cu_{\rm res}}

\newcommand{\out}{\operatorname{out}}
\newcommand{\inn}{\operatorname{in}}
\newcommand{\coker}{\operatorname{coker}}
\newcommand{\Wedge}{{\sf \Lambda}}
\newcommand{\id}{\operatorname{id}}
\newcommand{\ind}{\operatorname{ind}}
\newcommand{\sk}[1]{\left\langle #1\right\rangle}
\newcommand{\dcl}[1]{\id_{#1}+\trcl(#1)}
\newcommand{\trace}{\operatorname{tr}}
\newcommand{\ch}{{\cal H}}
\newcommand{\cf}{{\cal F}}
\newcommand{\cs}{{\cal S}}
\newcommand{\cv}{{\cal V}}
\newcommand{\sa}{{\sf A}}
\newcommand{\ev}{{\rm ev}}
\newcommand{\odd}{{\rm odd}}

\newcommand{\norm}[1]{\|#1\|}
\newcommand{\hsnorm}[1]{\|#1\|_\hs}

\newcommand{\range}{\operatorname{range}}
\newcommand{\lop}[1]{{\cal L}_{#1}}
\newcommand{\rop}[1]{{\cal R}_{#1}}
\newcommand{\charge}{\operatorname{charge}}

\begin{document}

  \newread\xrf %
  \def\xrfname{\jobname.xrf} %
  \openin\xrf=\jobname.xrf %
  \ifeof\xrf\message{! XRF WARNING: did not find \jobname.xrf;}\else\input\jobname.xrf\fi %
  \immediate\closein\xrf %
  \newwrite\xrf %
  \immediate\openout\xrf=\jobname.xrf %

\markboth{D.A.-Deckert, D. D\"urr, F. Merkl, M. Schottenloher}
{Time Evolution of the External Field Problem in QED}

\pagestyle{myheadsfoots}

  \title{TIME-EVOLUTION
         OF THE EXTERNAL FIELD PROBLEM
         IN QED}

  \title{TIME EVOLUTION\\
         OF THE EXTERNAL FIELD PROBLEM\\
         IN QED}
  \author{D.-A. Deckert\thanks{deckert@math.lmu.de},
          D. D\"urr\thanks{duerr@math.lmu.de},
          F. Merkl\thanks{merkl@math.lmu.de},
          M. Schottenloher\thanks{schotten@math.lmu.de}\\\mbox{}\\
          \textit{Mathematisches Institut, Ludwigs-Maximilians-Universit\"at M\"unchen}\\
          \textit{Theresienstra\ss e 39, D-80333 M\"unchen, Germany.}}

\date{\today}

\maketitle

\begin{abstract}
We construct the time-evolution for the second quantized Dirac equation subject 
to a smooth, compactly supported, time dependent electromagnetic 
potential and identify the degrees of freedom involved. Earlier works on this (e.g. Ruijsenaars) observed the Shale-Stinespring condition and showed that the one-particle time-evolution can be lifted to Fock 
space if and only if the external field had zero magnetic components. We scrutinize the idea, observed earlier by Fierz and Scharf, that the time-evolution can be implemented between time varying Fock spaces. 
In order to define these Fock spaces we are led to consider classes of reference vacua and polarizations.
We show that this implementation is up 
to a phase independent of the chosen reference vacuum or polarization and that all induced transition probabilities are well-defined and unique.
\end{abstract}

\keywords{Second Quantized Dirac Equation; External Field; Polarization Classes; Time-Varying Fock Spaces.}

\tableofcontents

\section{Introduction}

{
\subsection{The Problem and State of the Art}

This paper purports the the second-quantized time-evolution of Dirac fermions subject to a classical time-dependent external electromagnetic potential 
${\sf A}:\R^4\to\R^4$. In the early works
\cite{shale:65, Ruijsenaars:77-1,Ruijsenaars:77-2,Bellisard:75,Bellisard:76} it was recognized that the construction of such a time-evolution in the presence of an external potential which has non-zero magnetic components turns out to be impossible on one fixed Fock space. Let us briefly  explain the nature of the problem and state the classical results.

It is well known that the spectrum of the free Dirac operator $H^0 = -i\alpha\cdot \nabla+\beta m$\nomenclature{$H^0$}{Free Dirac hamiltonian} is $(-\infty,-m]\cup[+m,+\infty)$ and, thus, allows for wave functions associated with ``negative energy''. Throughout this work we use Planck units $\hbar=c=1$. The two components of the spectrum give rise to a splitting of the one-particle Hilbert space $L_2(\R^3,\C^4)$\nomenclature{$L_2(\R^3,\C^4)$}{The space of square integrable $\C^4$ valued functions on $\R^3$}, i.e. the space of square integrable $\C^4$ valued functions on $\R^3$, into two spectral subspaces $L_2(\R^3,\C^4)=\ch_- \oplus \ch_+ $. While a wave function $\psi\in\ch_+$ can be interpreted to describe the dynamics of electrons with positive kinetic energy the interpretation of the negative energy wave functions is not straightforward as we do not seem to see particles of negative kinetic energy in nature. Moreover, there is no mechanism in quantum mechanics to prevent transitions from the positive to the negative spectral subspace so we can not simply regard those negative energy wave functions as unphysical. In order to solve this problem Dirac very early came up with a physical theory for his equation, the so-called \emph{Dirac Sea} or \emph{Hole Theory}:
\begin{quote}
  Admettons que dans l'Univers tel que nous le connaissons, les \'etats d'energie n\'egative soient presque tous occup\'es par des \'electrons, et que la distribution ainsi obtenue ne soit pas accessible \`a notre observation \`a cause de son uniformit\'e dans toute l'etendue de l'espace. Dans ces conditions, tout \'etat d'energie n\'egative non occup\'e repr\'esentant une rupture de cette uniformit\'e, doit se r\'ev\'evler \`a observation comme une sorte de lacune. Il es possible d'admettre que ces lacunes constituent les positrons.
  \begin{flushright}
    \footnotesize{P.A.M. Dirac, Th\'eorie du Positron (1934), in Selected Papers on Quantum Electrodynamics, Ed. J. Schwinger, Dover Pub. (1958)}
    \end{flushright}
\end{quote}
It is assumed that all negative energy states are occupied by electrons which then constitute the Dirac sea. Due to its uniformity the Dirac sea is hidden from our observation and, thus, \emph{physically inaccessible}. What can be observed are the holes in the Dirac sea created by Dirac sea electrons which made transitions to the positive energy spectrum. The holes are called positrons.
\begin{quote}
  The exclusion principle will operate to prevent a positive-energy
  electron ordinarily from making transitions to states of negative
  energy. It will be possible, however, for such an electron to drop
  into an unoccupied state of negative energy. In this case we
  should have an electron and positron disappearing simultaneously,
  their energy being emitted in the form of radiation. The converse
  process would consist in the creation of an electron and a
  positron from electromagnetic radiation.
\begin{flushright}
  \footnotesize{P.A.M.
    Dirac , Theory of the positron,
    in: The Principles of Quantum Mechanics, Oxford (1930)}
\end{flushright}
\end{quote}
Dirac's theory predicted the existence and properties of positrons, pair creation and pair annihilation, which shortly later were verified by Anderson \cite{Anderson:36}.

In the language of quantum field theory the Dirac sea is represented in the so-called second quantization procedure by the ``vacuum vector'' on which two types of creation operators act. Those creating electrons and those creating positrons. This way one implements Dirac's idea that one only considers the ``net description of  particles: electrons and positrons'' and neglects what is going on ``deep down in the sea'', assuming that nothing physically relevant happens in there. The Hilbert space for this many particle system is the \emph{Fock space} built by successive applications of creation operators on the vacuum.

As Dirac already pointed out in \cite{Dirac:34}, the Dirac sea, however, is inaccessible and the choice of $\ch_-$ in the presence of an external field is not obvious at all. In addition, Dirac's invention and likewise quantum field theory are plagued by a serious problem: As soon as an electromagnetic field $\sa=(\sa_\mu)_{\mu=0,1,2,3}=(\sa _0,-\vec\sa )$ enters the Dirac equation, i.e. as soon as ``interaction is turned on'',  one has generically transitions of negative energy wave functions to positive energy wave functions, i.e. pair creation and pair annihilation. For a mathematical proof of pair creation in the adiabatic regime see \cite{PicklDuerr:08-1, PicklDuerr:08-2}. While pair creation and annihilation is an observed phenomenon it nevertheless has mathematically a devastating side effect. Pictorially speaking, the negative energy states are ``rotated'' by the external field and thus develop components in the positive energy subspace. Thus the Dirac sea containing  infinitely many particles generically produces under the influence of an external field infinitely many electrons as soon as the field acts. Such a state does not anymore belong to the Fock space and there is no reason to hope that in general a lift of the one-particle Dirac time-evolution to this Fock space exists.

This problem manifests itself in the classical results as follows. Under reasonable assumptions on the external potential $\sa=(\sa_\mu)_{\mu=0,1,2,3}=(\sa _0,-\vec\sa )$ the one-particle Dirac Hamiltonian $H^{\sa(t)} =-i\alpha \cdot(\nabla-\vec\sa(t) )+\beta m + \sa(t) ^0$ generates a unitary time-evolution $U^{\sf A}:\ch\to\ch$ on the Hilbert space $\ch=L_2(\R^3,\C^4)$ of square integrable, $\C^{4}$ valued functions on $\R^{3}$.  Having Dirac's sea idea \cite{Dirac:34, dirac_discussion_1934} in mind one introduces a splitting of this Hilbert space $\ch$ into a Hilbert direct sum $\ch = \ch_{+}\oplus\ch_-$, where $\ch_{\pm}$ are the spectral subspaces of the free Dirac Hamiltonian $H^0$ and $P_{\pm}$ the corresponding orthogonal projectors. The ``states'' of $\ch_-$ are assumed to be filled with electrons, historically referred to as Dirac sea.  In modern quantum field theory the notion of the Dirac sea is replaced by the so called vacuum. In order to extract finite and physical meaningful expressions from this infinite particle picture Dirac's idea is to focus only on the ``net balance'' between the initial Dirac sea and the time-evolved Dirac sea while neglecting what is going on ``deep down in the seas''. Transitions between $\ch_{-}$ and $\ch_{+}$ are thought to describe pair creation and annihilation \cite{Dirac:34, dirac_discussion_1934}; for a mathematically rigorous proof of the creation of a pair of asymptotically free moving electron and positron in an adiabatically changing strong field see \cite{PicklDuerr:08-2,PicklDuerr:08-1}. The Fock space $\cf(\ch_+,\ch_-)=\Wedge \ch_+\otimes \Wedge \overline{\ch_-}$ serves as a mathematical setup for this infinite particle picture. One intends to lift the one-particle Dirac time-evolution $U^{\sf A}$ on $\ch$ to this Fock space. The Shale-Stinespring Theorem \cite{shale:65} gives the following necessary and sufficient condition for the existence of such a lift. For times $t_{0},t_{1}\in\R$ the one-particle Dirac time-evolution $U^{\sf A}(t_1,t_0)$ can be lifted to a second-quantized time-evolution on $\cf(\ch_+,\ch_-)$ if and only if the off-diagonal terms $U_{+-}^{\sf A}(t_{1},t_{0}):=P_+U^{\sf A}(t_{1},t_{0})P_-$ and $U_{-+}^{\sf A}(t_{1},t_{0}):=P_-U^{\sf A}(t_1,t_0)P_+$ are both Hilbert-Schmidt operators. Such a lift, if it exists, is unique up to a phase. Ruijsenaars \cite{Ruijsenaars:77-1,Ruijsenaars:77-2} supplied the physical implications of the Shale-Stinespring Theorem: the operators $U_{+-}^{\sf A}(t_{1},t_{0})$ and $U_{-+}^{\sf A}(t_{1},t_{0})$ are Hilbert-Schmidt operators for all times $t_{0},t_{1}$ if and only if $\vec{\sf A}=0$, a somehow devastating result. Not only it means there are no lifts of the one-particle Dirac time-evolution for external potentials with non-zero magnetic components, it also means that gauge transformations which add non-zero spatial components to the external potential cannot be implemented.

Let us give an intuition for the Shale-Stinespring condition. We regard the time-evolution $U^\sa(t_1,t_0):\ch_-\oplus\ch_+\to\ch_-\oplus\ch_+$
in matrix form:
\begin{align}
  U^\sa(t_1,t_0)=\begin{pmatrix}
    U^\sa_{++}(t_1,t_0) & U^\sa_{+-}(t_1,t_0)\\
    U^\sa_{-+}(t_1,t_0) & U^\sa_{--}(t_1,t_0)
  \end{pmatrix}.
\end{align}
The non-diagonal terms describe
pair creation and annihilation.
In leading order, for $U^\sa(t_1,t_0)$ close to the identity and neglecting multiple pair creation,
the squared Hilbert-Schmidt norm
\begin{align}\label{eqn:pair creation rate}
  \|U^\sa_{+-}(t_1,t_0)\|^2_\hs=
 \sum_{n\in\N}\|U^\sa_{+-}(t_1,t_0)\varphi_n\|^2_{\ch}
\end{align}
for any orthonormal basis $(\varphi_n)_{n\in\N}$ of $\ch_-$ may be interpreted as the probability of creating one pair from the Dirac sea. In this sense the Shale-Stinespring condition ensures that the pair creation probabilities are well-defined.

Metaphorically speaking: The negative energy states in $\ch_{-}$ are  ``rotated'' by the interaction term $-i\alpha\cdot\vec{\sf{A}}+\sf{A}^{0}$ of the Dirac Hamiltonian $H^{\sf A}$ and develop components in the positive energy subspace $\ch_{+}$ as soon as the field acts. While the term $\sf{A}^{0}$ induces only a mild rotation, the rotation induced by $-i\alpha\cdot\vec{\sf{A}}$ is strong due to the presence of the $\alpha$ matrix. The catastrophe of ill-defined pair creation probabilities happens as long as the field is acting. When the field is switched off, most of spinors are however rotated back into the ``free Dirac sea''. Therefore focusing on the scattering matrix only one expects that the off-diagonal of the S-matrix  $P_\pm S^\sa P_\mp$ consists of  Hilbert-Schmidt operators. Hence, a lift of the S-matrix to Fock space exists \cite{Bellisard:75,Bellisard:76}. This lift is, as we said, only unique up to a phase. In the scattering situation the initial Dirac sea is ``more or less''  restored, so that ingoing states in Fock space are transformed to outgoing states in the \emph{same} Fock space.

Since the S-matrix captures the asymptotic time-evolution it is desirable to interpolate the asymptotic free time-evolutions of scattering by a full time-evolution when the external field acts. Due to the catastrophic pair creation discussed above one must adjust the sea, i.e. the vacuum, so that the most of the spinors remain ``sea-vectors''. Therefore, one considers a second quantized Dirac time-evolution in the presence of an external field with respect to time-dependent reference polarizations in contrast to one fixed polarization $\ch_{-}$; we shall call all closed subspaces of $\ch$ with infinite dimension and codimension \emph{polarizations}. One implements the second quantized time-evolution on time-varying Fock spaces instead of only one fixed Fock space. Such an implementation was already described in \cite{fierz-scharf-79}, and this idea was further developed by Mickelsson \cite{mickelsson:98}. Mickelsson gives a time-dependent unitary transformation of the Dirac Hamiltonian $H^{\sf A}$ such that its off-diagonal parts become Hilbert-Schmidt operators. Furthermore, he identifies the missing phase of the second quantized time-evolution up to a remaining freedom. A related but different approach to this phase is described in \cite{scharf:86}.
}

{

\subsection{What this Paper is about}\label{sec:towards a solution}

The obstacle in implementing the second-quantized time-evolution on time-varying Fock spaces is that there is no canonical choice of polarization. To illustrate this consider $H^A $, the Dirac operator for a fixed, time independent four-vector potential $A$. The spectrum is in general not anymore as simple as in the free case and there is no canonical way of defining a polarization since a splitting splitting into subspaces $L_2(\mathbb{R}^3,\mathbb{C}^4)={\ch}_-^A \oplus {\ch}_+^A$ is reasonably arbitrary. The choice of polarization becomes in particular interesting when the external potential $\sa$ is time-dependent. (Four-vector potentials defined on space-time $\R^{4}$ are denoted by the sans serif letter $\sa$, while defined on the space $\R^{3}$ they are denoted by the italic letter $A$.) Suppose that at time $t_0$ the field is zero and at 
a later time the field is switched on. To better understand the issue of the choice of polarization, observe that choosing $U^\sa 
(t,t_0)\cal H_-$ as the polarization at time $t$ would not allow for the description of pair creation: starting from a Dirac sea in $\ch_{-}$ at time $t_{0}$ all one-particle wave functions will remain in the sea $U^\sa(t,t_0)\ch_{-}$ forever. Also, this choice of polarization depends not only on the field $\sa$ at time $t$ but also on the whole history $(\sa(s))_{s\leq t}$. We shall show below that for another field $\widetilde\sa$ with $\widetilde \sa(t_0)=\sa
(t_0)$ and $\widetilde \sa(t)=\sa(t)$ the orthogonal 
projectors onto $U^\sa (t,t_0)\ch_-$ and $U^{\widetilde\sa}(t,t_0)\ch_-$ differ by a Hilbert-Schmidt operator. Moreover, as discussed in \cite{fierz-scharf-79}, all apparent choices of polarizations like the negative spectral subspace of $H^{\sa}$ which allow for pair-creation are not Lorentz invariant.

This suggests that a particle/anti-particle picture can presumably not be based on spectral considerations. Instead of choosing specific polarizations we consider equivalence classes of polarizations. It turns out that an appropriate equivalence relation ``$ \approx$'' between polarizations is given by the condition that the difference of the corresponding orthogonal projectors is a Hilbert-Schmidt operator. This is in accordance with: First, the intuition described along with (\ref{eqn:pair creation rate}) which implies that transition amplitudes stay well-defined. And second, if $\sa(t_{0})=0$, the equivalence class $C(t)=[U^\sa(t,t_0)\ch_{-}]_{\approx}$, $t>t_{0}$, turns out to depend only on the external potential $\sa(t)$ at time $t$ but not on the history $(\sa(s))_{t_{0}< s< t}$. A specific choice of polarization in this equivalence class is then mathematically a choice of reference frame with respect to which we represent the second-quantized Dirac time-evolution. This brings us to the content of this work:

\begin{enumerate}[1.]
	\item We show that these polarization classes $C(t)$ are uniquely identified by the spatial (magnetic) components $\vec \sa(t)$ of the field $\sa$ at time $t$; see Theorem \pref{ident polclass}, Subsection \pref{sec:Ident Pol classes}. This generalizes the case of $\vec \sa(t)=0$ regarded in \cite{Ruijsenaars:77-1} to general $\vec \sa(t)$. 
	\item We give a simple representative $e^{Q^{\sa (t)}}\ch_-\in C(t)$ for each polarization class in terms of a simple and explicit operator $Q^{\sa(t)}$ which naturally appears as the key object in the  variant of the Born series of $U^\sa $ that we use in Subsection \pref{time-evolution}.
	\item We implement the Dirac time-evolution as unitary maps between between time-varying Fock spaces, in Theorem \pref{thm:second quant dirac}, Subsection \pref{sec:second quantized time-evolution}. This implementation is unique up to a phase.
	\item We conclude with a brief discussion of gauge transformations of the external field; see Theorem \pref{thm:gauge}, Subsection \pref{sec:gauge}.
\end{enumerate}

A next step would be to derive the polarization charge current within this framework which also must be defined in a neat way so that it accounts only for the ``net description'' comparing two Dirac seas as mentioned above.

Our work in this field of QED was mainly inspired by 
Dirac's original idea \cite{Dirac:34,dirac_discussion_1934}, the work of Fierz and Scharf \cite{fierz-scharf-79}, Scharf's book \cite{scharf:95} as well as Pressley and Segal's book \cite{pressley:86} and also the work of Mickelsson et al.
\cite{langmann:96,mickelsson:98}. Furthermore, we would like to call attention to an approach to QED by Finster \cite{finster_action_2009, finster_regularized_2008, finster_formulation_2009,finster_principle_2006} known under the name: ``The Fermionic Projector''. Though mathematically different, his approach also revisits Dirac's original idea \cite{dirac_discussion_1934,Dirac:34} in a serious way. The main difference between the Fermionic Projector approach and ours is that we do not seek a distinguished polarization or, in other words, a unique vacuum. Moreover, we use a Fock space description which by construction allows for superposition and entanglement which at the moment seems to be elaborate using ferminonic projectors.

\paragraph{The Setup.}  The purpose of this paragraph is to give a heuristic description of how we construct the second 
quantized time-evolution of the Dirac Hamiltonian in the presence of a time-dependent, external field. What is described in this Subsection will be rigorously introduced and proven in Sections \pref{sec:wedge spaces} and \pref{time-evolution_ex}. The definitions and assertions will later be formulated in a general form.

Since we aim at a description depending only on polarization classes instead of specific polarizations we resort to a representation of the Fock space which is different to the standard one ($\cf(\ch_{-},\ch_{+})$). We shall refer to it as the \emph{infinite wedge product spaces}. Although our results can be rephrased in the standard Fock space language, the infinite wedge product formalism, in our opinion, is closer to Dirac's original idea \cite{Dirac:34, dirac_discussion_1934} and opens up a more transparent view on the nature of the second-quantized Dirac time-evolution and on the role of the Dirac sea (i.e. Fock vacuum).

We construct Dirac seas
concretely as infinite wedge products like in \cite{Dirac:34}: given a polarization $V\subset\ch$ and for that an 
orthonormal basis $\varphi=(\varphi_n)_{n\in\N}$ that spans $V$ the alternating 
product of all $\varphi_n$, $n\in\N$, is supposed to represent a Dirac sea belonging to polarization $V
$. We introduce an equivalence class $\cs=\cs(\varphi)$ of other 
representatives, namely of all sequences $\psi=(\psi_n)_{n\in\N}$ in $\ch$ such 
that the $\N\times \N$-matrices
\begin{align}
   (\sk{\psi_n,\psi_m})_{n,m\in\N} && \text{and} && (\sk{\psi_n,\varphi_m})_{n,m
\in\N},
\end{align}
$\sk{\cdot,\cdot}$ denoting the inner product on $\ch$, differ from the unity 
matrix only by a matrix in the trace class and thus have a determinant. In this 
case we write $\psi\sim\phi$. This is our notion of Dirac seas being asymptotically equal 
``deep down in the sea". We define the following bracket:
\begin{align}\label{eqn:inn prod}
  \sk{\psi,\chi}:= \det(\sk{\psi_n,\chi_m})_{n,m\in\N}=\lim_{k\to\infty}\det
(\sk{\psi_n,\chi_m})_{n,m=1,\ldots, k},\quad \psi,\chi\in \cs.
\end{align}
With this at hand one constructs a Hilbert space $\cf_{\cs}=\cf_{\cs(\varphi)}$, 
where the bracket gives rise to the inner product. We refer to $\cf_\cs$ as the 
\emph{infinite wedge space}. By this construction, see Definition \pref{def 
dach}, a sequence $\psi\in \cs$ is mapped to the wedge product $\Wedge\psi=
\psi_1\wedge\psi_2\wedge\psi_3\wedge\ldots\in\cf_\cs$. The rigorous construction 
of wedge spaces is carried out in Subsection \pref{sec:construction}; in Section \pref{comparison} we also discuss the relationship of $\cf_{\cs(\varphi)}$ 
to the standard Fock space. It is important to note that $\Wedge\varphi$ has no meaning as ``the one and only'' Dirac sea since $\cf_\cs$ depends only on the equivalence class $\cs=\cs(\varphi)$. In fact, changing the reference $\varphi$ within the same equivalence class $\cs$ can be viewed as a Bogolyubov transformation. 

The equivalence relation $\approx$ between two polarizations will be refined as follows: For two polarizations $V,W$ we define $V\approx_0 W$ to mean $V\approx W$ 
and that $V$ and $W$ have the same ``relative charge''. Intuitively the 
``relative charge'' has the following meaning: Consider two states $\Wedge
\varphi$ and $\Wedge \psi$ where $\varphi$ and $ \psi$ are orthonormal bases of 
$V$ and $W$, respectively. Then the relative charge is the difference of the 
electric charges of the physical sates represented by $\Wedge\varphi$ and $
\Wedge\psi$, respectively. Mathematically the relative charge is defined in 
terms of Fredholm indices in Definition \pref{charge}. The use of the Fredholm 
index to describe the relative charge is quite frequent in the literature; see 
e.g. \cite{pressley:86, langmann:96} as well as in the work of Hainzl et al. \cite{hainzl:05}. The relation $\approx_0$ is 
also an equivalence relation on the set of polarizations, and one finds an 
intimate connection between this equivalence relation $\approx_0$ on the set of 
polarizations and the equivalence relation $\sim$ on the set Dirac seas: Two 
equivalent Dirac seas span two equivalent polarizations and for every two 
polarizations $W\approx_0 V$ such that $\varphi$ spans $V$ there is a 
Dirac sea $\Wedge\psi\in\cf_{\cs(\varphi)}$ such that $\psi\sim\varphi$ and 
$\psi$ spans $W$. Consequently, every wedge space can be associated with a
polarization class with respect to $\approx_0$. Details are given in Section 
\pref{sec:construction}.

On the other hand, assuming $\varphi$ spans $V$, not all Dirac seas $\Wedge\psi$ 
such that $\psi$ spans $W\approx_0 V$ are in $\cf_{\cs(\varphi)}$ because one 
can obviously find an orthonormal basis $\psi$ of $W$ for which $(\sk{\psi_n,
\varphi_m})_{n,m\in\N}$ differs from the identity by more than a trace class 
operator. Because of this we consider below operations (the operations from 
the right) that mediate between all wedge spaces belonging to the same 
polarization class with respect to $\approx_0$. These operations are needed to 
later define the physically relevant transition probabilities.

On any element of $\cs$  the action of any unitary map $U$ on $\ch$ is then 
naturally defined by having it act on each factor of the wedge product. 
Consequently we have a (left) operation on any $\cf_{\cs}$, namely $\lop U:\cf_
{\cs}\to\cf_{US}$, such that
\begin{align}
  \lop U\left(\psi_1\wedge\psi_2\wedge\psi_3\wedge\ldots\right)=U\psi_1\wedge U
\psi_2\wedge U\psi_3\wedge\ldots,\quad \psi\in \cs,
\end{align}
which then incorporates a ``lift'' of $U$ as a unitary map from one wedge space 
to another. Now, this can of course also be done for the one-particle time 
evolution $U=U^\sa (t,t_0)$ for fixed times $t_0$ and $t$. However, we need to 
find a way to relate the Dirac seas in $\cf_\cs$ to the ones in $\cf_{U\cs}$ by 
considering the ``net balance'' between them. As we discussed already what we mathematically have at hand are the polarization classes $C(t_{0})$ and $C(t)$ at times $t_0$ and $t$. We 
choose any two polarizations $V\in C(t_{0})$ and $W\in C(t)$ and orthonormal bases $\varphi(t_0)$ of $V$ and likewise $
\varphi(t)$ of $W$ and denote their equivalence classes with 
respect to $\sim$ by $\cs(\varphi(t_0))$ and $\cs(\varphi(t))$. This way 
physical ``in'' and ``out'' states can be described by elements in $\psi^
{\inn}\in\cf_{\cs
(\varphi(t_0))}$ and $\psi^
{\out}\in\cf_{\cs(\varphi(t))}$, respectively. But in general $U\cs
(\varphi(t_0))$ will not be equal $\cs(\varphi(t))$ so that $\lop U\psi^
{\inn}$ and $\psi^{\out}$ are likely to lie in different wedge spaces. However, we show that 
the polarization classes of $\cf_{U\cs(\varphi(t_0))}$ and $\cf_{\cs(\varphi
(t))}$ are the same. Therefore, the only difference between those two spaces may 
come from our specific choice of bases $\varphi(t_0)$ and $\varphi(t)$. In order 
to make them compatible we introduce another operation (from the right): For all 
unitary $\N\times\N$-matrices $R=(R_{nm})_{n,m\in\N}$, we define the operation 
from the right as follows. For $\psi\in \cs$, let $\psi R:=\left(\sum_{n\in\N}
\psi_n R_{nm}\right)_{m\in\N}$. In this way, every unitary $R$ gives rise to a 
unitary map $\rop R:\cf_{\cs}\to\cf_{\cs R}$, such that
\begin{align}\label{eqn:right op}
  \rop R \Wedge\psi=\Wedge(\psi R),\quad \psi\in \cs.
\end{align}
By construction the operations from the left and from the right commute. 
We show that two such unitary matrices $R,R'$ yield $\cf_{\cs R}=\cf_{\cs R'}$ 
if and only if $R^{-1}R'$ has a determinant. Furthermore, in Subsection \pref{time-evolution} we prove that there always exists a unitary matrix $R$ for which $U\cs(t_0)R=
\cs(t)$. Now we have all we need to compute the transition probabilities:
\begin{align}\label{eqn:trans amp}
  |\sk{\psi^{\out},\rop R\circ \lop U\psi^{\inn}}|^2.
\end{align}
The matrix $R$ is not unique because for any $R'$ with $\det R'=1$ one 
has $U\cs(t_0)R=U\cs(t_0)RR'=\cs(t)$ so that the arbitrariness in $R'$ gives 
rise to a phase. However, this has clearly no effect on the transition probabilities. 

As an example let us consider $\psi^{\inn}$ and $\psi^{\out}$ being Hartree-Fock states: Let $V\in C(t_{0})$ and $W\in C(t_{1})$ be appropriate polarizations describing the one-particle wave functions present in a given experimental setup. Furthermore, let $\varphi$ and $\psi$ to be orthogonal bases in $V$ and $W$, respectively, and $\psi^{\inn}=\Wedge\varphi$ as well as $\psi^{\out}=\Wedge\phi$. Using (\pref{eqn:trans amp}), (\pref{eqn:inn prod}) and the notation $|A|^{2}=A^{*}A$ one can express the transition probability as
\begin{align}
	|\sk{\Wedge\psi,\rop R\circ \lop U\Wedge\varphi}|^2&=
	|\det \sk{\psi_{n},(U\varphi R)_{m}}_{n,m}|^{2}\\
	&=\det |P_{W}UP_{V}|^{2}\big|_{V\to V}=\det \left(1-|P_{W^\perp}UP_{V}|^{2}\right)\big|_{V\to V}.
\end{align}
where we have used that $R$ and $U$ are unitary. This determinant is well-defined whenever its argument differs from the identity only by a trace class operator. Hence, the above expression is only well-defined for $P_{W^{\perp}}UP_{V}$ being a Hilbert-Schmidt operator. The leading order of this determinant is given by $1-\|P_{W^\perp}UP_{V}\|_{I_{2}}$, which agrees with the intuition given along  (\pref{eqn:pair creation rate}). The operations from the left and from the right are introduced in Subsection \pref{sec:left and right op} while in Subsection \pref{sec:lift cond} we identify the conditions under which $R$ exists.

The next step would be to derive the charge current of the created pairs. It must be also defined in a neat way so that it accounts only for the ``net description'' and not for a whole Dirac sea. For doing that one would e.g. need to analyze the behavior of the phase, cf. \cite{scharf:95}, which will be the content of our subsequent paper.
}

\section{Infinite Wedge Spaces}\label{sec:wedge spaces}

\subsection{Construction}
\label{sec:construction}

In this section we give a rigorous construction of infinite wedge products which we
described in the introduction. Throughout this work the notion \emph{Hilbert space} stands for \emph{separable, infinite dimensional, complex Hilbert space}. Let $\ch$\nomenclature{$\ch$}{One-particle Hilbert space} and $\ell$\nomenclature{$\ell$}{Index Hilbert space} be Hilbert spaces with corresponding scalar products $\sk{\cdot,\cdot}$\nomenclature{$\sk{\cdot,\cdot}$}{Scalar products in respective Hilbert spaces}. For a typical example think of $\ch=L_2(\R^3,\C^4)$ and $\ell=\ell_2(\N)$\nomenclature{$\ell_2(\N)$}{The space of square summable sequences in $\C$}, the space of square summable sequences in $\C$. The space $\ell$ will only play the role of an index space. We refer to $\ch$ as the \emph{one-particle} Hilbert space. Furthermore, we denote the space of so-called \emph{trace class} operators on $\ell$, i.e. bounded operators $T$ on $\ell$ for which $\|T\|_{\trcl} :=\trace \sqrt{T^*T}$\nomenclature{$\nomdsep\cdot\nomdsep_{\trcl}$}{Trace class norm} is finite, by $\trcl(\ell)$, the superscript $*$\nomenclature{$T^*$}{The Hilbert space adjoint of the operator $T$} denoting the Hilbert space adjoint. We say a bounded linear operator $T$ on a Hilbert space $\ell$ \emph{has a determinant}\nomenclature{$\det T$}{The Fredholm determinant of $T$} if it differs from the identity operator $\id_\ell$\nomenclature{$\id_\ell$}{Identity operator on some space $\ell$} on $\ell$ only by a trace class operator, i.e. $T-\id_\ell\in \trcl(\ell)$\nomenclature{$\trcl=\trcl(\ell)$}{Space of trace class operators on $\ell$}; see \cite{gohberg-90}. We also need the space of Hilbert-Schmidt operators, i.e. the space of bounded operators $T:\ell\to\ch$ such that the Hilbert-Schmidt norm $\|T\|_{\hs}:=\sqrt{\trace T^*T}$ is finite. The space of Hilbert-Schmidt operators is denoted by $\hs=\hs(\ell,\ch)$\nomenclature{$\hs=\hs(\ell,\ch)$}{The space of Hilbert-Schmidt operators $T:\ell\to\ch$}\nomenclature{$\nomdsep\cdot\nomdsep_\hs$}{Hilbert-Schmidt norm}, and we write $\hs(\ch)=\hs(\ch,\ch)$.

At first let us define the notions: polarizations, polarization classes and the set of Dirac seas from the introduction.
\begin{our_definition}[Polarizations and Polarization Classes]
\begin{enumerate}
\item
Let $\pol(\ch)$\nomenclature{$\pol(\ch)$}{The set of all polarizations $V\subset\ch$} denote the set of all closed, linear
subspaces $V\subset \ch$ such that $V$ and $V^\perp$
are both infinite dimensional. Any $V\in \pol(\ch)$\nomenclature{$V,W,$ etc.}{Typical polarizations, i.e. elements of $\pol(\ch)$} is called a
{\em polarization} of $\ch$.
For $V\in \pol(\ch)$, let $P_V:\ch\to V$\nomenclature{$P_V:\ch\to V$}{Orthogonal projector of $\ch$ on the subspace $V$} denote the orthogonal projection
of $\ch$ onto $V$.
\item
For $V, W\in \pol(\ch)$, $V\approx W$\nomenclature{$V\approx W$}{$\Leftrightarrow P_V-P_W\in\hs(\ch)$ for $V,W\in \pol(\ch)$} means
$P_V-P_W\in \hs(\ch)$.
\end{enumerate}
\end{our_definition}
The space $\pol(\ch)$ is a kind of Grassmann space of all infinite dimensional closed subspaces with infinite dimensional complement. Obviously, the relation $\approx$ is an equivalence relation on $\pol(\ch)$. Its equivalence classes $C\in \pol(\ch)/{\approx}$\nomenclature{$C,C',$ etc.}{Typical elements of polarization classes} are called {\it polarization classes}. Its basic properties are collected in the following lemma. We denote by $|_{X\to Y}$\nomenclature{$\nomsep_{X\to Y}$}{Restriction to the map $X\to Y$} the restriction to a map from $X$ to $Y$.
\begin{our_lemma}[Properties of $\approx$]
\label{lemma approx}
For $V,W\in \pol(\ch)$, the following are equivalent:
\begin{enumerate}[(a)]
\item
$V\approx W$
\item $P_{W^\perp} P_V\in \hs(\ch)$ and
$P_W P_{V^\perp}\in \hs(\ch)$
\item \label{lemma_approx_c}The operators
$P_V P_WP_V|_{V\to V}$ and $P_W P_VP_W|_{W\to W}$
both have determinants.
\item The operators $P_V P_WP_V|_{V\to V}$ and
$P_{V^\perp} P_{W^\perp}P_{V^\perp}|_{V^\perp\to V^\perp}$
both have determinants.
\item \label{lemma approx e}$P_W|_{V\to W}$ is a Fredholm operator and $P_{W^\perp}|_{V\to W^\perp}\in\hs(V)$.
\end{enumerate}
\end{our_lemma}
\begin{proof}
\mbox{}
\begin{list}{}{}
  \item[(a)$\Rightarrow$(b):] Let $V,W\in \pol(\ch)$ fulfill $P_V-P_W\in
    \hs(\ch)$. We conclude that
    \begin{align}
      P_{W^\perp}P_V&=(\id_\ch-P_W)P_V=(P_V-P_W)P_V\in \hs(\ch) \;\text{and}\\
      P_WP_{V^\perp}&=P_W(\id_\ch-P_V)=-P_W(P_V-P_W)\in \hs(\ch).
    \end{align}
\item[(b)$\Rightarrow$(c):]
  Assuming (b), we conclude
  \begin{align}
    P_V-P_VP_WP_V&=
    P_VP_{W^\perp} P_V=(P_{W^\perp} P_V)^*(P_{W^\perp} P_V)\in \trcl(\ch) \;\text{and}\\
    P_W-P_WP_VP_W&=
    P_WP_{V^\perp} P_W=(P_WP_{V^\perp})^*(P_WP_{V^\perp})\in \trcl(\ch).
  \end{align}
  This implies  $(P_V-P_VP_WP_V)|_{V\to V}\in \trcl(V)$
  and $(P_W-P_WP_VP_W)|_{W\to W}\in \trcl(W)$ and thus the claim (c).
\item[(c)$\Rightarrow$(d):]
  Assuming (c), we need to show that
  $P_{V^\perp} P_{W^\perp}P_{V^\perp}|_{V^\perp\to V^\perp}$
  has a determinant. Indeed:
  As $P_WP_VP_W|_{W\to W}$ has a determinant,
  we know that
  \begin{align}
  (P_{V^\perp}P_W)^*(P_{V^\perp}P_W)=P_WP_{V^\perp}P_W=P_W-P_WP_VP_W
  \in {\trcl}(\ch)
  \end{align}
  and thus $P_{V^\perp}P_W\in \hs(\ch)$.
  This implies
  \begin{align}
  P_{V^\perp}-P_{V^\perp} P_{W^\perp}P_{V^\perp}
  =
  P_{V^\perp} P_WP_{V^\perp}=(P_{V^\perp}P_W)(P_{V^\perp}P_W)^*\in { \trcl}(\ch).
  \end{align}
  The claim $(P_{V^\perp} P_{W^\perp}P_{V^\perp}|_{V^\perp\to V^\perp})
  \in\dcl{V^\perp}$ follows.
\item[(d)$\Rightarrow$(b):] Assuming (d), we know
  \begin{align}
  &(P_WP_{V^\perp})^*(P_WP_{V^\perp})=
  P_{V^\perp}P_WP_{V^\perp}=P_{V^\perp}-P_{V^\perp}P_{W^\perp}
  P_{V^\perp}\in \trcl(\ch) \;\text{and}\\
  &(P_{W^\perp}P_V)^*(P_{W^\perp}P_V)=
  P_VP_{W^\perp}P_V=P_V-P_VP_WP_V\in \trcl(\ch).
  \end{align}
  This implies the claim (b).
\item[(b)$\Rightarrow$(a):] Assuming $P_{W^\perp} P_V\in \hs(\ch)$ and
  $P_W P_{V^\perp}\in \hs(\ch)$, we conclude
  that
  \begin{align}
  P_V-P_W&=(P_V-P_WP_V)-(P_W-P_WP_V)\nonumber \\
  &=P_{W^\perp}P_V-P_WP_{V^\perp}\in  \hs(\ch).
  \end{align}
\item[(b)$\Rightarrow$(e):] We write the identity on $\ch$ in matrix form
    \begin{align}
      \id_\ch:V\oplus V^\perp\to W\oplus W^\perp, && (x,y)\mapsto \begin{pmatrix}
        P_W|_{V\to W} & P_W|_{V^\perp\to W}\\
        P_{W^\perp}|_{V\to W^\perp} & P_{W^\perp}|_{V^\perp\to W^\perp}
      \end{pmatrix}\begin{pmatrix}
        x\\
        y
      \end{pmatrix}.
    \end{align}
    Assuming (b) we know that the non-diagonal operators $P_{W^\perp}|_{V\to W^\perp}$ and $P_{W}|_{V^\perp\to W}$ are Hilbert-Schmidt operators. Subtracting the non-diagonal from the identity we get a new map
    \begin{align}
      Q:V\oplus V^\perp\to W\oplus W^\perp, && (x,y)\mapsto \begin{pmatrix}
        P_W|_{V\to W} & 0\\
        0 & P_{W^\perp}|_{V^\perp\to W^\perp}
      \end{pmatrix}\begin{pmatrix}
        x\\
        y
      \end{pmatrix}
    \end{align}
    which is by construction a perturbation of the identity by a compact operator and, thus, a Fredholm operator. However, this holds if and only if both $P_W|_{V\to W}$ and $P_{W^\perp}|_{V^\perp\to W^\perp}$ are Fredholm operators which implies (e).
\item[(e)$\Rightarrow$(b):] Assuming (e) we compute
    \begin{align}
      0&=P_{V^\perp}|_V=P_{V^\perp}(P_W+P_{W^\perp})|_V\nonumber\\
      &=P_{V^\perp}|_W\;P_W|_{V\to W} + P_{V^\perp}|_{W^\perp}\;P_{W^\perp}|_{V\to W^\perp}
    \end{align}
    from which follows that $P_{V^\perp}|_W\;P_W|_V$ is a Hilbert-Schmidt operator since $P_{V^\perp}|_{W^\perp}$ is a bounded operator and by assumption $P_{W^\perp}|_V\in\hs(V)$. Furthermore, by assumption $P_W|_{V\to W}$ is a Fredholm operator so that $P_{V^\perp}|_W\;P_W|_{V\to W}\in\hs(V)$ yields $P_{V^\perp}|_W\in\hs(W)$. Finally, we have $P_{W^\perp}|_V=P_{W^\perp}P_V|_V$ and $P_{V^\perp}|_W=P_{V^\perp}P_W|_W$ so that $P_{W^\perp}P_V, P_{V^\perp}P_W\in\hs(\ch)$ which implies the claim (b).
\end{list}
\end{proof}

Note that in general $P_{W^\perp} P_V\in \hs(\ch)$ is not equivalent to  $P_W P_{V^\perp}\in \hs(\ch)$. As an example, take $V$ and $W$ such that $V\subset W$ and $V$ has infinite codimension in $W$; compare with condition (\ref{lemma approx e}) of the above lemma. Condition (\ref{lemma approx e}) appears in Chapter 7 of \cite{pressley:86} where an equivalence class $C\in\pol(\ch)/{\approx}$ is endowed with the structure of a complex manifold modeled on infinite dimensional separable Hilbert spaces. Consequently, the space $\pol(\ch)$ is a complex manifold -- the Grassmann manifold of $\ch$ -- which decomposes into the equivalence classes $C\in\pol(\ch)/{\approx}$ as open and closed submanifolds.

Where exactly the one-particle Hilbert space $\ch$ is cut into parts by a choice of a polarization in a polarization class will determine the relative charge between two Dirac seas. Within one polarization class the charge may only differ by an integer from one chosen polarization to another. Given $V,W\in \pol(\ch)$ with $V\approx W$,  we know from Lemma \pref{lemma approx}(e) that $P_W|_{V\to W}$ and $P_V|_{W\to V}$ are Fredholm operators. So we are led to the definition of the \emph{relative charge}:
\begin{our_definition}[Relative Charge]\label{charge}
For $V,W\in \pol(\ch)$ with $V\approx W$,
we define the {\em relative charge}
of $V,W$ to be the Fredholm index of $P_{W}|_{V\to W}$\nomenclature{$|_V$}{The according map restricted to the set $V$}\nomenclature{$\charge(V,W)$}{Relative charge between two polarizations, i.e. the Fredholm index $\ind(P_W|_{V\to W})$}\nomenclature{$\ind T$}{Fredholm index of $T$}:
\begin{align}
\charge(V,W)&:=\ind(P_W|_{V\to W})
=\dim\ker(P_W|_{V\to W})
- \dim\ker(P_W|_{V\to W})^*
\nonumber\\&
=\dim\ker(P_W|_{V\to W})
- \dim\coker(P_W|_{V\to W})
.
\end{align}
\end{our_definition}
Let $\cu(\ch,\ch')$\nomenclature{$\cu(\ch,\ch')$}{Set of unitary maps $U:\ch\to\ch'$} be the set of unitary operators $U:\ch\to\ch'$. We collect some basic properties of the relative charge:
\begin{our_lemma}[Relative Charge Properties]
\label{lemma charge}
Let $C\in \pol(\ch)/{\approx}$ be a polarization class and $V,W,X\in C$. Then the following hold:
\begin{enumerate}
\item $\charge(V,W)=-\charge(W,V)$
\item $\charge(V,W)+
\charge(W,X)=\charge(V,X)$
\item
Let $\ch'$ be another Hilbert space and $U\in \cu(\ch,\ch')$.
Then
$\charge(V,W)=\charge(UV,UW)$.
\item
Let $U\in \cu(\ch,\ch)$ such that $UC=C$. Then
$\charge(V,UV)=\charge(W,UW)$.
\end{enumerate}
\end{our_lemma}
\begin{proof}
\begin{enumerate}
\item[(a)] $P_W|_{V\to W}$ and $P_V|_{W\to V}$ are Fredholm operators
with
\begin{align}
&\charge(W,V)+\charge(V,W)=
\ind(P_V|_{W\to V})+\ind(P_W|_{V\to W})
\nonumber\\&
=\ind(P_VP_W|_{V\to V})
=\ind(P_VP_WP_V|_{V\to V})=0,
\end{align}
as $P_VP_WP_V|_{V\to V}$ is a perturbation of the identity map on $V$ by a compact operator.
\item[(b)]
As $P_W$ and $P_X$ differ only by a compact operator, we get
\begin{align}
&\charge(V,W)+\charge(W,X)
=
\ind(P_W|_{V\to W})+\ind(P_X|_{W\to X})
\nonumber\\&
=\ind(P_XP_W|_{V\to X})=\ind(P_XP_X|_{V\to X})=\charge(V,X).
\end{align}
\item[(c)]
This follows immediately, since unitary transformations do not change the
Fredholm index.
\item[(d)]
We know $UV\approx V\approx W\approx UW$ by assumption.
Using parts (a), (b) and (c) of the lemma, this implies
\begin{align}
&
\charge(V,UV)=\charge(V,W)+\charge(W,UW)+\charge(UW,UV)
\nonumber\\&
=\charge(V,W)+\charge(W,UW)+\charge(W,V)=\charge(W,UW).
\end{align}
\end{enumerate}
\mbox{}
\end{proof}
With the notion of relative charge we refine the polarization classes further into classes of polarizations of equal relative charge:
\begin{our_definition}[Equal Charge Classes]
For $V,W\in \pol(\ch)$, $V\approx_0 W$\nomenclature{$V\approx_0 W$}{$\Leftrightarrow V\approx W$ and $\ind(V,W)=0$ for $V,W\in \pol(\ch)$} means $V\approx W$ and $\charge(V,W)=0$.
\end{our_definition}
By Lemma \pref{lemma charge} (Relative Charge Properties) the relation $\approx_0$ is an equivalence relation on $\pol(\ch)$. This finer relation is better adapted for the lift of unitary one-particle operators like the Dirac time-evolution which conserve the charge.

Next, we introduce the mathematical representation of the Dirac seas:
\begin{our_definition}[Dirac Seas]
\label{def-M}
\begin{enumerate}[(a)]
  \item Let $\seas(\ch)=\seas_\ell(\ch)$\nomenclature{$\seas(\ch)=\seas_\ell(\ch)$}{All possible Dirac seas, e.g. bounded operators $\Phi:\ell\to\ch$ such that $\range \Phi\in\pol(\ch)$ and $\det\Phi^*\Phi$ exists.} be the
    set of all linear, bounded operators $\Phi:\ell\to \ch$ such that $\range \Phi\in\pol(\ch)$ and $\Phi^* \Phi:\ell\to\ell$ has a determinant, i.e. $\Phi^* \Phi\in\dcl\ell$.
  \item
    Let $\iseas(\ch)=\iseas_\ell(\ch)$\nomenclature{$\iseas(\ch)=\iseas_\ell(\ch)$}{Only Dirac seas out of $\seas(\ch)$ which also are isometries} denote the
    set of all linear isometries $\Phi:\ell\to \ch$ in $\seas_\ell(\ch)$.
  \item For any $C\in\pol(\ch)/{\approx}_0$ let $\ocean(C)=\ocean_\ell(C)$\nomenclature{$\ocean(C)=\ocean_\ell(C)$}{The set of all $\Phi\in\iseas_\ell(\ch)$ such that $\range\Phi\in C$} be the set of all $\Phi\in\iseas_\ell(\ch)$ such that $\range\Phi\in C$.
\end{enumerate}
\end{our_definition}
Figuratively speaking, an ocean consists of a collection of related seas. To connect to the introduction in Subsection \pref{sec:towards a solution} consider the following example: In the case of $\ell=\ell_2(\N)$ we encode this map in an orthonormal basis $(\varphi_n)_{n\in\N}$ of $V$ such that for the canonical basis $(e_n)_{n\in\N}$ in $\ell^2$ one has $\Phi e_n=\varphi_n$ for all $n\in\N$.

The set $\seas(\ch)$ can naturally be structured by the relation introduced now:
\begin{our_definition}[Relation between Dirac Seas]\label{def:dirac seas equiv}
For $\Phi,\Psi\in\seas(\ch)$,
$\Phi\sim \Psi $\nomenclature{$\Phi\sim\Psi$}{$\Leftrightarrow \Psi^*\Phi\in 1+\trcl(\ell)$} means $\Phi^* \Psi \in\dcl\ell$, i.e.
$\Phi^*\Psi $ has a determinant.
\end{our_definition}
In the forthcoming Corollary \pref{cor:equivalencerel} we show that $\sim$ is an equivalence relation. For its proof we need the following lemma, which will also be frequently used later because it allows us to work for most purposes with Dirac seas in $\iseas(\ch)$ instead of $\seas(\ch)$:
\begin{our_lemma}[Isometries are good enough]
\label{polarlemma}
For every $\Psi\in \seas(\ch)$ there exist $\Upsilon\in \iseas(\ch)$
and $R\in \dcl\ell$ { which fulfill} $\Psi=\Upsilon R$, $\Upsilon^* \Psi=R\ge 0$, $\Upsilon\sim \Psi$, and $R^2=\Psi^*\Psi$.
\end{our_lemma}
\begin{proof}
Let $\Psi\in \seas(\ch)$. The operator $\Psi^*\Psi:\ell\to\ell$ has a determinant
and is hence a Fredholm operator. In particular, $\ker (\Psi^*\Psi)=
\ker \Psi$ is finite dimensional.
Let $\Psi=VR$ be the polar decomposition of $\Psi$, with $R=\sqrt{\Psi^*\Psi}$ and
$V:\ell\to \ch$ being a partial isometry with $\ker V=
(\range R)^\perp=\ker \Psi$.
Then $V$ and $\Psi$ have the same range, and this range has infinite
codimension in $\ch$. Since $\ker V$ has finite dimension,
we can extend the restriction of $V$ to $(\ker V)^\perp$
to an isometry $\Upsilon:\ell\to \ch$. We get:
$\Upsilon^* \Psi= V^* \Psi=V^* VR=R\ge 0$
and $\Upsilon R=VR=\Psi$. Now, as $R^2=\Psi^*\Psi$
has a determinant,
its square root $R$ has also a determinant. This implies $\Upsilon\sim \Psi$.\\
\mbox{}
\end{proof}
\begin{our_corollary}\label{cor:equivalencerel}
The relation $\sim$ is an equivalence relation on $\seas(\ch)$.
\end{our_corollary}
\begin{proof}
By definition of $\seas(\ch)$, the relation $\sim$ is reflexive.
To show symmetry, take
$\Phi,\Psi \in \seas(\ch)$ with $\Phi\sim \Psi $. We conclude
$\Psi ^* \Phi-\id_\ell= (\Phi^* \Psi -\id_\ell)^* \in \trcl(\ell)$ and thus
$\Psi \sim \Phi$.
To show transitivity, let $\Phi,\Psi ,\Gamma \in \seas(\ch)$ with $\Phi\sim \Psi $ and $\Psi \sim \Gamma $.
By Lemma \pref{polarlemma} (Isometries are good enough), take $\Upsilon \in\iseas(\ch)$ and $R \ge 0$
corresponding to $\Psi $.
Let $P:\ch\to \ch$ denote the orthogonal projection having the same
range as $\Upsilon $, and let $P^c=\id_\ch-P$ denote the
complementary projection. In particular, one has $P=\Upsilon \Upsilon ^*$.
Then
\begin{equation}
\label{eins}
\Phi^*\Gamma = \Phi^*P \Gamma +\Phi P^c\Gamma = (\Phi^*\Upsilon )(\Gamma ^* \Upsilon )^*+\Phi^*P^c\Gamma.
\end{equation}
Now, since $\Phi\sim \Psi $ we know
that $\Phi^*\Psi =\Phi^*\Upsilon R $ has a determinant. Since $R $ has also a determinant,
we conclude that $\Phi^*\Upsilon $ has a determinant, too.
Using $\Psi\sim\Gamma$, the same argument shows that $\Gamma ^*\Upsilon $ has a determinant,
and thus { $(\Phi^*\Upsilon )(\Gamma^*  \Upsilon )^*$} has a determinant.
Next we show that $P^c\Gamma $ is a Hilbert-Schmidt operator.
Indeed, $(P^c\Gamma )^*(P^c\Gamma )=\Gamma ^*P^c\Gamma =\Gamma ^*\Gamma -\Gamma ^*P \Gamma
=\Gamma ^*\Gamma -(\Gamma ^* \Upsilon )(\Gamma ^* \Upsilon )^*$ is a difference of two
operators having a determinant, since $\Gamma ^*\Gamma $ and $\Gamma ^* \Upsilon $ both have determinants. Hence, $(P^c\Gamma )^*(P^c\Gamma )\in \trcl(\ell)$,
which implies $P^c\Gamma \in \hs(\ell)$.
The same argument, applied to $\Phi$ instead of $\Gamma $, shows that
$P^c\Phi\in \hs(\ell)$. We conclude
$\Phi^*P^c \Gamma =(P^c\Phi)^*(P^c\Gamma )\in \trcl(\ell)$.
Using (\pref{eins}) this yields that $\Phi^*\Gamma $ has a determinant, as
the sum of an operator having a determinant and a trace class operator. This proves that $\Phi\sim\Gamma$.
\end{proof}
For $\Phi\in \seas(\ch)$, the equivalence class of $\Phi$ with respect to $\sim$
turns out to form an affine space. The following definition and lemma characterize these equivalence classes. These properties will later be used to show that the wedge spaces to be constructed (in forthcoming Definition \pref{def dach} (Infinite Wedge Spaces)) are separable spaces.
\begin{our_definition}[Dirac Sea Classes]\label{def:equiclass}
Let $\Phi\in\seas(\ch)$.
\begin{enumerate}
  \item Let $\cs(\Phi)\subset\seas(\ch)$\nomenclature{$\cs(\Phi)$}{The equivalence class w.r.t. $\sim$} denote the equivalence class of $\Phi$ with respect to $\sim$.
  \item For bounded linear, operators $L:\ell\to \ch$, we define $\|L\|_\Phi:=\|\Phi^* L\|_\trcl+\|L\|_\hs$\nomenclature{$\nomdsep\cdot\nomdsep_\Phi$}{The norm $\nomdsep\cdot\nomdsep_\Phi:=\nomdsep\cdot\nomdsep_\trcl+\nomdsep\cdot\nomdsep_\hs$} and the vector space
      \begin{align*}
        \cv(\Phi):=\{L:\ell\to\ch \;|\; L\text{ is linear and bounded with }\|L\|_\Phi<\infty\}.
      \end{align*}
      \nomenclature{$\cv(\Phi)$}{For all $\Phi\in\seas(\ch)$ we have $\cs(\Phi)=\Phi+\cv(\Phi)$}
\end{enumerate}
\end{our_definition}
\begin{our_lemma}[Dirac Sea Class Properties]\label{lem:equiclass_characterization}
Let $\Phi\in\seas(\ch)$.
\begin{enumerate}[(a)]
  \item It holds that $\cs(\Phi)=\Phi+\cv(\Phi)$.
  \item For $\Psi\in \seas(\ch)$ with $\Phi\sim\Psi$, one has $\cv(\Phi)=\cv(\Psi)$, and
    the norms $\|\cdot\|_\Phi$ and $\|\cdot\|_\Psi$ are equivalent.
\end{enumerate}
\end{our_lemma}
\begin{proof}
\begin{enumerate}[(a)]
\item Take $\Psi\sim \Phi$. By definition,
$\Phi^*\Psi\in \dcl\ell$ and
$\Phi^*\Phi\in \dcl\ell$. The difference yields
$\Phi^*(\Psi-\Phi)\in \trcl(\ell)$.
Similarly, $\Psi^*\Psi\in \dcl\ell$ and
$\Psi^*\Phi\in \dcl\ell$.
Combining all this, we get $(\Psi-\Phi)^*(\Psi-\Phi)\in \trcl(\ell)$,
and hence $\Psi-\Phi \in \hs(\ell,\ch)$. This shows $\Psi-\Phi\in \cv(\Phi)$.

Conversely, take $B\in \cv(\Phi)$. We set $\Psi=\Phi+B$.  First we show that $\range\Psi\in\pol(\ch)$, i.e. that it is closed and has infinite dimension and codimension. To do this we use the following general fact: A Fredholm operator between two Hilbert spaces maps closed, infinite dimensional, and infinite codimensional subspaces, respectively, to closed, infinite dimensional, and infinite codimensional subspaces, respectively. Consider
\begin{align}
  \widetilde\Phi:\ell\oplus\range\Phi^\perp\to\ch,&& (x,y)&\mapsto\Phi x+ y\\
  \widetilde\Psi:\ell\oplus\range\Phi^\perp\to\ch,&& (x,y)&\mapsto\Phi x+B x+ y
\end{align}
with the direct sum is understood as orthogonal direct sum. Since $\range\Phi$ is in $\pol(\ch)$ and therefore closed, the map $\widetilde\Phi$ is onto. Furthermore, $\Phi^*\Phi\in\dcl\ch$ is a perturbation of the identity by a compact operator and therefore a Fredholm operator. In particular, this implies $\dim\ker\Phi=\dim\ker\Phi^*\Phi<\infty$. Thus, $\widetilde\Phi$ is also a Fredholm operator. Now, $\widetilde \Psi$ is a perturbation of $\widetilde\Phi$ by the compact operator $(x,y)\mapsto Bx$ and therefore is Fredholm operator, too. Since $\ell\oplus 0$ is closed, infinite dimensional, infinite and codimensional, so is $\range\Psi=\widetilde\Psi(\ell\oplus 0)$.

Using $\Phi\in\seas(\ch)$ and the definition of $\cv(\Phi)$,
we get $\Psi^*\Psi=\Phi^*\Phi+\Phi^* B+(\Phi^* B)^*+B^* B
\in (\id_\ell+\trcl)+\trcl+\trcl+\hs\hs
=\dcl\ell$. This shows $\Psi\in \seas(\ch)$.
Furthermore,
$\Phi^* \Psi=\Phi^* \Phi+\Phi^* B\in (\id_\ell+\trcl)
+\trcl=\dcl\ell$ holds. This yields $\Psi\sim \Phi$.

\item Since $\Phi\sim\Psi$ there is a $L\in \cv(\Phi)$ such that $\Phi=\Psi+L$. Let $M\in \cv(\Psi)$. Using the triangle inequality in $\trcl(\ell)$ and
$\|L^* M\|_\trcl\le \|L\|_\hs \|M\|_\hs$, we get
  \begin{align}
    \|M\|_\Phi&=\|\Phi^*M\|_\trcl+\|M\|_\hs
\le \|\Psi^*M\|_\trcl+\|L^*M\|_\trcl+\|M\|_\hs\nonumber
 \\
&
\leq (1+\|L\|_\hs)(\|\Psi^*M\|_\trcl + \|M\|_\hs)
 = (1+\|L\|_\hs)\|M\|_\Psi.
  \end{align}
In the same way, we get $\|M\|_\Psi\leq (1+\|L\|_\hs)\|M\|_\Phi$.
\end{enumerate}
\end{proof}
The equivalence classes of $\seas(\ch)/{\sim}$ and $\pol(\ch)/{\approx}_0$ go hand in hand quite naturally as the following lemma shows.
\begin{our_lemma}[Connection between $\sim$ and $\approx_0$]\label{lem:connection sim approx}
  Given $C\in\pol(\ch)/{\approx}_0$ and $\Phi\in\ocean(C)$ we have
  \begin{align*}
    C=\{\range \Psi\;|\;\Psi\in\iseas(\ch)\text{ such that }\Psi\sim\Phi\}.
  \end{align*}
\end{our_lemma}
\begin{proof}
  Let $C':=\{\range \Psi\;|\;\Psi\in\iseas(\ch)\text{ such that }\Psi\sim\Phi\}$ and $V:=\range\Phi$.
  \begin{list}{}{}
    \item[$C'\subseteq C$:]
      Let $W\in C'$, then there is a $\Psi\sim\Phi$ such that $\range\Psi=W$.
      One has
      \begin{align}
        P_VP_WP_V|_{V\to V} = \Phi(\Phi^*\Psi\Psi^*\Phi)\Phi^*|_{V\to V}.
      \end{align}
      But $\Phi^*\Psi,\Psi^*\Phi\in\id_\ell+\trcl(\ell)$, hence, $\Phi^*\Psi\Psi^*\Phi\in\id_\ell+\trcl(\ell)$ and $\Phi^*|_V$ is unitary, so we conclude that $P_VP_WP_V|_{V\to V}$ has a determinant. Analogously, we get that $P_WP_VP_W|_{W\to W} = \Psi(\Psi^*\Phi\Phi^*\Psi)\Psi^*$
      has a determinant because, again, $\Psi^*|_W$ is unitary.
      Lemma \pref{lemma approx} (Properties of $\approx$) then states $V\approx W$. We still need to show $V\approx_0 W$. Therefore, consider  $\charge(W,V)=\ind(P_V|_{W\to V})$  and $P_V|_{W\to V}=\Phi\Phi^*\Psi\Psi^*|_{W\to V}$. Since $\Phi^*\Psi\in\id_\ell+\trcl(\ell)$ and $\Psi^*|_W$ is unitary, $\Psi^*P_V|_{W\to V}\Psi=\Psi^*\Phi\Phi^*\Psi\in\id_\ell+\trcl(\ell)$, which is a perturbation of the identity by a compact operator. Therefore $\ind(P_V|_{W\to V})=0$. Hence, we have shown that $V\approx_0 W$ and therefore $W\in C$.
    \item[$C'\supseteq C$:]
      Let $W\in C$, then $W\approx V$ and $\charge(V,W)=0$. We need to find an isometry $\Psi\sim\Phi$ such that $\range \Psi=W$. We make a polar decomposition of $P_W\Phi$.
      By Lemma \pref{lemma approx}(\ref{lemma approx e}) we know that $\range P_W|_V$ is closed. There is a partial isometry $U:\ell\to\range P_W\Phi=\range P_W|_V=\overline{\range P_W\Phi}\subset \ch$ with $\ker U=\ker P_W\Phi$ such that $P_W\Phi=U|P_W\Phi|$ where $|P_W\Phi|$ is given by the square root of the positive semi-definite operator $(P_W\Phi)^*(P_W\Phi)$. Furthermore, $(P_W\Phi)^*(P_W\Phi)=\Phi^* P_VP_WP_V\Phi\in\id_\ell+\trcl(\ell)$ holds by Lemma \pref{lemma approx}. Hence, this operator is a Fredholm operator. That means also that $\ker\big((P_W\Phi)^*(P_W\Phi)\big)=\ker P_W\Phi$ is finite dimensional. Moreover, $0=\charge(V,W)=\ind(P_W|_{V\to W})$ implies that $\dim \ker P_W\Phi=\dim W/(\range P_W|_V)=\dim W\cap(\range P_W|_V)^\perp$. Thus, there is another partial isometry $\widetilde U:\ell\to\ch$ of finite rank such that $\widetilde U|_{\ker P_W\Phi}$ maps $\ker P_W\Phi$ unitarily onto $W\cap(\range P_W\Phi)^\perp$ and $\widetilde U|_{(\ker P_W\Phi)^\perp}=0$. We set $\Psi:=U+\widetilde U:\ell\to\ch$, and get $\Psi^*\Psi=U^*U+{\widetilde U}^*\widetilde U=1$ and therefore $\Psi\in\iseas(\ch)$. By construction, $\range \Psi=W$ holds. Furthermore, we have the identities
      \begin{align}
        U^*\Phi=U^*P_W\Phi=U^*U|P_W\Phi|=|P_W\Phi|.
      \end{align}
      This operator has a determinant since $|P_W\Phi|\ge 0$ and
      \begin{align}
        |P_W\Phi|^2=\Phi^*P_VP_WP_V\Phi\in\id_\ell+\trcl(\ell)
      \end{align}
      hold. The last identity follows since $P_VP_WP_V|_{V\to V}$ has a determinant by Lemma \pref{lemma approx}. On the other hand, $\widetilde U^*\Phi$ has finite rank since $\widetilde U$ does. Hence, $\Psi^*\Phi=U^*\Phi+\widetilde U^*\Phi\in\id_\ell+\trcl(\ell)$, i.e. $\Psi\sim\Phi$, which means that $W\in C'$.
   \end{list}
\end{proof}
Now we begin with the construction of the infinite wedge spaces for each equivalence class of Dirac seas $\cs\in\seas(\ch)/{\sim}$. We follow the standard linear algebra method: First, we construct with the elements of $\cs$ a space of formal linear combinations $\C^{(S)}$ which we equip with a semi-definite sesquilinear form that in turn induces a semi-norm. Completion with respect to this semi-norm yields the infinite wedge space of $\cs$.
\begin{our_construction}[Formal Linear Combinations]\label{const:proto}
\begin{enumerate}
\item
For any set $\cs$, let $\C^{(\cs)}$ denote the set of all
maps $\alpha:\cs\to \C$ for which $\{\Phi\in \cs\;|\;\alpha(\Phi)\neq 0\}$ is
finite. For $\Phi\in \cs$, we define $[\Phi]\in \C^{(\cs)}$\nomenclature{$[\Phi],[\Psi], etc.$}{For any $\Phi\in \cs$, $[\Phi]$ denotes the map $\cs\to\C$ in $\C^{(\cs)}$ fulfilling $[\Phi](\Phi)=1$ and $[\Phi](\Upsilon)=0$ for $\Phi\neq\Upsilon\in \cs$} to be the map fulfilling $[\Phi](\Phi)=1$ and $[\Phi](\Psi)=0$ for $\Phi\neq\Psi\in \cs$.
Thus, $\C^{(\cs)}$\nomenclature{$\C^{(\cs)}$}{The space of formal $\C$-linear combinations of elements of $\cs$} consists of all finite formal linear
combinations $\alpha=\sum_{\Psi\in \cs}\alpha(\Psi)[\Psi]$\nomenclature{$\alpha,\beta$, etc.}{Typical elements of $\C^{(\cs)}$}
of elements of $\cs$ with coefficients in $\C$.
\item
Now, let $\cs\in\seas(\ch)/{\sim}$ as in Definition \pref{def:equiclass} (Dirac Sea Classes). We define the map
$\sk{\cdot,\cdot}:\cs\times \cs\to \C$, $(\Phi,\Psi)\to\sk{\Phi,\Psi }:=\det(\Phi^*\Psi )$.
Note that this is well defined since for $\Phi,\Psi \in \cs$ the fact $\Phi\sim \Psi $
implies that $\Phi^*\Psi $ has a determinant.
\item
Taking $\cs$ as before,
let $\sk{\cdot,\cdot}:\C^{(\cs)}\times \C^{(\cs)}\to \C$
denote the sesquilinear extension of $\sk{\cdot,\cdot}:\cs\times \cs\to \C$,
defined as follows:
For $\alpha,\beta\in \C^{(\cs)}$,
\begin{equation}
\sk{\alpha,\beta}=\sum_{\Phi\in \cs}\sum_{\Psi \in \cs} \overline{\alpha(\Phi)}\beta(\Psi )
\det(\Phi^*\Psi ).
\end{equation}
Here, the bar denotes the complex conjugate\nomenclature{$\overline z$}{Complex conjugate of $z$}. Note that the sums consist of at most finitely many nonzero summands. In particular we have $\sk{[\Phi],[\Psi ]}=\sk{\Phi,\Psi }$ for $\Phi,\Psi \in \cs$.
\end{enumerate}
\end{our_construction}
\begin{our_lemma}
The sesquilinear form $\sk{\cdot,\cdot}:\C^{(\cs)}\times \C^{(\cs)}\to \C$
is hermitean and positive semi-definite, i.e. $\sk{\alpha,\beta}=\overline{\sk{\beta,\alpha}}$
and $\sk{\alpha,\alpha}\ge 0$ hold for all
$\alpha,\beta\in \C^{(\cs)}$.
\end{our_lemma}
\begin{proof}
For $\Phi,\Psi \in \cs$, we have
\begin{equation}
\sk{\Phi,\Psi }=\det(\Phi^*\Psi )=\overline{\det(\Psi ^*\Phi)}=\overline{\sk{\Psi ,\Phi}}
\end{equation}
This implies that  $\sk{\cdot,\cdot}:\C^{(\cs)}\times \C^{(\cs)}\to \C$
is hermitean. Let $\alpha\in \cs$. We get
\begin{align}
\label{step1}
\sk{\alpha,\alpha}
=\sum_{\Phi\in \cs}\sum_{\Psi \in \cs} \overline{\alpha(\Phi)}\alpha(\Psi )
\det(\Phi^*\Psi).
\end{align}
Let $(e_i)_{i\in\N}$ be an orthonormal basis in $\ell$.
In the following, we abbreviate $\N_m=\{1,\ldots,m\}$.
Fredholm determinants are approximated
by finite-dimensional determinants (see Section VII.3, Theorem 3.2
in \cite{gohberg-90}), therefore
\begin{equation}
\label{step2}
\det(\Phi^*\Psi)=
\lim_{m\to\infty}\det(\sk{e_i,\Phi^*\Psi e_j})_{i,j\in\N_m}
.
\end{equation}
Let $(f_k)_{k\in\N}$ be an orthonormal basis of $\ch$.
For every $i,j\in\N$, we get
\begin{equation}
\sk{e_i,\Phi^*\Psi e_j}=\lim_{n\to\infty}
\sum_{k=1}^n\sk{\Phi e_i,f_k}\sk{f_k,\Psi e_j},
\end{equation}
and, hence, for every $m\in\N$
\begin{align}
\det(\sk{e_i,\Phi^*\Psi e_j})_{i,j\in\N_m}
&=\lim_{n\to\infty}
\det\left(\sum_{k=1}^n\sk{\Phi e_i,f_k}\sk{f_k,\Psi e_j}\right)_{i,j\in\N_m}
\nonumber
\\&=\lim_{n\to\infty}
\sum_{\substack{I\subseteq \N_n\\|I|=m}}
\det(\sk{\Phi e_i,f_k})_{\substack{k\in I\\,i\in\N_m} }
\det(\sk{f_k,\Psi e_j})_{\substack{k\in I,\\ j\in \N_m}}
\end{align}
Substituting this in  (\pref{step1}) and (\pref{step2}), we conclude
\begin{align}
\sk{\alpha,\alpha}
&=\lim_{m\to\infty}\lim_{n\to\infty}
\sum_{\Phi\in \cs}\sum_{\Psi \in \cs}
\sum_{\substack{I\subseteq \N_n\\|I|=m}}
\overline{\alpha(\Phi)}
\alpha(\Psi )
\overline{\det(\sk{f_k,\Phi e_i})_{\substack{k\in I,\\i\in\N_m}}}
\det(\sk{f_k,\Psi e_j})_{\substack{k\in I,\\ j\in \N_m}}
\nonumber
\\&=
\lim_{m\to\infty}\lim_{n\to\infty}
\sum_{\substack{I\subseteq \N_n\\|I|=m}}
\left|
\sum_{\Phi\in \cs}
\alpha(\Phi)\det(\sk{f_k,\Phi e_i})_{\substack{k\in I,\\i\in\N_m}}
\right|^2
\ge 0
.
\end{align}
\end{proof}
\begin{our_definition}\label{def:semi norm null space}
Let
$\|\cdot\| :\C^{(\cs)}\to\R$, $\alpha\mapsto\|\alpha\|=\sqrt{\sk{\alpha,\alpha}}$
denote the semi-norm associated to $\sk{\cdot,\cdot}$, and $N_\cs=\{\alpha\in\C^{(\cs)}\;|\;\sk{\alpha,\alpha}=0\}$ denote the null space of $\C^{(\cs)}$ with respect to $\|\cdot\|$.
\end{our_definition}
This null space $N_\cs$ is quite large. The following lemma identifies a few elements of this null space and is also the key ingredient to Corollary \pref{cor null space} (Null space) and therewith to Lemma \pref{lem:equivalent_right_ops} (Uniqueness up to a Phase).
\begin{our_lemma}
\label{lemma null space}
For $\Phi\in \cs$ and $R\in \dcl\ell$,
one has $\Phi R\in \cs$ and
$[\Phi R]-(\det R) [\Phi]\in N_\cs$.
\end{our_lemma}
\begin{proof}
First, we observe that
$(\range (\Phi R))^\perp\supseteq (\range \Phi)^\perp$
is infinite-dimensional.
Since $\Phi^*\Phi\in \dcl\ell$ and $R\in \dcl\ell$,
we have $\Phi^*(\Phi R)\in \dcl\ell$ and
$(\Phi R)^*(\Phi R)=R^*(\Phi^*\Phi)R\in \dcl\ell$.
This shows $\Phi R\in \seas(\ch)$ and $\Phi\sim \Phi R$, and thus $\Phi R\in \cs$.
We calculate:
\begin{align}
\|[\Phi R]-(\det R)[\Phi]\|^2
=&\det((\Phi R)^*(\Phi R))-(\det R) \det((\Phi R)^* \Phi)\nonumber\\
&-\overline{\det R}
\det(\Phi^*\Phi R)+|\det R|^2\det(\Phi^*\Phi)
\nonumber\\=&
2|\det R|^2\det(\Phi^*\Phi)-2|\det R|^2\det(\Phi^*\Phi)
=0.
\end{align}
\mbox{}
\end{proof}
Now we have everything needed to define some key objects in this work: The \emph{infinite wedge spaces}. These spaces shall make up the playground for the second quantized Dirac time-evolution:
\begin{our_definition}[Infinite Wedge Spaces]
\label{def dach}
Let $\cf_\cs$ be the completion of $\C^{(\cs)}$
with respect to the semi-norm $\|\cdot\|$. We refer to $\cf_\cs$ as infinite wedge space over $\cs$. Let $\iota:\C^{(\cs)}\to\cf_\cs$ denote the canonical map.
The sesquilinear form $\sk{\cdot,\cdot}:\C^{(\cs)}\times\C^{(\cs)}\to\C$
induces a scalar product
$\sk{\cdot,\cdot}:\cf_\cs\times\cf_\cs
\to \C$. Let $\mathsf{\Lambda}:\cs\to \cf_\cs$ denote the canonical map
$\mathsf{\Lambda} \Phi=\iota([\Phi])$, $\Phi\in \cs$.
\end{our_definition}
Note that $\iota[N_\cs]=\{0\}$.
Hence, the null space is automatically factored out during the
completion procedure. In fact, the null space of the canonical map $\iota:\C^{(\cs)}\to\cf_\cs$
equals $\ker\iota=N_\cs$.
Thus we can rewrite Lemma
\pref{lemma null space} in the following way:
\begin{our_corollary}[Null space]
\label{cor null space}
For $\Phi\in \cs$ and $R\in \dcl\ell$, one has $\Wedge(\Phi R)=(\det R) \Wedge \Phi$.
\end{our_corollary}
Combining Corollary \pref{cor null space} above
with Lemma \pref{polarlemma} (Isometries are good enough), we get the following:
For every $\Phi\in \cs$ there are $\Upsilon\in \cs\cap \iseas(\ch)$ and $R\in\dcl\ell$ with
$r=\det R\in\R^+_0$
such that $\Wedge\Phi=r \Wedge \Upsilon$.
As a consequence, $\{\Wedge \Psi\;|\; \Psi\in \cs\cap\iseas(\ch)\}$ spans a dense subspace
of  $\cf_\cs$. The scalar product $\sk{\cdot,\cdot}$
gives $\cf_\cs$ the structure of a separable
Hilbert space:
\begin{our_lemma}[Separability]\label{separable}
The inner product space $(\cf_\cs,\sk{\cdot,\cdot})$ is separable.
\end{our_lemma}
\begin{proof}
It suffices to show that there exists a countable dense subset of
$\Wedge \cs$ with respect to the norm $\|\cdot\|_{\mathcal F_\cs}$ in $\mathcal F_\cs$.
Choose $\Phi\in \cs$.
then by Lemma \pref{lem:equiclass_characterization} (Dirac Sea Class Properties) we know that $\cs=\Phi+\cv(\Phi)$.  Now, the set of operators of finite rank is dense and separable in
$(\cv(\Phi),\|\cdot\|_\Phi)$. Hence, we can choose a countable,
dense subset $D$ in $(\cv(\Phi),\|\cdot\|_\Phi)$.
We show now that $\Wedge(\Phi+D)$ is dense in $\Wedge\cs$
with respect to the norm $\|\cdot\|_{\mathcal F_\cs}$.
Let $\Psi=\Phi+L\in \cs$ with
$L\in \cv(\Phi)$.
We find a sequence  $(L_n)_{n\in\N}$ in $D$ with
 $\|L_n-L\|_{\Phi}\to 0$ for $n\to\infty$ and define $\Psi_n:=\Phi+L_n$. One then obtains the following estimate for all large $n$:
  \begin{align}
    \|\mathsf{\Lambda}\Psi-\mathsf{\Lambda}\Psi_n\|^2_{\mathcal F_\cs} &= \sk{\mathsf{\Lambda}\Psi-\mathsf{\Lambda}\Psi_n,\mathsf{\Lambda}\Psi-\mathsf{\Lambda}\Psi_n}_{\mathcal F_\cs}\nonumber\\
    &=\det(\Psi^*\Psi)-\det(\Psi^*\Psi_n)-\det(\Psi_n^*\Psi)+\det(\Psi_n^*\Psi_n)\nonumber\\
    &\leq \xconstl{c:dreizehn}(\Psi)\left(\|\Psi^*(\Psi-\Psi_n)\|_{\trcl}+\|\Psi_n^*(\Psi-\Psi_n)
    \|_{\trcl}\right)
  \end{align}
   by local Lipschitz continuity of the Fredholm determinant with respect to the norm in $\trcl(\ell)$; see \cite[Theorem 3.4 p. 34]{Simon:05}. The constant $\xconstr{c:dreizehn}(\Psi)<\infty$ depends only on $\Psi$. Next, the triangle inequality applied to the second term gives
  \begin{align}
\nonumber
    \ldots &\leq \xconstr{c:dreizehn}(\Psi) \left(2\|\Psi^*(\Psi-\Psi_n)\|_{\trcl}+\|(\Psi-\Psi_n)^*(\Psi-\Psi_n)\|_{\trcl}\right)\\
\nonumber
    &\leq 2\xconstr{c:dreizehn}(\Psi) \|\Psi-\Psi_n\|_\Psi=
2\xconstr{c:dreizehn}(\Psi) \|L-L_n\|_\Psi
\\&\le \xconstl{c:norm faktor}(\Psi,\Phi) \|L-L_n\|_\Phi\
\xrightarrow{n\to\infty}{} 0
  \end{align}
  for some constant $\xconstr{c:norm faktor}(\Psi,\Phi)<\infty$ depending only on $\Psi$ and $\Phi$ since the norms $\|\cdot\|_\Psi$ and $\|\cdot\|_\Phi$ are equivalent by Lemma \pref{lem:equiclass_characterization} (Dirac Sea Class Properties).
This shows that $\Wedge(\Phi+D)$
is a countable, dense subset of $\Wedge(\cs)$.
\end{proof}
The following diagram summarizes the setup:
\newdir{ >}{!/2.5pt/@{}*:(1,-.2)@^{>}*:(1,+.2)@_{>}}
\begin{align*}
  \xymatrix{
    & & & \ar@{..>}[d]|{\text{\tiny chooses}} \fbox{Nature}\\
    \ch \ar@{|->}[r]^-{\text{\tiny take}}_-{\text{\tiny splittings}} & \pol(\ch) \ar[r]^-{[\cdot]_{\approx_0}} & \pol(\ch)/{\approx}_0\; \ar@{~ >}[r]^-\ni & C \ar@{~ >}[d]_-{\text{\tiny take}}^-{\text{\tiny{bases}}}\\
    \fbox{Human} \ar@{..>}[r]^-{\text{\tiny chooses}} & \cs \ar@{|->}[d]^-{\Wedge \text{ \tiny construction}} & \ar@{~ >}[l]_-\in \ocean_\ell(C)/{\sim} & \ar[l]_-{[\cdot]_\sim} \ocean_\ell(C)\\
    & \cf_\cs
  }
\end{align*}
Note that, by Lemma \pref{lem:connection sim approx} (Connection between $\sim$ and $\approx_0$), $\cf_\cs$ carries the whole information of the polarization class $C\in\pol(\ch)/{\approx}_0$; however, it depends on a choice of basis. In this sense we say that the wedge space $\cf_\cs$ belongs to the polarization class $C$.

\subsection{Operations from the Left and from the Right}\label{sec:left and right op}
Having constructed the infinite wedge spaces $\cf_\cs$ for each $\cs\in\seas(\ch)/{\sim}$ we now introduce two types of operations on them which are the tools needed in the next subsection. In the following let $\ch'$, $\ell'$ be also two Hilbert spaces.
\begin{our_construction}[The Left Operation]\label{constr:leftop}
\mbox{}
\begin{enumerate}
\item
The following operation from the left is well-defined:
\begin{equation*}
\cu(\ch,\ch')\times \seas_\ell(\ch)\to \seas_\ell(\ch'), \quad(U,\Phi)\mapsto U\Phi.
\end{equation*}
\item
This operation from the left is compatible with the
equivalence relation $\sim$ in the following sense:
For $U\in \cu(\ch,\ch')$ and $\Phi,\Psi \in\seas_\ell(\ch)$, one has $\Phi\sim \Psi $
if and only if
$U\Phi\sim U\Psi $ in $\seas_\ell(\ch')$. Thus,
the action of $U$ on $\seas_\ell(\ch)$
from the left induces also an operation from the
left on equivalence classes modulo $\sim$ as follows.
For $\cs\in \seas_\ell(\ch)/{\sim}$ and $U\in \cu(\ch,\ch')$,
\begin{align*}
U\cs =\{U\Phi\;|\;\Phi\in \cs\}\in \seas_\ell(\ch')/{\sim}.
\end{align*}
\item
For $U\in \cu(\ch,\ch')$ and $\cs\in\seas_\ell(\ch)/{\sim}$,
the induced operation
$\lop U:\C^{(\cs)}\to \C^{(US)}$\nomenclature{$\lop U$}{The left operation $\C^{(\cs)}\to\C^{(US)}$ induced by $U$},
given by
\begin{align*}
\lop U\left(\sum_{\Phi\in \cs}\alpha(\Phi)[\Phi]\right)
=\sum_{\Phi\in \cs}\alpha(\Phi)[U\Phi],
\end{align*}
is an isometry with respect to
the hermitean forms $\sk{\cdot,\cdot}$ on $\C^{(\cs)}$ and on
$\C^{(US)}$.
In particular one has $\lop U[N_\cs]\subseteq N_{US}$.
\item
For every $U\in \cu(\ch,\ch')$,
the operation from the left $\lop U:\C^{(\cs)}\to \C^{(US)}$
induces a
unitary map $\lop U:\cf_\cs\to \cf_{US}$\nomenclature{$\lop U$}{Left operation $\cf_\cs\to\cf_{US}$ induced by $U$},
characterized by
$\lop U(\mathsf{\Lambda} \Phi)=\mathsf{\Lambda}(U\Phi)$ for $\Phi\in \cs$.
This operation is functorial in the following sense.
Let $\ch''$ be another Hilbert space.
For $U\in \cu(\ch,\ch')$, $V\in \cu(\ch',\ch'')$ and $\cs\in\seas_\ell(\ch)/{\sim}$,
one has $\lop U\lop V=\lop{UV}:\cf_\cs
\to\cf_{UVS}$
and $\lop{\id_\ch}=\id_{\cf_\cs}$.
\end{enumerate}
\end{our_construction}
In complete analogy to the operation from the left, we introduce next
an operation from the right. Let $\ell'$ be another Hilbert space, and let $\glm(\ell',\ell)$\nomenclature{$\glm(\ell',\ell)$}{The set of all bounded invertible linear operators $R:\ell'\to \ell$ with the property $R^*R\in \dcl{\ell'}$}
denote the set of all bounded invertible linear operators $R:\ell'\to \ell$ with the property $R^*R\in \dcl{\ell'}$. Note that $\glm(\ell):=\glm(\ell,\ell)$ is a group with respect to composition.
\begin{our_construction}[Operation from the Right]\label{def:rightop}
\mbox{}
\begin{enumerate}
\item
The following operation from the right is well-defined:
\begin{equation*}
\seas_\ell(\ch)\times \glm(\ell',\ell)\to\seas_{\ell'}(\ch),
\quad
(\Phi,R)\mapsto \Phi R.
\end{equation*}
\item
This operation from the right is compatible with the
equivalence relations $\sim$: For $\Phi,\Psi \in\seas_\ell(\ch)$ and
$R\in \glm(\ell',\ell)$, one has
$\Phi\sim \Psi $ if and only if $\Phi R\sim \Psi R$ in $\seas_{\ell'}(\ch)$.
Thus, the operation of $R$ from the right induces also an
operation from the right on equivalence classes modulo $\sim $
as follows: For $\cs\in \seas_\ell(\ch)/{\sim}$ and $R\in\glm(\ell',\ell)$,
\begin{equation*}
\cs R=\{\Phi R\;|\;\Phi\in \cs\}\in \seas_{\ell'}(\ch)/{\sim}.
\end{equation*}
\item
For $\cs\in\seas_\ell(\ch)/{\sim}$ and $R\in \glm(\ell',\ell)$
the induced operation from the right
$\rop R:\C^{(\cs)}\to \C^{(\cs R)}$\nomenclature{$\rop R$}{The operation from the right $\C^{(\cs)}\to\C^{(\cs)}$ induced by $R$},
given by
\begin{align*}
\rop R\left(\sum_{\Phi\in \cs}\alpha(\Phi)[\Phi]\right)
=\sum_{\Phi\in \cs}\alpha(\Phi)[\Phi R],
\end{align*}
is an isometry up to scaling with respect to
the hermitean forms $\sk{\cdot,\cdot}$ on $\C^{(\cs)}$ and on
$\C^{(\cs R)}$.
More precisely, one has for all $\alpha,\beta\in \C^{(\cs)}$:
\begin{align*}
\sk{\rop R\alpha,\rop R\beta}
= \det(R^*R) \sk{\alpha,\beta}.
\end{align*}
In particular one has $\rop R[N_\cs]\subseteq N_{\cs R}$.
\item
For every $R\in\glm(\ell',\ell)$,
the operation $\rop R:\C^{(\cs)}\to \C^{(\cs R)}$
induces a bounded linear
map, again called  $\rop R:\cf_\cs\to \cf_{\cs R}$\nomenclature{$\rop R$}{The operation from the right $\cf_\cs\to \cf_{\cs R}$ induced by $R$},
characterized by
$\rop R(\mathsf{\Lambda} \Phi)=\mathsf{\Lambda}(\Phi R)$ for $\Phi\in \cs$.
Up to scaling, this map is unitary. More precisely, for $\Phi,\Psi\in
\cf_\cs$, one has
\begin{equation*}
\sk{\rop R\Phi,\rop R\Psi}=\det (R^*R)\sk{\Phi,\Psi}.
\end{equation*}
The operation $\rop R$ is
contra-variantly functorial in the following sense.
Let $\ell''$ be another Hilbert space.
For $Q\in \glm(\ell'',\ell')$, $R\in \glm(\ell',\ell)$
and $\cs\in{\seas(\ch,\ell)}/{\sim}$,
one has $\rop Q\rop R=\rop {RQ}:\cf_\cs
\to\cf_{\cs RQ}$
and $\rop {\id_\ell}=\id_{\cf_\cs}$.
\end{enumerate}
\end{our_construction}
The associativity of composition $(U\Phi)R=U(\Phi R)$ immediately yields:
\begin{our_lemma}[Left and Right Operations Commute]
The operations from the left and from the right commute:
For $U\in \cu(\ch,\ch')$, $R\in \glm(\ell',\ell)$,
and $\cs\in \seas_\ell(\ch)/{\sim}$, one has
$\lop U\rop R=\rop R\lop U:\cf_\cs
\to\cf_{U\cs R}$.
\end{our_lemma}
We conclude this subsection with a last lemma that states an important property of the infinite wedge spaces. Essentially, it says that for any $R\in\glm(l)$ such that $R$ has a determinant we have $\cf_\cs=\cf_{\cs R}$. We introduce $\slg(\ell)$\nomenclature{$\slg(\ell)$}{The set of all operators $R\in\dcl{\ell}$ with the property $\det R=1$} to denote the set of all operators $R\in\dcl{\ell}$ with the property $\det R=1$.
\begin{our_lemma}[Uniqueness up to a Phase]\label{lem:equivalent_right_ops}
\begin{enumerate}
\item
For all $R\in\glm(\ell)$ and $\cs\in\seas_\ell(\ch)/{\sim}$, one has
$\cs=\cs R$ if and only if $R$ has a determinant.
In this case,
$\rop R(\Psi)=(\det R)\Psi$ holds for all
$\Psi\in \cf_\cs$.
As a special case, if $R\in\slg(\ell)$,
then $\rop R:\cf_\cs\to\cf_\cs$ is the identity map.
\item
For all $Q,R\in \glm(\ell',\ell)$ and $\cs \in \seas_\ell(\ch)/{\sim}$, we have
$\cs R=SQ$ if and only if $Q^{-1}R\in \glm(\ell')$ has a determinant.
In this case, one has for all $\Psi\in \cf_\cs$:
\begin{equation*}
\rop R\Psi=\det (Q^{-1}R)\rop Q\Psi
\end{equation*}
\end{enumerate}
\end{our_lemma}
\begin{proof}
\begin{enumerate}
\item[(a)]
Given $R\in\glm(\ell)$ and $\cs\in\seas_\ell(\ch)/{\sim}$,
take any $\Phi\in \cs$. Then, as $\Phi^*\Phi$ has a determinant,
$\Phi^*\Phi R$ has a determinant if and only if $R$ has a determinant.
This is equivalent to $\Phi\sim \Phi R$ and to $\cs=\cs R$.
In this case, Lemma \pref{cor null space} (Null space) implies
$\rop R\Psi=(\det R)\Psi$
for all $\Psi\in\cf_\cs$.
\item[(b)]
Let $\Phi\in \cs\cap \iseas_\ell(\ch)$. Then $\cs R=SQ$ holds if and only if
$\Phi R \sim \Phi Q$, i.e. if and only if
$Q^*R=(\Phi Q)^*\Phi R$ has a determinant. Since $Q^*Q$ has a determinant
and is invertible, this is equivalent
to $Q^{-1}R\in \id_{\ell'}+\trcl(\ell')$.
Using part (a), for any $\Psi\in \cf_\cs$,  we have in this case:
$\rop R\Psi= \rop {Q^{-1}R}\rop Q\Psi=
\det (Q^{-1}R)\rop Q\Psi$.
\end{enumerate}
\end{proof}

\subsection{Lift Condition}\label{sec:lift cond}
Given two Hilbert spaces $\ch$ and $\ch'$ and two polarization classes $C\in\pol(\ch)/{\approx}_0$ and $C'\in\pol(\ch)/{\approx}_0$ we now identify conditions under which a unitary operator $U:\ch\to\ch'$ can be lifted to a unitary map between two wedge spaces.

By Lemma \pref{lemma approx} (Properties of $\approx$) it is clear how any unitary $U:\ch\to\ch'$ acts on polarization classes, and we do not prove the following simple lemma:
\begin{our_lemma}[Action of $\cu$ on Polarization Classes]
The natural operation
\begin{align*}
\cu(\ch,\ch')\times \pol(\ch)\to \pol(\ch'),\quad (U,V)\mapsto UV=\{Uv\;|\;v\in V\}
\end{align*}
is compatible with the equivalence relations $\approx$ in the
following sense: For $U\in \cu(\ch,\ch')$ and $V,W\in \pol(\ch)$, one has $V\approx W$
if and only if $UV\approx UW$. As a consequence, this operation from the left induces a natural
operation on polarization classes $\cu(\ch,\ch')\times (\pol(\ch)/{\approx})\to \pol(\ch')/{\approx},
\quad (U,[V]_{\approx})\mapsto [UV]_{\approx}$.
\end{our_lemma}
In order to describe charge-preserving time-evolutions $U\in\cu(\ch,\ch')$, the following subclass of $\cu(\ch,\ch')$, the restricted set of unitary operators, will be convenient:
\begin{our_definition}[Restricted Set of Unitary Operators]\label{def:ures0}
Given the polarization classes $C\in \pol(\ch)/{\approx_0}$ and $C'\in \pol(\ch')/{\approx_0}$ we define\nomenclature{$\ur^0(\ch,C;\ch',C')$}{The set of operators $U\in\cu(\ch,\ch')$ such that for all $V\in C$ one has $UV\in C'$}
\begin{align*}
  \ur^0(\ch,C;\ch',C'):&=\{U\in \cu(\ch,\ch')\;|\;\text{ for all } V\in C
  \;\text{holds }UV\in C'\}\nonumber\\
  &=\{U\in \cu(\ch,\ch')\;|\;\text{there exists } V\in C
  \;\text{such that }UV\in C'\}.
\end{align*}
As a special case, we yield a group $\ur^0(\ch,C):=\ur^0(\ch,C;\ch,C)$\nomenclature{$\ur^0(\ch,C)$}{$:=\ur^0(\ch,C;\ch,C)$}.
\end{our_definition}
Note that for a third Hilbert space $\ch''$ with a polarization class $C''\in \pol(\ch'')/{\approx}_0$, one has $\ur^0(\ch',C';\ch'',C'')\ur^0(\ch,C;\ch',C')= \ur^0(\ch,C;\ch'',C'')$.

For unitary operations that change the relative charge of two polarizations by $c\in\Z$, there is a natural generalization of $\ur^0$ to $\ur^c$; this is discussed in a different context in Section \pref{comparison}.

Now we have all what is needed to prove the main result of this section: The following theorem is our version of the classical Shale-Stinespring theorem \cite{shale:65}, and hence not completely new. The connection is explained in Section \pref{comparison}.
\begin{our_theorem}[Lift Condition]
\label{unser abstrakter Shale Stinespring}
For given polarization classes $C\in\pol(\ch)/{\approx}_0$ and $C'\in\pol(\ch')/{\approx}_0$, let $\cs\in \ocean_\ell(C)/{\sim}$ and $\cs'\in \ocean_\ell(C')/{\sim}$. Then, for any unitary map $U:\ch\to \ch'$, the following are equivalent:
\begin{enumerate}
\item
There is $R\in U(\ell)$ such that $U\cs R=\cs'$, and hence
$\rop R\lop U$ maps $\cf_\cs$ to $\cf_{\cs'}$.
\item[(a')]
There is $R\in \glm(\ell)$ such that $U\cs R\sim \cs'$.
\item[(b)]
$U\in \ur^0(\ch,C; \ch', C')$.
\end{enumerate}
\end{our_theorem}
\begin{proof}
We take $\Phi\in\cs$, $\Phi'\in\cs'$ and set $V=\range\Phi$, $V'=\range\Phi'$.
\begin{list}{}{}
\item[$(a)\Rightarrow(b):$]
Take $R\in U(\ell)$ such that $U\cs(\Phi)R=\cs(\Phi')$. In particular,
$U\Phi R\sim \Phi'$, and hence $\Phi'^*U\Phi R\in \dcl\ell$.
This implies
\begin{equation}
(\Phi'^*U\Phi R)^*\Phi'^*U\Phi R\in \dcl\ell.
\end{equation}
Because $U\Phi R:\ell\to UV$ is unitary and $\Phi'\Phi'^*=P_{V'}$, we
conclude that $P_{UV}P_{V'}P_{UV}=P_{UV}\Phi'\Phi'^*P_{UV}|_{UV\to UV}$ has a determinant.
Similarly,
\begin{equation}
\Phi'^*P_{UV}\Phi^*=
\Phi'^*U\Phi R(\Phi'^*U\Phi R)^*\in \dcl\ell
\end{equation}
implies that $P_{V'}P_{UV}P_{V'}|_{V'\to V'}$ has also a determinant.
Together this yields $UV\approx V'$ by Lemma \pref{lemma approx} (Properties of $\approx$).

Furthermore, because of $U\Phi R\sim \Phi'$, we know that
$\Phi'^*U\Phi R$ is a Fredholm operator with
index $0$. Since $\Phi R:\ell\to V$ and $\Phi':\ell\to V'$ are unitary,
$P_{V'}|_{UV\to V'}$ is also a Fredholm operator with index 0,
i.e. $\charge(UV,V')=0$. This shows $UV\approx_0 V'$,
and the claim $U\in \ur^0(\ch,C; \ch',C')$ follows.
\item[$(b)\Rightarrow(a')$:]
We abbreviate $A=P_{V'}|_{UV\to V'}$.
The assumption $U\in \ur^0(\ch,C; \ch', C')$ implies
$A^*A\in \id_{UV}+\trcl(UV)$, and $A$ is a Fredholm operator with
index $\ind A=0$.
Using that $\Phi:\ell\to V$ and $\Phi':\ell\to V'$ are unitary maps,
we rewrite this in the form
$(\Phi'^* U\Phi)^*\Phi'^* U\Phi\in \dcl\ell$,
and $\Phi'^* U\Phi$ is a Fredholm operator with
$\ind(\Phi'^* U \Phi)=0$.
We now use a polar decomposition of $\Phi'^* U \Phi$ in the
form $\Phi'^* U \Phi=B Q$, where $B:\ell \to \ell$
is positive semi-definite and
$Q:\ell\to\ell$ is unitary. Note that we can take $Q$ to be {\it unitary},
not only a partial isometry, as  $\Phi'^* U \Phi$ has the Fredholm index $0$.
Taking $R=Q^{-1}$, we get $\Phi'^* U \Phi R=B$. Now $B^2=B^*B$
has a determinant because $Q^*B^*BQ=(\Phi'^* U\Phi)^*\Phi'^* U\Phi$
has a determinant. Since $B\ge 0$, this implies that $B$ has
also a determinant.
We conclude  $U \Phi R\sim \Phi'$.
\item[$(a')\Rightarrow(a)$:]
We take $R\in \glm(\ell)$ with $U\Phi R\sim\Phi'$.
By polar decomposition, we write $R$ in the form $R=R'Q$, where
$R':\ell\to\ell$ is unitary and $Q:\ell\to\ell$ is invertible, positive
definite, and has a determinant. As
$\Phi'^*U\Phi R=\Phi'^*U\Phi R'Q$ and $Q$ both have determinants,
$\Phi'^*U\Phi R'$ has also a determinant. This shows
$U\Phi R'\sim\Phi'$ and hence $\cs(U\Phi R')=U\cs(\Phi)R'=\cs(\Phi')$.
In particular, $\rop R\lop U$ maps $\cf_{\cs(\Phi)}$ to
$\cf_{U\cs(\Phi)R'}=\cf_{\cs(\Phi')}$.
\end{list}
\end{proof}
For $U=\id_\ch$ we immediately get:
\begin{our_corollary}[Orbits in $\ocean$]
  Given $C\in\pol(\ch)/{\approx_0}$ and $\cs\in\ocean(C)/{\sim}$ we have
  \begin{align*}
    \ocean(C)/{\sim}=\{\cs R\;|\; R\in\cu(\ell)\}.
  \end{align*}
\end{our_corollary}
This is the counterpart to Lemma \pref{lem:connection sim approx} (Connection between $\sim$ and $\approx_0$) which stated that for every polarization class $C$, the equivalence class with respect to $\sim$ of any single element $\Phi\in\ocean(C)$ suffices to recover $C$. For every $S\in\ocean(C)/{\sim}$ we constructed a wedge space $\cf_\cs$. Now, the above corollary states that all these wedge spaces $\{\cf_\cs\;|\;\cs\in\ocean(C)/{\sim}\}$ are related to each other by unitary operations from the right.

\begin{figure}[h]
  \begin{center}
    {\includegraphics[scale=0.3]{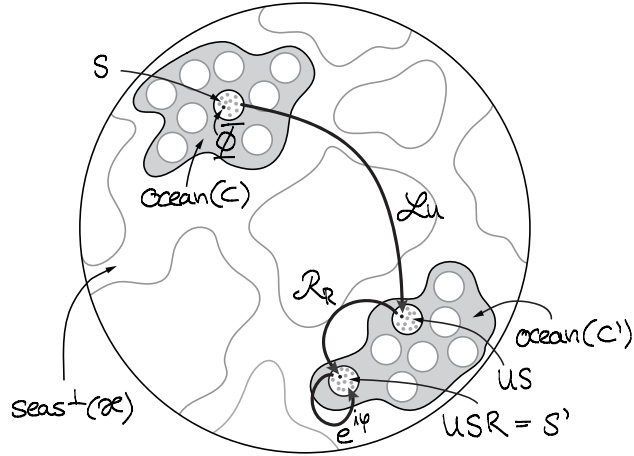}}
    \caption{\label{fig:seas} A sketch of how the equivalence classes are related.}
  \end{center}
\end{figure}

Also together with Lemma \pref{lem:equivalent_right_ops} (Uniqueness up to a Phase) one gets:
\begin{our_corollary}[Uniqueness of the Lift up to a Phase]\label{col:uniqueness up to a phase}
  Given $C,C',\cs,\cs'$ as in Theorem \pref{unser abstrakter Shale Stinespring} (Lift Condition), let $U\in \ur^0(\ch,C; \ch', C')$. Take an $R\in \cu(\ell)$ as in Theorem \pref{unser abstrakter Shale Stinespring} (Lift Condition). Then the elements of the set
  \begin{align*}
    \{\rop Q \rop R \lop U\;|\; Q\in \cu(\ell)\cap\left(\id_\ell+\trcl(\ell)\right)\}=\{e^{i\varphi}\rop
    R \lop U\;|\;\varphi\in\R\}
  \end{align*}
  are the only unitary maps from $\cf_\cs$ to $\cf_{\cs'}$ in the set $\{\rop T \lop U\;|\;T\in\cu(\ell)\}$.
\end{our_corollary}
In this sense we refer to the lift $\lop U \rop R$ as being unique up to a phase. A typical situation is this: Consider for example the one-particle Dirac time-evolution $U:\ch\to\ch$ and assume that $U\in\ur^0(\ch,C;\ch,C')$ for two given polarization classes $C,C'\in \pol(\ch)/{\approx}_0$; we justify this assumption in Section \pref{time-evolution_ex} below. We choose $\Phi,\Phi'\in\seas^\perp(\ch)$ such that $\range\Phi\in C$ and $\range\Phi'\in C'$. By Lemma \pref{lem:connection sim approx} (Connection between $\sim$ and $\approx_0$) it follows that $\cs=\cs(\Phi)\in\ocean(C)/{\sim}$ and $\cs'=\cs(\Phi)\in\ocean(C')/{\sim}$ hold. The elements of the associated  wedge spaces $\cf_{\cs}$ and $\cf_{\cs'}$ represent the ``in" and ``out" states, respectively. Theorem \pref{unser abstrakter Shale Stinespring} (Lift Condition) and Corollary \pref{col:uniqueness up to a phase} (Uniqueness of the Lift up to a Phase) assure for the $\cs,\cs'$ there is an $R\in \cu(\ell)$ such that
\begin{align*}
  \xymatrix{
    \cf_{\cs} \ar[r]^{\lop U} & \cf_{U\cs} \ar[r]^-{\rop R} & *[r]{\cf_{U\cs R}} & *[l]{=\cf_{\cs'}} \ar@(dr,ur)[]_{e^{i\varphi}} & *[r]{\varphi\in\R}
  }.
\end{align*}
We have illustrated this situation in Figure \pref{fig:seas}.

\subsection{Comparison with Standard Fock Spaces} \label{comparison}

 In this subsection we sketch the relation of infinite wedge spaces to standard Fock spaces omitting complete proofs.

\subsubsection{Connection with Constant Charge Sectors of Fock Spaces}

As before, let $\ch$ be a complex separable Hilbert space. We pick an orthonormal basis $\varphi = (\varphi_j)_{j\in\Z}$ of $\ch$ which will be fixed throughout this Subsection \pref{comparison}. An \emph{infinite form} is a \emph{formal} expression of the form
\begin{equation}\label{inf form}
\psi=\varphi_{j_1}\wedge \varphi_{j_2} \wedge \ldots = \bigwedge_{n\in\N} \varphi_{j_n},
\end{equation}
where $(j_n)_{n\in\N}$ is a strictly increasing sequence with the property that $j_{n+1} =j_n+1 $ for all $n$ larger than a suitable $n_1\in\N$. Let $B = B(\varphi)$ be the set of all such forms. These forms $\psi\in B(\varphi)$, $\psi= \varphi_{j_1}\wedge \varphi_{j_2}\wedge\ldots$, are infinite formal exterior products of basis elements $\varphi_j$, where only finitely many $\varphi_j$ with $j<0$ occur and all except for finitely many of the $\varphi_k$ with $k>0$ occur. This description is Dirac's original picture of his sea \cite{Dirac:34}.
\begin{our_definition} \label{charge}  The charge $C(\psi)$ of a form $\psi\in B$ is the value $c\in \Z$ with $j_n=n-c$ for suitably large $n$. Denote $B_c=B_c(\varphi):= \{\psi \;|\; C(\psi)= c\}$  for $c\in \Z$.
\end{our_definition}
 Now one can define a natural Fock space (attached to $\varphi$) as the uniquely determined Hilbert space generated by $B$ and having $B$ as an orthonormal basis. However, we want to get directly into contact with the infinite wedge spaces of Subsections \pref{sec:construction} and \pref{sec:towards a solution}. To do this we observe that a form $\psi\in B(\varphi)$ can also be regarded  as to be the sequence $\psi=(\psi_n)_{n\in\N}$ with $\psi_n=\varphi_{j_n}$. The condition $C(\psi)=0$ means  $\psi_n=\varphi_n$ for $n\geq n_1$ and therefore implies that the $\N\times\N$ matrices $( \sk {\psi_n,\varphi_m} )_{m,n\in\N}$ and $(\sk {\psi_n,\psi_m} )_{m,n\in\N}$ differ from the identity only by a trace class operator and thus have a determinant. Hence, $B_0(\varphi)$ regarded as a set of sequences in $\ch$ is a subset of the equivalence class $\cs(\varphi)$, which we defined in Subsection \pref{sec:towards a solution} of the introduction. We abbreviate $\cf_{\cs(\varphi)}$ by $\cf_\varphi$ here, and in the following. By a slight abuse of notation the form $\psi$ can be regarded as to be the element $\Wedge\psi$ of $\cf_\varphi$. With this identification $B_0\subset \cf_\varphi$ is orthonormal since $\sk{\psi,\psi'}= \det (\sk{\varphi_{j_n}, \varphi_{j'_n}})_{n\leq n_2}$ whenever $\psi'=\varphi_{j'_1}\wedge \varphi_{j'_2} \wedge \ldots \in B_0$ with $j'_n = n$ for $n \geq n'_1 \geq n_1$. Using similar techniques as in Lemma \pref{separable} (Separability) one can show the following statement which we do not prove here:
\begin{our_lemma}
\label{endlich dicht}
$B_0(\varphi)\subset  \cf_\varphi$ spans a dense subspace of $\cf_\varphi$ and therefore is an orthonormal basis.
\end{our_lemma}
This result extends to arbitrary charges $c\in\Z.$ For each $c\in\Z$ the sector $\cf_{\varphi,c}$ is defined in analogy to $\cf_\varphi$: In the notation of Sect. \pref{sec:construction} let $\Phi_c\in \seas_\ell(\ch)$, where $\ell=\ell_2(\N)$, be defined by $\Phi_c(e_i):=\varphi_{i-c}$ for $i\in\N$. We obtain the Hilbert space  $\cf_{\varphi,c}:= \cf_{\cs(\Phi_c)}$ (see Definition \pref{def dach} (Infinite Wedge Spaces)) as the Fock space sector of charge $c$. Note that $\cf_{\varphi,0} = \cf_\varphi$.
In the same way as before $B_c(\varphi)$ is an orthonormal basis of $\cf_{\varphi,c}$. We call the Hilbert space sum
\begin{align}
  \cf_\varphi^\infty := \bigoplus_{c\in\Z} \cf_{\varphi,c}
\end{align}
to be the (full) \emph{Fock space associated} to $\varphi$ with the sectors $\cf_{\varphi,c}$ of charge $c$. Observe, that $B=B(\varphi)$ is an orthonormal basis of $\cf^\infty_\varphi$. The given basis $(\varphi_j)_{j\in\Z}$ induces a polarization $\ch=\ch_{+}\oplus \ch_{-}$ of $\ch$, where $\ch_{+}$ respectively $\ch_{-}$ is the closed subspace generated by $\{\varphi_j \;|\; j\in \Z\,,\, j\leq 0\,\}$ respectively $ \{ \varphi_j \;|\; j\in \Z\,,\, j > 0\,\}$.

Let $\bigwedge ^\bullet \ch_+$ and $\bigwedge ^\bullet \ch_-$ be their exterior algebras and let $\cf(\ch_+,\ch_-) :=\bigwedge ^\bullet \ch_+ \otimes \bigwedge ^\bullet \overline {\ch_-}$ be the \emph{standard Fock space}, in both cases in the Hilbert space sense. Then $B(\varphi)$ can be interpreted as a special listing of an orthonormal basis of the standard Fock space $\cf(\ch_+,\ch_-)$. Namely,  to an element $\psi = \bigwedge_{n=1}^\infty \varphi_{j_n}$ (in the sense of equation (\pref{inf form})) corresponds the vector
\begin{align}
  i_\varphi(\psi) := \varphi_{j_1}\wedge \ldots \wedge \varphi_{j_M} \otimes \varphi_{i_1}\wedge \ldots  \wedge \varphi_{i_N} \in  \ch_+^{\wedge M} \otimes \overline {\ch_-}^{\wedge N},
\end{align}
where $j_1 < j_2 < \ldots < j_M \leq 0 \leq j_{M+1}$ and $\{i_1, \ldots, i_N\} = \N\setminus \{j_n \;|\; n> M\}\,,\, i_1 < \ldots <i_N.$  Note that $C(\psi) = M-N$ in this case. We obtain:
\begin{our_proposition} The subset $i_\varphi (B)\in \cf(\ch_+,\ch_-)$  is an orthonormal basis of $\cf(\ch_+,\ch_-)$. Consequently, $\psi \mapsto i_\varphi(\psi)\,,$ induces an isometric isomorphism
\begin{align}
  i_\varphi:  \cf_\varphi^\infty \to \cf(\ch_+,\ch_-)
\end{align}
of Hilbert  spaces. The $c$-sector $\cf^\infty_{\varphi,c}$ is mapped by $i_\varphi$ onto the $c$-sector
\begin{align}
  \cf_c(\ch_+,\ch_-) := \bigoplus_{M,N\in\N,M= c+N} \ch_+^{\wedge M} \otimes  \overline {\ch_-}^{\wedge N}
\end{align}
of $\cf(\ch_+,\ch_-)\,,$ and the vacuum $\Omega_\varphi =\varphi_1\wedge \varphi_2\wedge\ldots \in \cf_\varphi$ is mapped to the vacuum $\Omega = 1\otimes 1\in \cf_{\varphi,0}(\ch_+,\ch_-)$.
\end{our_proposition}

\subsubsection{Creation Operators}

For every $\chi\in \ch$ the \emph{creation operator} $a^*_\chi$ on the $c$-sector $\cf_{\varphi,c}$ is the map $a^*_\chi: \cf_{\varphi,c} \to \cf_{\varphi,c+1}$ induced by
\begin{align}
   a^*_{\chi}: \varphi_{j_1} \wedge \varphi_{j_2} \wedge \ldots \mapsto \chi\wedge \varphi_{j_1} \wedge \varphi_{j_2}\wedge\ldots
\end{align}
and extended appropriately. Then $a^*_\chi$ is a well-defined linear operator
$a^*_\chi : \cf_{\varphi,c}\to \cf_{\varphi, c+1}$ of  norm $\|\chi\|$  depending complex linearly on $\chi$ which induces the norm $\|\chi\|$ operator $a^*_\chi: \cf_\varphi^\infty \to \cf_\varphi^\infty$ on the full Fock space. If we use for the standard creation operator on the standard Fock space $\cf(\ch_+,\ch_-)$ the same symbol $a^*_\chi: \cf(\ch_+,\ch_-) \to \cf(\ch_+,\ch_-)$ we obtain the commutativity  $i_\varphi\circ a^*_\chi=a^*_\chi\circ i_\varphi$:
\begin{align}
  \begin{split}
    \label{diagram phi}
    \xymatrix{\cf_\varphi^\infty \ar[r]^{a^*_\chi} \ar[d]_{i_\varphi} & \cf_\varphi^\infty \ar[d]^{i_\varphi}\\
                    \cf(\ch_+,\ch_-) \ar[r]^{a^*_\chi}  & \cf(\ch_+,\ch_-)}
  \end{split}
\end{align}
The commutativity in Diagram (\pref{diagram phi}) holds similarly for the \emph{annihilation operator} $a_\chi: \cf_\varphi^\infty \to \cf_\varphi^\infty$ which is the adjoint of $a^*_\chi$. Note that for all $\chi,\chi'\in \ch$, $a_\chi\,,\, a^*_{\chi'}$ satisfy the canonical anti-commutation relations (CAR)
\begin{equation} \label{CAR}
\{a_\chi,a^*_{\chi'}\}=a_\chi a^*_{\chi'}+ a^*_{\chi'}a_\chi = \sk{\chi,\chi'}.
\end{equation}
The actions of the creation and annihilation operators on $\cf_\varphi^\infty$ yield another description of the space $\cf_\varphi$.
 Each $\psi=\psi_1\wedge\psi_2\wedge\ldots\in \Wedge\cs(\varphi)\subset \cf_\varphi$ can be expressed as the following limit:
 \begin{align}
   \psi= \lim_{n\to\infty} a^*_{\psi_1}\ldots a^*_{\psi_n}a_{\varphi_n}\dots a_{\varphi_1}\Omega_\varphi
\end{align}
with the vacuum $\Omega_\varphi = \varphi_1\wedge \varphi_1\ldots \wedge \varphi_n\wedge \ldots $ of $\cf_\varphi$. Using the commutativity of Diagram (\pref{diagram phi}) and the corresponding one for $a_\chi$ this property establishes:
\begin{equation}
i_\varphi(\psi)= \lim_{n\to\infty} a^*_{\psi_1}\ldots a^*_{\psi_n}a_{\varphi_n}\dots a_{\varphi_1}\Omega.
\end{equation}

\subsubsection{Connection to the Shale-Stinespring Criterion}
\label{Shale}

In general, Fock spaces arise as representation spaces of the CAR algebra $\mathfrak A= \mathfrak A(\ch)$ of a given Hilbert space $\ch$. $\mathfrak A$ is the natural $C^*$-algebra with 1 generated by the elements of $\ch$ and respecting the relations (\pref{CAR}).

Let $\ch=\ch_+\oplus \ch_-$ a polarization and $\varphi = (\varphi_n)_{n\in \Z}$ an adapted orthonormal basis, i.e. $(\varphi_n)_{n\in\N}$ is an orthonormal basis  of $\ch_-$ and $(\varphi_n)_{n\leq0}$ is an orthonormal basis of $\ch_+$. Then the annihilation and creation operators   induce a representation
\begin{align}
  \pi_\varphi : \mathfrak A(\ch) \to B(\cf^\infty_\varphi)
\end{align}
by $\pi_\varphi (\chi) := a_\chi + a_\chi^*: \cf_\varphi^\infty \to \cf_\varphi^\infty\,,\, \chi \in \ch$. These representations are irreducible.\\

The question answered by the theorem of Shale-Stinespring \cite{shale:65} is the following: Given two orthonormal basis $\varphi, \varphi'$ (or two polarizations) of $\ch$, when are the representations $ \pi_\varphi\,,\, \pi_{\varphi'}$ are equivalent? That is, when does there exist a unitary (intertwining) operator $T: \cf_\varphi^\infty \to \cf_{\varphi'}^\infty$ such that $T\circ \pi_\varphi (\chi) = \pi_{\varphi'}(\chi)\circ T$ for all $\chi\in \ch$?
\begin{align}
  \begin{split}
    \xymatrix{\cf_\varphi^\infty \ar[r]^{T} \ar[d]_{\pi_\varphi(\chi)} & \cf_{\varphi'}^\infty \ar[d]^{\pi_{\varphi'}(\chi)}\\
                    \cf_{\varphi}^\infty \ar[r]^{T}  & \cf_{\varphi'}^\infty}
  \end{split}
\end{align}
This question is closely related to the \textit{implementation} problem: Given a unitary operator $U \in \cu(\ch)$, when does there exist a unitary operator $U^\sim \in U(\cf_\varphi^\infty)$ such that $ U^\sim\circ \pi _\varphi (\chi) = \pi_\varphi (U\chi)\circ U^\sim$ for all $\chi \in \ch$?
\begin{align}
  \begin{split}
    \xymatrix{\cf_\varphi^\infty \ar[r]^{U^\sim} \ar[d]_{\pi_\varphi(\chi)} & \cf_{\varphi}^\infty \ar[d]^{\pi_{\varphi}(U\chi)}\\
                    \cf_{\varphi}^\infty \ar[r]^{U^\sim}  & \cf_{\varphi}^\infty}
  \end{split}
\end{align}
If the first question can be solved for $\varphi' = U\varphi$, i.e. for $\varphi'_n=U(\varphi_n)$, then $U^\sim :=  \lop{U^{-1}} \circ T $, see Construction \pref{constr:leftop}, is an answer to the second problem, where $\lop U: \cf_\varphi^\infty \to \cf_{U\varphi}^\infty$ is the natural left action induced by $U$. This holds since $\lop U\circ \pi_\varphi(U\chi) = \pi_{U\varphi}(\chi) \circ \lop U = \pi_{\varphi'}(\chi)\circ \lop U$, hence
$$U^\sim\circ \pi_\varphi(\chi) =  \lop {U^{-1}} \circ T \circ \pi_\varphi(\chi) =
\lop{U^{-1}}  \circ \pi_{U\varphi(\chi)} \circ T = \pi_\varphi (U\chi)\circ \lop{U^{-1}} \circ T =\pi_\varphi(U\chi)\circ U^\sim.$$
\begin{align}
  \begin{split}
    \xymatrix{\cf_\varphi^\infty \ar[r]^{T} \ar[d]_{\pi_\varphi(\chi)} & \cf_{\varphi'}^\infty \ar[d]^{\pi_{\varphi'}(\chi)} \ar[r]^{\lop {U^{-1}}} & \cf_\varphi^\infty \ar[d]^{\pi_\varphi(U\chi)}\\
                    \cf_{\varphi}^\infty \ar[r]^{T}  & \cf_{\varphi'}^\infty \ar[r]^{\lop {U^{-1}}} & \cf_\varphi^\infty}
  \end{split}
\end{align}
Conversely, considering the second question for the unitary map $U$ defined by $U\varphi_i = \varphi'_i\,,\,i\in\Z$, the implementation $U^\sim$ composed with the left action $\lop U: \cf_\varphi \to \cf_{\varphi'} $ yields an intertwining operator thus answering the first question.

The answer to each of the two questions is that $U$ has to be in the restricted unitary group; see \cite{shale:65}.

We obtain this result in a slightly different situation with our methods in the following simple way. We ask for an implementation of the unitary map $U\in \cu(\ch)$ with the additional requirement that the implementation $U^\sim$ should not change the charge, i.e. $U^\sim (\cf_{\varphi,c}) \subset \cf_{\varphi,c}$, in particular $U^\sim \in U(\cf_\varphi)$.

Let us, first of all, consider the left operation $\lop U: \cf_\varphi^\infty \to \cf_{U\varphi}^\infty$ for a given unitary $U:\ch\to \ch$. Altering $U\varphi$ to $\varphi' := U\varphi R$ by the operation from the right induced by an appropriate $R$, see Construction \pref{def:rightop}, we obtain  a unitary map $\rop R\circ \lop U: \cf_\varphi^\infty \to \cf_{\varphi'}^\infty$.
This approach gives an implementation $U^\sim = \rop R\circ \lop U$ if and only if $R\in \cu(\ell)$
can be chosen such that $\cf_\varphi^\infty = \cf_{\varphi'}^\infty$ or, equivalently such that $U\cs(\varphi)R=\cs(\varphi)$.
According to Theorem \pref{unser abstrakter Shale Stinespring} (Lift Condition) this holds together with the charge conservation exactly when $U$ is in the group $U_{\mathrm{res}}^0(\ch,[\ch_-]_{\approx_0})$, see Definition \pref{def:ures0}. As a result we have:
\begin{our_theorem} The following conditions are equivalent for $U\in U(\ch)$:
\begin{itemize}
\item $U$ has an implementation $U^\sim : \cf_\varphi \to \cf_\varphi$ (and in the same way  $U^\sim \in U(\cf_0(\ch_+,\ch_-))$).
\item $U$ has an implementation $U^\sim : \cf_\varphi^\infty \to \cf_\varphi^\infty$  with $U^\sim (\cf_{\varphi,c}) \subset \cf_{\varphi,c}$ for all $c\in \Z$ (and in the same way  $U^\sim \in U(\cf(\ch_+,\ch_-))$).
\item $U\in U_{\mathrm{res}}^0(\ch_+,[\ch_-]_{\approx_0})$.
\end{itemize}
\end{our_theorem}
Disregarding the charge condition one can show the original Shale-Stinespring theorem \cite{shale:65} in a similar straight-forward way. 

\section{Application to the External Field Problem in QED}\label{sec:external 
field}
\label{time-evolution_ex}
We now come to the one-particle Dirac time-evolution in an external
four-vector field $\sa\in {\mathcal C}^\infty_c(\R^4,\R^4)$\nomenclature{${\mathcal C}^\infty_c(\R^4,\R^4)$}{Infinitely often continuously differentiable $\R^4$
valued functions on $\R^4$ with compact support}, i.e. the set of infinitely
often differentiable $\R^4$ valued functions on $\R^4$ with
compact support. Recall the discussion at the end of Subsection
\pref{sec:lift cond}:
In order to apply Theorem \pref{unser abstrakter Shale Stinespring} (Lift Condition) to the one-particle Dirac time-evolution $U^\sa(t_1,t_0)$ for fixed
$t_0,t_1\in\R$ and in
this way to obtain a lift to unitary maps from one wedge space to another (the second quantized time-evolution) we need to show that $U^\sa(t_1,t_0)\in\ur^0(\ch,C(t_0);\ch,C(t_1))$ for appropriate $C(t_0),C(t_1)\in\pol(\ch)/{\approx}_0$.  To ensure this condition holds is the main content of this last section.

This section is structured as follows: In the first Subsection, we show that for
any $t_0,t_1\in\R$ there exist $C(t_0),C(t_1)\in\pol(\ch)/{\approx}_0$,
depending only on $\sa(t_0)$ and $\sa(t_1)$, respectively,
such that $U^\sa(t_1,t_0)\in\ur^0\big(\ch,C(t_0);\ch,C(t_1)\big)$.
In the second Subsection we identify the polarization classes $C(t)$
uniquely by the magnetic components of $\sa(t)$ for all $t\in\R$.
The third Subsection
combines these results with Section \pref{sec:wedge spaces} and shows the
existence of the second quantized Dirac time-evolution for the external field
problem in QED.
Finally, the fourth Subsection concludes with the analysis of
second quantized gauge transformations as unitary maps between
varying Fock spaces.

\subsection{One-Particle Time-Evolution}
\label{time-evolution}

Throughout this section we work with $\ch=L_2(\R^3,\C^4)$, with
$\R^3$ being interpreted as momentum space.
The free Dirac equation in momentum representation is given by
\begin{align}\label{eqn:freediraceq}
i\frac{d}{dt}
\psi^0(t)&=H^0\psi^0(t)
\end{align}
for $\psi^0(t)\in\operatorname{domain}(H^0)\subset\ch$, where $H^{0}$ is the multiplication operator with
\begin{align}
H^0(p)&=\alpha\cdot p +\beta m=\sum_{\mu=1}^3\alpha^\mu p_\mu+\beta m,\quad p\in\R^3\label{eqn:hamiltonian_formula}
\end{align}
and the $\C^{4\times 4}$ Dirac matrices $\beta$ and $\alpha^\mu$, $\mu=1,2,3$, fulfill
\begin{align}\label{eqn:dirac matrices}
  \begin{split}
    \beta^2 &= 1\\
    ({\alpha^\mu})^2 &= 1
  \end{split} &&
  \begin{split}
    \{\alpha^\mu,\beta\}&:=\alpha^\mu\beta+\beta\alpha^\mu = 0\\
    \{\alpha^\mu,\alpha^\nu\}&:=\alpha^\mu\alpha^\nu+\alpha^\nu\alpha^\mu = 2\delta^{\mu \nu}.
  \end{split}
\end{align}
Which specific representation of this matrix algebra with hermitean matrices
is used does not affect any of the following arguments. For convenience we introduce also
\begin{align*}
  \alpha^0 = 1\in\C^{4\times4}.
\end{align*}
$H^0$ is a self-adjoint multiplication operator which generates a
one-parameter group of unitary operators
\begin{align}\label{eqn:u0}
  U^0(t_1,t_0)=U^0(t_1-t_0):=\exp(-i(t_1-t_0)H^0)
\end{align}
on $\ch$ for all $t_0,t_1\in\R$. The matrix $H^0(p)$ has double eigenvalues $\pm E(p)$, where $E(p)=\sqrt{|p|^2+m^2}>0$, $p\in\R^3$. Therefore, the
spectrum of the free Dirac operator
is $\sigma(H^0)=(-\infty,-m]\cup[+m,+\infty)$ and the corresponding free spectral projectors $P_\pm$ are multiplication operators with the matrices
\begin{equation}\label{eqn:projector_formula}
P_\pm(p)=\frac12\left(1\pm\frac{H^0(p)}{E(p)}\right).
\end{equation}
We define $\ch_\pm:=P_\pm\ch$ for which $\ch=\ch_-\oplus\ch_+$. For any linear operator $L$ on $\ch$ and signs $\sigma,\tau\in\{+,-\}$ we write
$L_{\sigma\tau}=P_\sigma L P_\tau$. Furthermore, $L_\ev=L_{++}+L_{--}$
denotes the even (diagonal) part, and
$L_\odd=L_{+-}+L_{-+}$ for the odd (non-diagonal) part of $L$.
If $L$ has an integral kernel $(q,p)\mapsto L(p,q)$, the kernel of
$L_{\sigma\tau}$ is given by $(p,q)\mapsto L_{\sigma\tau}(p,q)=
P_\sigma(p)L(p,q) P_\tau(q)$.

Now, let $\sa=(\sa_\mu)_{\mu=0,1,2,3}\in {\mathcal C}_c^\infty(\R^4,\R^4)$ be a smooth, compactly supported, external four-vector field. We denote its time slice at time $t\in\R$ by $\sa(t)=(\R^3\ni x\mapsto (\sa_\mu(t,x))_{\mu=0,1,2,3})$.
The Dirac equation with the external field $\sa$ in momentum representation is then given by
\begin{align}\label{eqn:Dirac}
i\frac{d}{dt}
\psi(t)&=H^{\sa(t)}\psi(t) =\left(H^0+iZ^{\sa(t)}\right)\psi(t)
\end{align}
where for $A=(A_\mu)_{\mu=0,1,2,3}=(A_0,-\vec A)\in {\mathcal C}_c^\infty(\R^3,\R^4)$, the operator
$Z^A$ on $\ch$ is defined as follows:
\begin{align}
i Z^A&= e\sum_{\mu=0}^3\alpha^\mu \widehat A_\mu,
\end{align}
denoting the elementary charge by $e$.
Here we understand $\widehat A_\mu$, $\mu=0,1,2,3$, as convolution operators
\begin{equation}\label{eqn:fourier transform A}
(\widehat A_\mu\psi)(p)=
\int_{\R^3}\widehat A_\mu(p-q)\psi(q)\,dq,\quad p\in\R^3,
\end{equation}
 for $\psi\in\ch$ and $\widehat A_\mu$ being the Fourier transform of $A_\mu$ given by
\begin{equation}
\label{Fourier transform}
\widehat A_\mu(p)=\frac{1}{(2\pi)^3}\int_{\R^3}e^{-ipx}A_\mu(x)\,dx.
\end{equation}
Therefore, in momentum representation, $Z^A$
is an integral operator with integral kernel
\begin{equation}
(p,q)\mapsto Z^A(p-q)=-ie\sum_{\mu=0}^3\alpha^\mu \widehat A_\mu(p-q),\quad p,q\in\R^3.
\label{eqn:def Z}
\end{equation}
The Dirac equation with external field $\sa$ gives also rise to a family of unitary operators $(U^\sa(t_1,t_0))_{t_0,t_1\in\R}$ on $\ch$ which fulfill
\begin{align}
\label{ua}
\frac{\partial}{\partial t_1}U^\sa(t_1,t_0)&=-iH^{\sa(t_1)}U^\sa(t_1,t_0),\\
\label{ub}
\frac{\partial}{\partial t_0}U^\sa(t_1,t_0)&=iU^\sa(t_1,t_0)H^{\sa(t_0)}
\end{align}
on the appropriate domains,
such that for every solution $\psi(t)$ of equation (\pref{eqn:Dirac}) one
has $\psi(t_1)=U^\sa(t_1,t_0)\psi(t_0)$;
see \cite{thaller:92}.

We now introduce key objects of this work:
\begin{our_definition}[Induced Polarization Classes]\label{def:indpolclasses}
For $A\in {\mathcal C}_c^\infty(\R^3,\R^4)$, we define the
integral operator $Q^A:\ch\to \ch$
by its integral kernel, also denoted by $Q^A$:
\begin{align}\label{eqn:operator_Q}
\R^3\times\R^3\ni
(p,q)\mapsto Q^A(p,q)&:=\frac{Z_{+-}^A(p,q)-Z_{-+}^A(p,q)}{i(E(p)+E(q))}\\
\mbox{with }
Z^A_{\pm\mp}(p,q)&:=
P_\pm(p) Z^A(p-q) P_\mp(q)\nonumber.
\end{align}
Furthermore, we define  the polarization class $C(0):=[\ch_-]_{\approx_0}$
belonging to the negative spectral space $\ch_-$ of the free Dirac operator
$H^0$, and therewith the polarization classes
\begin{align}
  C(A):=e^{Q^{A}}C(0)=\{e^{Q^{A}}V\;|\;V\in C(0)\}.
\end{align}
\end{our_definition}
The operators $Q^A$ are bounded and skew-adjoint, and thus, the operators $e^{Q^{A}}$ are unitary. They will appear naturally in the
iterative scheme that we use to control
the time-evolution,
and their origin will become clear as we go along
(Lemma \pref{lem:partial integration}).

We now state the main result of this section,
using the notation of Section \pref{sec:wedge spaces}.
\begin{our_theorem}[Dirac Time-Evolution with External Field]
\label{Hilbert-Schmidt}
For all four-vector potentials $\sa\in {\mathcal C}^\infty_c(\R^4,\R^4)$
and times $t_1,t_0\in\R$ it is true that
\begin{equation*}
U^\sa (t_1,t_0)\in \ur^0\big(\ch,C(\sa(t_0));\;\ch,C(\sa(t_1))\big).
\end{equation*}
\end{our_theorem}
We do {\it not} focus on finding the weakest regularity conditions
on the external four-vector potential $\sa$ under
which this theorem holds, although much weaker conditions will suffice.
Actually, the theorem and also its proof remain valid
for four-vector potentials $\sa$ in the following class
${\mathcal A}\supset {\mathcal C}_c^\infty(\R^4,\R^4)$:
\begin{our_definition}[Class of External Four-Vector Potentials]
\label{def: class A}
Let ${\mathcal A}$ be the class of four-vector potentials
$\sa=(\sa_\mu)_{\mu=0,1,2,3}:\R^4\to\R^4$ such that for all $\mu=0,1,2,3$, $m=0,1,2$ and $p=1,2$ the integral
\begin{align}
\int_{\R}\left\|\frac{d^m}{dt^m}\widehat \sa_\mu(t)\right\|_p\,dt
\end{align}
exists and is finite.
Here $\widehat\sa_\mu(t)$ denotes the Fourier transform of a time slice
$\sa_\mu(t)$ with respect to the {\it spatial} coordinates.
\end{our_definition}
This class of four-vector potentials has also been considered by Scharf
in his analysis of the second-quantized scattering
operator in an external potential (Theorem 5.1 in \cite{scharf:95}).
We remark that the class ${\mathcal A}$
does {\em not} contain the Coulomb potential, not even when one truncates
it at large times.

Since quite some
computation is involved in the proof of the above theorem,
we split it up into a series of small lemmas,
to separate technicalities from ideas. Here is the skeleton of
the proof:
\begin{align*}
  \xymatrix@1{
    \txt{%
      Theorem \pref{Hilbert-Schmidt}\\
      (Dirac Time-Evolution\\
       with External Field)}
    &
    &
    \ar[ll]
    \txt{%
      Lemma \pref{lem:gronwall}\\
      (Gr\"onwall Argument)}\\
    \ar@(u,dl)[urr]
    \ar[u]
    \txt{%
      Lemma \pref{Lemma HS}\\
      ($\hs$ Estimates)}
    &
    \ar@(u,dl)[ur]
    \txt{%
      Lemma \pref{lem:bornseries}\\
      (Fixed Point Form\\of the Dirac Equation)} &
    \ar[u]
    \txt{%
      Lemma \pref{lem:partial integration}\\
      (Partial Integration)}
  }
\end{align*}
The key ideas are worked out in Lemma \pref{lem:gronwall} (Gr\"onwall Argument). The other lemmas have a more technical character.

In the following, when dealing with a given external vector potential
$\sa\in{\mathcal A}$,
we abbreviate $U(t_1,t_0)=U^\sa(t_1,t_0)$, $H(t)=H^{\sa(t)}$,
$Z(t)=Z^{\sa(t)}$, and $Q(t)=Q^{\sa(t)}$.
We start with putting things
together:

\begin{proof}[Proof of Theorem \pref{Hilbert-Schmidt} (Dirac Time-Evolution with External Field)]
   By Lemma \pref{lemma approx}(b), we need only to show that for
some $V\in C(\sa(t_1))$ and
some $W\in C(\sa(t_0))$ it is true that
  \begin{align}\label{eqn:hscond}
    P_{V^\perp} U(t_1,t_0)P_W, P_{V} U(t_1,t_0)P_{W^\perp}\in \hs(\ch).
  \end{align}
  Let us choose $V=e^{Q(t_1)}\ch_-\in C(\sa(t_1))$ and
$W=e^{Q(t_0)}\ch_-\in C(\sa(t_0))$.
  Then, claim (\pref{eqn:hscond}) is equivalent to
  \begin{align}
    e^{Q(t_1)} P_\pm e^{-Q(t_1)}U(t_1,t_0)e^{Q(t_0)} P_\mp e^{-Q(t_0)} \in \hs(\ch).
  \end{align}
  Since $e^{Q(t_1)}$ and $e^{-Q(t_0)}$ are both unitary operators,
this claim is equivalent to
  \begin{align}
    P_\pm e^{-Q(t_1)}U(t_1,t_0)e^{Q(t_0)}
   P_\mp \in \hs(\ch).
  \end{align}
Now $Q(t)$ is a bounded operator, and
  Lemma \pref{Lemma HS} ($\hs$ Estimates) states that $Q^2(t)\in\hs(\ch)$ for any time
  $t\in\R$.
Therefore, by expanding $e^{\pm Q(t)}$ in its series, we find that
  \begin{align}
    e^{\pm Q(t)}-(\id_{\ch}\pm Q(t))\in\hs(\ch)
\label{first order eQ}
.
  \end{align}
Hence it suffices to prove
  \begin{align}
    P_\pm (\id_{\ch}-Q(t_1))U(t_1,t_0)(\id_{\ch}+Q(t_0)) P_\mp \in \hs(\ch)
.
  \end{align}
  This is just the claim (\pref{eqn:PQUQP formula}) of
  Lemma \pref{lem:gronwall} (Gr\"onwall Argument)
  and concludes the proof.
\end{proof}

The following stenographic notation will be very convenient:
For families of operators $A=(A(t_1,t_0))_{t_1\ge t_0}$
and $B=(B(t_1,t_0))_{t_1\ge t_0}$,
indexed by time intervals $[t_0,t_1]\subset\R$, we set
\begin{equation*}
AB=\left(\int_{t_0}^{t_1} A(t_1,t)B(t,t_0)\,dt\right)_{t_1\ge t_0},
\end{equation*}
whenever this is well-defined.
Furthermore, if $C=(C(t))_{t\in\R}$ and $D=(D(t))_{t\in R}$
denote families of operators indexed by time points, we abbreviate
$AC=(A(t_1,t_0)C(t_0))_{t_1\ge t_0}$,
$CA=(C(t_1)A(t_1,t_0))_{t_1\ge t_0}$, and
$CD=(C(t)D(t))_{t\in\R}$.
The operator norm on bounded operators on
$\ch$ is denoted by $\|\cdot\|$. We set
\begin{align}
\begin{split}
\norm{A}_{\infty}&:=\sup_{s,t\in\R:\,s\ge t}
\norm{A(s,t)},
\\
\norm{C}_{1}&:=\int_{\R} \norm{C(t)}\,dt,
\end{split}
&&
\begin{split}
\norm{A}_{\hs,\infty}&:=\sup_{s,t\in\R:\,s\ge t}
\hsnorm{A(s,t)},
\\
\norm{C}_{\hs,\infty}&:=\sup_{t\in\R}
\hsnorm{C(t)},
\end{split}
\label{def norms}
\end{align}
whenever these quantities exist. Recall Definition \pref{def:indpolclasses} (Induced Polarization Classes)
of the operators $Q(t)=Q^{\sa(t)}$, and let
$(Q'(t):\ch\to\ch)_{t\in \R}$ denote their time derivative,
defined by using the time derivative of the corresponding kernels
\begin{align}
\label{eqn: def Q'}
  (p,q)\mapsto Q'(t,p,q):=
\frac{\partial}{\partial t} Q^{\sa(t)}(p,q) ,\quad p,q\in\R^3, t\in\R.
\end{align}

We now provide the lemmas in the above diagram: For our purposes, the following
fixed point form (\pref{fixa}) of the Dirac equation
is technically more convenient to handle than the Dirac equation in
its differential form (\pref{eqn:Dirac}), as the fixed point equation
gives rise to iterative approximation methods and deals only
with bounded operators. We could have used it as our starting point.

\begin{our_lemma}[Fixed Point Form of the Dirac Equation]\label{lem:bornseries}
The one-particle Dirac time-evolution $U$ fulfills the
fixed point equation
  \begin{align}
    \label{fixa}
    U=U^0+U^0ZU.
  \end{align}
\end{our_lemma}
we only sketch its proof:
\begin{proof}
Using the Dirac equation in the form
(\pref{ua}-\pref{ub}), we get for $t_0,t_1\in\R$ on an appropriate domain:
\begin{align}
\label{d U0 U}
\frac{\partial}{\partial t}
[U^0(t_1,t)U(t,t_0)]=-iU^0(t_1,t)[H^{\sa(t)}-H^0]U(t,t_0)
=U^0(t_1,t)Z(t)U(t,t_0).
\end{align}
Note that although $H^{\sa(t)}$ and $H^0$ are unbounded operators,
their difference $iZ(t)$ is a bounded operator.
Integrating (\pref{d U0 U}),
and using $U(t,t)=\id_\ch=U^0(t,t)$, we get
\begin{align}
U(t_1,t_0)=U^0(t_1,t_0)+\int_{t_0}^{t_1} U^0(t_1,t)Z(t)U(t,t_0)\,dt.
\end{align}
This equation recast in our stenographic notation
is the fixed point equation (\pref{fixa}) for $U$.
\end{proof}
Iterating this fixed point equation leads to the well-known Born series.

\begin{our_lemma}[Gr\"onwall Argument]\label{lem:gronwall}
For all $t_0,t_1\in \R$, the following holds:
  \begin{align}\label{eqn:PQUQP formula}
    P_\pm (\id_{\ch}-Q(t_1))U(t_1,t_0)(\id_{\ch}+Q(t_0)) P_\mp \in \hs(\ch)
  \end{align}
\end{our_lemma}

\begin{proof}
Without loss of generality and to simplify the notation,
we treat only the case $t_1\ge t_0$.
Let
\begin{align}
\label{def-R}
R:=(\id_{\ch}-Q)U(\id_{\ch}+Q).
\end{align}
The strategy is to expand $R$ in a series and to check the Hilbert-Schmidt properties of the non-diagonal part term by term.
Lemma \pref{lem:bornseries} (Fixed Point Form of the Dirac Equation)
states that the Dirac time-evolution $U$ fulfills the fixed-point
equation $U=U^0+U^0ZU$ (equation (\pref{fixa}) below); recall that $U^0$ is the free Dirac time-evolution introduced in (\pref{eqn:u0}).
Iterating this fixed point equation once yields
\begin{equation*}
U=U^0+U^0ZU^0+U^0ZU^0ZU.
\end{equation*}
Before going into the details, let us explain informally
some ideas behind the subsequent proof.
The first-order term $U^0 ZU^0$
appears over and over again. Therefore one may expect that its properties will be inherited by all other orders within the perturbation series.
We therefore take a closer look at this term in Lemma
\pref{lem:partial integration}
(Partial Integration).
Equation (\pref{partial integration}) in this lemma states
  \begin{equation*}
  U^0ZU^0=QU^0-U^0Q-U^0Q'U^0+U^0Z_\ev U^0.
  \end{equation*}
One finds that the non-diagonal part $(U^0ZU^0)_\odd$
does in general not consist of Hilbert-Schmidt operators
because of the first two terms $QU^0-U^0Q$ on the right hand side,
which are the boundary terms of the partial integration.
However, we show now that the transformation induced by $Q$
remedies these terms such that the non-diagonal part of $R$
consists of Hilbert-Schmidt operators.

Substituting the formula (\pref{partial integration}), cited above, into
the fixed point equation $U=U^0+U^0ZU$, we get
\begin{align}
U&=U^0+QU^0-U^0Q-U^0Q'U^0+U^0Z_\ev U^0
\nonumber\\&\quad
+QU^0ZU-U^0QZU-U^0Q'U^0ZU+U^0Z_\ev U^0ZU
\nonumber\\
&=U^0+QU-U^0Q-U^0Q'U+U^0Z_\ev U-U^0QZU.
\end{align}
We rewrite this as
\begin{equation}
\label{term}
(\id_{\ch}-Q)U=U^0(\id_{\ch}-Q)+U^0(-Q'+Z_\ev-QZ)U.
\end{equation}
Multiplying (\pref{term}) with $\id_{\ch}+Q$ from the right
and using the equation
\begin{equation*}
U(\id_{\ch}+Q)=(\id_{\ch}+Q)R+Q^2U(\id_{\ch}+Q),
\end{equation*}
which follows from the Definition (\pref{def-R}) of $R$,
we get
\begin{align}
R&=U^0(\id_{\ch}-Q^2)+U^0(-Q'+Z_\ev-QZ)U(\id_{\ch}+Q)
\nonumber\\&
=U^0(-Q'+Z_\ev-QZ)(\id_{\ch}+Q)R
\nonumber\\&
\quad +U^0(\id_{\ch}-Q^2)+U^0(-Q'+Z_\ev-QZ)Q^2 U(\id_{\ch}+Q).
\label{fixR}
\end{align}
We view  (\pref{fixR}) also as a fixed point equation for $R$.
In order to control the Hilbert-Schmidt norm of the non-diagonals
of $R$, we solve this fixed point equation for $R$ by iteration.
Using the abbreviation
\begin{align}
\label{def-F}
&F:=(-Q'+Z_\ev-QZ)(\id_{\ch}+Q),
\\
\label{def-G}
&G:=-U^0Q^2+U^0(-Q'+Z_\ev-QZ)Q^2 U(\id_{\ch}+Q),
\end{align}
we rewrite
(\pref{fixR}) as $R=U^0FR+U^0+G$ and define recursively
for $n\in\N_0$:
\begin{align}
R^{(0)}:=0, && R^{(n+1)}:=U^0FR^{(n)}+U^0+G.
\end{align}
Although our main interest is to control the Hilbert-Schmidt norm
$\onorm{R(t_1,t_0)}$, we need also some control of the $R^{(n)}$
in the operator norm.
We show first that $\norm{R^{(n)}-R}_\infty\to 0$
as $n\to\infty$.
We have for all $n\in\N_0$
\begin{align}
R^{(n+1)}-R=U^0F(R^{(n)}-R),
\end{align}
which implies
\begin{align}
R^{(n)}-R=(U^0F)^n(R^{(0)}-R)=-(U^0F)^nR.
\end{align}
Now for $s\ge t$, we know $\norm{U^0(s,t)F(t)}= \norm{F(t)}$,
because $U^0(s,t)$ is unitary.
Let  $t_1\ge t_0$. Using the abbreviation
\begin{equation}
I(t_1,t_0):=\{(s_1,\ldots,s_n)\in\R^n\;|\;t_1>s_n>\ldots>s_1>t_0\},
\end{equation}
we get
\begin{align}
&\norm{[R^{(n)}-R](t_1,t_0)}=\norm{[(U^0F)^nR](t_1,t_0)}
\nonumber
\\&
\le  \int_{I(t_0,t_1)}
\norm{F(s_n)}
\norm{F(s_{n-1})}
\ldots
\norm{F(s_1)}
\norm{R(s_1,t_0)}
\,ds_1\ldots ds_n
\nonumber\\&
\le
\frac{\norm{F}_{1}^n}{n!}\norm{R}_{\infty}
\stackrel{n\to\infty}{\longrightarrow}0;
\label{conv}
\end{align}
we use here the bounds
$\norm{F}_{1}<\infty$ and
$\norm{R}_{\infty}<\infty$ from (\pref{bound F and R}) in
Lemma \pref{Lemma HS}, below.
Note that the convergence in (\pref{conv}) is uniform in
the time variables $t_0$ and $t_1$.
This proves the claim
\begin{equation}
\norm{R^{(n)}-R}_\infty\stackrel{n\to\infty}{\longrightarrow}0.
\label{Rn to R}
\end{equation}
As a consequence, we find
\begin{equation}
\label{bound sup Rn}
\sup_{n\in\N_0}\norm{R^{(n)}}_{\infty}
\le \sup_{n\in\N_0}\norm{R^{(n)}-R}_{\infty}+
\norm{R}_{\infty}<\infty.
\end{equation}
Now we split $F$ into its diagonal and non-diagonal parts:
$F=F_\ev+F_\odd$, where
\begin{align}
\label{eqn: Fev}
F_\ev&=Z_\ev-QZ_\odd-QQ'-QZ_\ev Q,
\\
\label{eqn: Fodd}
F_\odd&=Z_\ev Q-QZ_\ev-Q'-QZ_\odd Q;
\end{align}
recall that $Q$ is odd: $Q=Q_\odd$.
We calculate for $n\ge 1$:
\begin{align}
&R^{(n+1)}=U^0FR^{(n)}+U^0+G
\nonumber\\&=U^0F_\ev R^{(n)}+
U^0F_\odd R^{(n)}+U^0+G
\nonumber\\&
=U^0F_\ev R^{(n)}+
U^0F_\odd U^0 FR^{(n-1)}+U^0F_\odd G+U^0F_\odd U^0+U^0+G.
\label{Rn+1}
\end{align}
Estimating the Hilbert-Schmidt norm for the non-diagonals in each summand
on the right hand side in (\pref{Rn+1}) now gives:
\begin{align}
\onorm{R^{(n+1)}(t_1,t_0)}\le
&
\int_{t_0}^{t_1}
\norm{[U^0F_\ev](t_1,t)}\onorm{R^{(n)}(t,t_0)}\,dt
\nonumber\\&
+
\int_{t_0}^{t_1}
\hsnorm{[U^0F_\odd U^0](t_1,t)} \norm{[FR^{(n-1)}](t,t_0)}\,dt
\nonumber\\&
+\int_{t_0}^{t_1}
\norm{[U^0F_\odd](t_1,t)} \hsnorm{G(t,t_0)}\,dt
\nonumber\\&
+\hsnorm{[U^0F_\odd U^0](t_1,t_0)}
+\hsnorm{G(t_1,t_0)}
\nonumber\\&
\le
\int_{t_0}^{t_1}
\norm{F_\ev(t)}\onorm{R^{(n)}(t,t_0)}\,dt+\xconstl{c:gronwall},
\label{hs-recursion}
\end{align}
where we have abbreviated
\begin{align}
\nonumber
\xconstr{c:gronwall}:=&\norm{U^0F_\odd U^0}_{\hs,\infty}
\norm{F}_{1}
\sup_{n\in\N}\norm{R^{(n-1)}}_{\infty}
+\norm{F_\odd}_1 \norm{G}_{\hs,\infty}
\\&
+\norm{U^0F_\odd U^0}_{\hs,\infty}
+\norm{G}_{\hs,\infty}.
\end{align}
Lemma
\pref{Lemma HS} ($\hs$ Estimates) states that
$U^0F_\odd U^0$ and $G$ consist of Hilbert-Schmidt operators, with
$\norm{U^0F_\odd U^0}_{\hs,\infty}<\infty$
and $\norm{G}_{\hs,\infty}<\infty$. Furthermore, it also states that
$\|F\|_1<\infty$, which implies also
$\|F_\ev\|_1<\infty$ and $\|F_\odd\|_1<\infty$.
Combining these facts with
the bound (\pref{bound sup Rn}), it follows that
\begin{equation}
\xconstr{c:gronwall}<\infty.
\end{equation}
We claim that the following bound holds for all $n\ge 1$:
\begin{equation}
\onorm{R^{(n)}(t_1,t_0)}\le \xconstr{c:gronwall}\exp\left(\int_{t_0}^{t_1}
\norm{F_\ev(t)}\,dt\right).
\end{equation}
We prove it by induction.
For $n=1$, we have $R^{(1)}=U^0+G$. Using  $U^0_\odd=0$ and
$t_1\ge t_0$,
we conclude
\begin{equation}
\onorm{R^{(1)}(t_1,t_0)}\le\norm{G}_{\hs,\infty}\le \xconstr{c:gronwall}
\le\xconstr{c:gronwall}\exp\left(\int_{t_0}^{t_1}
\norm{F_\ev(t)}\,dt\right).
\end{equation}
For the induction step $n\leadsto n+1$,
we calculate, using the estimate (\pref{hs-recursion}) in the first step and
the induction hypothesis in the second step:
\begin{align}
&\onorm{R^{(n+1)}(t_1,t_0)}\le
\int_{t_0}^{t_1}
\norm{F_\ev(t)}\onorm{R^{(n)}(t,t_0)}\,dt+\xconstr{c:gronwall}
\nonumber\\&
\le
\xconstr{c:gronwall}\int_{t_0}^{t_1}
\norm{F_\ev(t)}\exp\left(\int_{t_0}^{t}
\norm{F_\ev(s)}\,ds\right)\,dt+\xconstr{c:gronwall}
\nonumber\\&
=
\xconstr{c:gronwall}\exp\left(\int_{t_0}^{t_1}
\norm{F_\ev(t)}\,dt\right)
.
\end{align}
Finally, we get $\norm{R^{(n)}_\odd}_{\hs,\infty}\le \xconstr{c:gronwall} e^{\norm{F_\ev}_1}<\infty$,
which is a uniform bound in $n$.
We now use following general fact, which follows from Fatou's lemma:
If $(L_n)_{n\in\N}$
is a sequence of Hilbert-Schmidt operators converging
to a bounded operator $L$ with respect to the operator norm,
then the following bound holds:
\begin{equation}
\|L\|_{\hs}\le\liminf_{n\to\infty}\|L_n\|_{\hs}.
\end{equation}
An application of this fact to the sequence
$(R^{(n)}_\odd(t_1,t_0))_{n\in\N}$, using
the uniform convergence stated in (\pref{Rn to R}),
yields the result:
\begin{equation}
\sup_{t_1\ge t_0}\onorm{[(\id_{\ch}-Q(t_1))U(t_1,t_0)(\id_{\ch}+Q(t_0))]}
=
\norm{R_\odd}_{\hs,\infty}\le \xconstr{c:gronwall} e^{\norm{F_\ev}_1}<\infty.
\end{equation}
This proves the claim (\pref{eqn:PQUQP formula}).
\end{proof}

\begin{our_lemma}[Partial Integration]\label{lem:partial integration}
The following integration-by-parts formula holds true:
  \begin{equation}
  U^0ZU^0=QU^0-U^0Q-U^0Q'U^0+U^0Z_\ev U^0\label{partial integration}.
  \end{equation}
\end{our_lemma}

\begin{proof}
We split $Z=Z_\ev+Z_\odd$ into even and odd pieces:
\begin{equation}
\label{evenodd}
U^0ZU^0=U^0Z_\odd U^0+U^0Z_\ev U^0
\end{equation}
Now, $U^0Z_\odd U^0=U^0Z_{+-}U^0+U^0Z_{-+}U^0$
consists of integral operators with the following integral
kernels: The component $U^0Z_{+-}U^0$ has the integral kernel
\begin{align}
&(p,q)\mapsto \int_{t_0}^{t_1}
e^{-i (t_1-t)H^0(p)}P_+(p)Z^{\sa(t)}(p-q)P_-(q) e^{-i (t-t_0)H^0(q)}\,dt
\nonumber\\&
=\int_{t_0}^{t_1}
e^{-i (t_1-t)E(p)}P_+(p)Z^{\sa(t)}(p-q)P_-(q) e^{+i (t-t_0)E(q)}\,dt
\nonumber\\&
=e^{-i t_1 E(p)}P_+(p)\int_{t_0}^{t_1}
e^{i t(E(p)+E(q))}Z^{\sa(t)}(p-q)\,dt\, P_-(q) e^{-i t_0 E(q)}
.
\label{w+-1}
\end{align}
Recall that the function $E:\R^3\to \R$ is defined by
$E(p)=+\sqrt{m^2+p^2}$.
The crucial point is that the frequencies $E(p)$ and $E(q)$ have equal signs;
they
do {\it not} partially cancel each other, giving rise to a highly oscillatory
integral at high momenta. Note that this works only for the odd part of $Z$.
Integrating by parts, the right hand side in (\pref{w+-1}) equals
\begin{align}
\ldots=&
\frac{P_+(p)Z^{\sa(t)}(p-q)P_-(q)}{i (E(p)+E(q))}
e^{i (t_1-t_0)E(q)}
-
e^{-i (t_1-t_0)E(p)}
\frac{P_+(p)Z^{\sa(t)}(p-q)P_-(q)}{i (E(p)+E(q))}
\nonumber\\&
-e^{-i t_1 E(p)}P_+(p)\int_{t_0}^{t_1}
\frac{e^{i t(E(p)+E(q))}}{i (E(p)+E(q))}
\frac{\partial}{\partial t} Z^{\sa(t)}(p-q)\,dt\, P_-(q) e^{-i t_0 E(q)}
.
\label{w+-}
\end{align}
Similarly, the integral kernel of the $-+$ component $U^0Z_{-+}U^0$
can be rewritten by an integration by parts as
\begin{align}
\frac{P_-(p)Z^{\sa(t)}(p-q)P_+(q)}{-i (E(p)+E(q))}
e^{-i (t_1-t_0)E(q)}
-
e^{i (t_1-t_0)E(p)}
\frac{P_-(p)Z^{\sa(t)}(p-q)P_+(q)}{-i (E(p)+E(q))}
\nonumber\\-e^{i t_1 E(p)}P_-(p)\int_{t_0}^{t_1}
\frac{e^{-i t(E(p)+E(q))}}{-i (E(p)+E(q))}
\frac{\partial}{\partial t} Z^{\sa(t)}(p-q)\,dt\, P_+(q) e^{i t_0 E(q)}
.
\label{w-+}
\end{align}
The sum of (\pref{w+-}) and (\pref{w-+}) is just the integral kernel of $QU^0+U^0Q-U^0Q'U^0$. Substituting this into (\pref{evenodd}) proves the claim
(\pref{partial integration}).
\end{proof}

Recall that the class ${\mathcal A}\supset C^\infty_c(\R^4,\R^4)$
of vector potentials was introduced in Definition \pref{def: class A} (Class of External Four-Vector Potentials).

\begin{our_lemma}[$\hs$ Estimates]\label{Lemma HS}
Assume that the external vector potential $\sa$
belongs to the class ${\mathcal A}$.
Then the operators
$U^0Z_\ev QU^0$, $U^0QZ_\ev U^0$, $U^0Q'U^0$, $Q^2$, $Q'Q$ and $QZQ$,
constructed with this potential $\sa$,
are Hilbert-Schmidt operators.
Furthermore, their Hilbert-Schmidt norm is uniformly bounded
in the time variables.
Finally, the family of operators
$F=(-Q'+Z_\ev-QZ)(\id_{\ch}+Q)$,
$G=-U^0Q^2+U^0(-Q'+Z_\ev-QZ)Q^2 U(\id_{\ch}+Q)$
and $R=(\id_{\ch}-Q)U(\id_{\ch}+Q)$,
introduced in (\pref{def-F}), (\pref{def-G}), and
(\pref{def-R}), respectively, fulfill the following bounds
in the Hilbert-Schmidt norm:
\begin{equation}
\label{bound U0FU0 and G}
\norm{U^0F_\odd U^0}_{\hs,\infty}<\infty
\quad\mbox{and}\quad
\norm{G}_{\hs,\infty}<\infty,
\end{equation}
and the following bounds
in the operator norm:
\begin{equation}
\label{bound F and R}
\norm{F}_{1}<\infty
\quad \mbox{ and }\quad
\norm{R}_{\infty}<\infty.
\end{equation}
\end{our_lemma}
\begin{proof}
Preliminarily, we estimate for any
$\sa\in\mathcal A$, $\mu=0,1,2,3$, $m=0,1$, and $n=1,2$, using the
fundamental theorem of calculus and averaging the starting point
$s$ uniformly over the unit interval:
\begin{align}
  \begin{split}\label{eqn:sup_norm}
    &\sup_{t\in\R}\left\|\frac{d^{m}}{dt^{m}}\widehat \sa_\mu(t)\right\|_n =
\sup_{t\in\R}\left\|
\int_0^1\left[
\frac{d^{m}}{ds^{m}}\widehat \sa_\mu(s)
+ \int_s^t \frac{d^{m+1}}{du^{m+1}}\widehat \sa_\mu(u)\,du\right]\,ds\right\|_n\\
    &\quad\leq \int_{\R}\left\|\frac{d^{m}}{dt^{m}}
\widehat \sa_\mu(t)\right\|_n\,dt
+ \int_{\R} \left\|\frac{d^{m+1}}{dt^{m+1}}\widehat \sa_\mu(t)\right\|_n\,dt
<\infty.
  \end{split}
\end{align}

At first let us examine the operators
$U^0Z_\ev QU^0$, $U^0QZ_\ev U^0$, $U^0Q'U^0$.
All of these operators have in common that the operator $Q$
or its derivative are sandwiched between two free time-evolution operators
$U^0$. The kernel of $Q$, equation (\pref{eqn:operator_Q}), appeared the
first time after a partial integration in the time variable,
Lemma \pref{partial integration} (Partial Integration), which gave rise to the factor
$[i(E(p)+E(q))]^{-1}$. The idea is that with another partial integration
in the time variable,
we will gain another such factor, giving enough decay to see
the Hilbert-Schmidt property of the kernel.

In order to treat a part of the cases simultaneously, let
$V$ denote $Z_\ev Q$, $QZ_\ev$, or $Q'$.
Note that in each of these cases,
for $t\in\R$, $V(t):\mathcal H\to\mathcal H$ is an odd integral
operator. We denote its integral kernel by
$(p,q)\mapsto V(t,p,q)$.
For any $t_0,t_1\in\R$, we have
\begin{align}
  \|(U^0 V U^0)(t_1,t_0)\|_2 &\leq \|(U^0 V_{+-} U^0)(t_1,t_0)\|_2
+ \|(U^0 V_{-+} U^0)(t_1,t_0)\|_2,\\
  \|(U^0 V_{\pm\mp} U^0)(t_1,t_0)\|_2
&= \left\|\int_{t_0}^{t_1}dt\; e^{\mp iE(p)(t_1-t)}
V_{\pm\mp}(t,p,q)e^{\pm iE(q)(t-t_0)}\right\|_{2,(p,q)}.
\label{v+-}
\end{align}
Using a partial integration, the last expression (\pref{v+-})
is estimated as follows.
\begin{align}
  \ldots &= \left\|\int_{t_0}^{t_1}dt\; \left[\frac{d}{dt}\frac{e^{\mp i[E(p)+E(q)]t}}{\mp i[E(p)+E(q)]}\right]V_{\pm\mp}(t,p,q)e^{\mp iE(p)t_1}e^{\pm iE(q)t_0}\right\|_{2,(p,q)}
\nonumber\\
  &\leq 2\sup_{t\in\R}\left\|\frac{V_{\pm\mp}(t,p,q)}{E(p)+E(q)}
\right\|_{2,(p,q)} +
\int_{\R} dt\; \left\|\frac{V'_{\pm\mp}(t,p,q)}{E(p)+E(q)}\right\|_{2,(p,q)}
\nonumber\\
  &=:f[V_{\pm\mp}]+g[V'_{\pm\mp}].
\label{f+g}
\end{align}
The first summand comes from the two boundary terms for $t=t_0$ and $t=t_1$.
In the following, we show that
$f[V_{\pm\mp}]$ and
$g[V'_{\pm\mp}]$ are finite. Then,
$U^0Z_\ev QU^0,U^0QZ_\ev U^0,U^0Q'U^0$ are in $\hs$
with a Hilbert-Schmidt norm uniformly bounded in the time variable.

\paragraph{Case $V=Z_\ev Q$:}  The 2-norm of the kernel of
$V_{\pm\mp}(t)$ is estimated as follows:
\begin{align}
\nonumber
  &|V_{\pm\mp}(t,p,q)| = \left|\int_{\R^3} dk\; \frac{Z_{\pm\pm}(t,p,k)Z_{\pm\mp}(t,k,q)}{E(k)+E(q)}\right|\\
\nonumber
  &= \left|\int_{\R^3} dk\; \sum_{\mu,\nu=0}^3\frac{P_\pm(p)\alpha^\mu P_\pm(k)\alpha^\nu P_\mp(q) \widehat \sa_\mu(t,p-k)\widehat \sa_\nu(t,k-q)}{E(k)+E(q)}\right|
\\
\nonumber
  &\leq \sum_{\mu,\nu=0}^3\int_{\R^3} dk\; |P_\pm(p)\alpha^\mu P_\pm(k)\alpha^\nu P_\mp(q)| \frac{|\widehat \sa_\mu(t,p-k)\widehat \sa_\nu(t,k-q)|}{E(k)+E(q)}\\
  &\leq \xconstr{c:sechs}\sum^3_{\mu,\nu=0}\int_{\R^3} dk\; \frac{|\widehat \sa_\mu(t,p-k)\widehat \sa_\nu(t,k-q)|}{E(k)+E(q)}
\label{bound Vpm}
\end{align}
with the constant
\begin{align}
  \xconstl{c:sechs} := \sum_{\mu,\nu=0}^3\sup_{p,k,q\in\R^3}|P_\pm(p)\alpha^\mu P_\pm(k)\alpha^\nu P_\mp(q)|<\infty;
\end{align}
note that $\sup_{p\in\R^3}|P_\pm(p)|<\infty$
holds, because $P_\pm$ are orthogonal projections.
An analogous argument for $V'$ yields
\begin{align}
\nonumber
  &|V'_{\pm\mp}(t,p,q)| = \left|\int_{\R^3}dk\; \frac{Z_{\pm\pm}'(t,p,k)Z_{\pm\mp}(t,k,q)+Z_{\pm\pm}(t,p,k)Z_{\pm\mp}'(t,k,q)}{E(k)+E(q)}\right|\\
  &\leq \xconstr{c:sechs}\sum^3_{\mu,\nu=0}\int_{\R^3}dk\; \frac{|\widehat \sa'_\mu(t,p-k)\widehat \sa_\nu(t,k-q)|+|\widehat \sa_\mu(t,p-k)\widehat \sa'_\nu(t,k-q)|}{E(k)+E(q)}.
\label{kernel V'}
\end{align}
With the bound (\pref{bound Vpm}), we compute
\begin{align}
\nonumber
  f[V_{\pm\mp}] &= 2\sup_{t\in\R}\left\|\frac{V_{\pm\mp}(t,p,q)}{E(p)+E(q)}\right\|_{2,(p,q)}\\
  &\leq 2\xconstr{c:sechs}\sum^3_{\mu,\nu=0}\sup_{t\in\R}\left\|\int_{\R^3}dk\;\frac{|\widehat \sa_\mu(t,p-k)\widehat \sa_\nu(t,k-q)|}{[E(p)+E(q)][E(k)+E(q)]}\right\|_{2,(p,q)}.
\end{align}
Lemma \pref{lem:integral estimates}(\pref{eqn:convolution_integrals_2}) (Integral Estimates),
applied to the present situation, states that
the norm in the last expression is bounded by $\xconstr{c:neun}\|\widehat \sa_\mu(t,\cdot)\|_1 \|\widehat \sa_\nu(t,\cdot)\|_2$ with a finite constant $\xconstr{c:neun}$.
Applying this yields
\begin{align}
  f[V_{\pm\mp}]&\leq 2\xconstr{c:sechs}\xconstr{c:neun}\sum_{\mu,\nu=0}^3\sup_{t\in\R} \|\widehat \sa_\mu(t,\cdot)\|_1 \|\widehat \sa_\nu(t,\cdot)\|_2.
\end{align}
The fact $\sa\in\mathcal A$ and inequality (\pref{eqn:sup_norm})
ensure that this expression is finite.

The second summand on the right hand side of (\pref{f+g}) is estimated
with the help of the bound (\pref{kernel V'}) as follows:
\begin{align}
\nonumber
  &g[V'_{\pm\mp}]=
\int_{\R}dt\; \left\|\frac{V'_{\pm\mp}(t,p,q)}{E(p)+E(q)}\right\|_{2,(p,q)}\\
\nonumber
  &\leq \xconstr{c:sechs}\sum^3_{\mu,\nu=0}\int_{\R}dt\; \left\|\int_{\R^3}dk\; \frac{|\widehat \sa'_\mu(t,p-k)\widehat \sa_\nu(t,k-q)|}{[E(p)+E(q)][E(k)+E(q)]}\right\|_{2,(p,q)}\\
  &\quad +\xconstr{c:sechs}\sum^3_{\mu,\nu=0}\int_{\R}dt\; \left\|\int_{\R^3}dk\; \frac{|\widehat \sa_\mu(t,p-k)\widehat \sa'_\nu(t,k-q)|}{[E(p)+E(q)][E(k)+E(q)]}\right\|_{2,(p,q)}.
\end{align}
Again by Lemma \pref{lem:integral estimates}(\ref{eqn:convolution_integrals_2}) (Integral Estimate) we then find
\begin{align}
  g[V'_{\pm\mp}(t)]&\leq \xconstr{c:sechs}\xconstr{c:neun}\sum^3_{\mu,\nu=0}\int_{\R}dt\; \left(\|\widehat \sa'_\mu(t)\|_1\|\widehat \sa_\nu(t)\|_2+\|\widehat \sa_\mu(t)\|_1\|\widehat \sa'_\nu(t)\|_2\right),
\end{align}
while the fact $\sa\in\mathcal A$ together with its consequence
(\pref{eqn:sup_norm})
ensure the finiteness of this expression.
Summarizing, we have shown that $\|U^0Z_\ev QU^0\|_{\hs,\infty}<\infty$.
\paragraph{Case $V=QZ_\ev $:} We reduce this case to the case $V=Z_\ev Q$,
which we treated already.
For any linear operator $A$ on $\ch$,
$\|A\|_\hs=\|A^*\|_\hs$ holds.
Using this and recalling that $Z$ is self-adjoint and $Q$ is
skew-adjoint, we compute
\begin{align}
  \|U^0QZ_\ev U^0\|_{\hs,\infty}=\|-(U^0Z_\ev QU^0)^*\|_{\hs,\infty}=
\|U^0Z_\ev QU^0\|_{\hs,\infty}
,
\end{align}
which we have already shown to be finite.

\paragraph{Case $V=Q'$:} In this case we get
\begin{align}\nonumber
  |V_{\pm\mp}(t,p,q)| &= \frac{|Z'_{\pm\mp}(t,p,q)|}{E(p)+E(q)} \leq \sum^3_{\mu=0} |P_\pm(p)\alpha^\mu P_\mp(q)|\frac{|\widehat\sa'_\mu(t,p-q)|}{E(p)+E(q)}\\
  &\leq  \xconstr{c:acht}\sum^3_{\mu=0} \frac{|\widehat\sa'_\mu(t,p-q)|}{E(p)+E(q)}
\end{align}
with the finite constant
\begin{align}
  \xconstl{c:acht} := \sum_{\mu=0}^3\sup_{p,q\in\R^3} |P_\pm(p)\alpha^\mu P_\mp(q)|,
\end{align}
A similar bound holds for the derivative
\begin{align}
  |V'_{\pm\mp}(t,p,q)| \leq  \xconstr{c:acht}\sum^3_{\mu=0} \frac{|\widehat \sa''_\mu(t,p-q)|}{E(p)+E(q)}.
\end{align}
Lemma \pref{lem:integral estimates}(\ref{eqn:convolution_integrals_1}) (Integral Estimates),
applied to the present situation, states the following bound:
\begin{equation}
\left\|\frac{\widehat\sa'_\mu(t,p-q)}{[E(p)+E(q)]^2}\right\|_{2,(p,q)}
\le \xconstr{c:neun}\|\widehat\sa'_\mu(t)\|_2.
\end{equation}
Using this yields the following estimate:
\begin{align}\nonumber
  f[V_{\pm\mp}(t)]&=2\sup_{t\in\R}\left\|\frac{V_{\pm\mp}(t,p,q)}{E(p)+E(q)}\right\|_{2,(p,q)}\leq 2\xconstr{c:acht}\sum^3_{\mu=0} \sup_{t\in\R} \left\|\frac{\widehat\sa'_\mu(t,p-q)}{[E(p)+E(q)]^2}\right\|_{2,(p,q)}\\
  &\leq 2\xconstr{c:acht}\xconstr{c:neun}\sum^3_{\mu=0} \sup_{t\in\R} \|\widehat\sa'_\mu(t)\|_2.
\end{align}
The fact that $\sa\in\mathcal A$ and inequality (\pref{eqn:sup_norm}) ensures the
finiteness of this expression. Furthermore, we estimate
\begin{align}
  g[V'_{\pm\mp}(t)]&=\int_{\R}dt\; \left\|\frac{V'_{\pm\mp}(t,p,q)}{E(p)+E(q)}\right\|_{2,(p,q)}\leq \xconstr{c:acht}\sum^3_{\mu=0}\int_{\R}dt\; \left\|\frac{\widehat\sa''_\mu(t,p-q)}{[E(p)+E(q)]^2}\right\|_{2,(p,q)}
\end{align}
Again Lemma \pref{lem:integral estimates}(\ref{eqn:convolution_integrals_1}) (Integral Estimates) gives that the last
expression is bounded as follows:
\begin{align}
  \ldots \leq \xconstr{c:acht}\xconstr{c:neun}\sum^3_{\mu=0}\int_{\R}dt\; \|\widehat\sa''_\mu(t)\|_2
\end{align}
which is also finite since $\sa\in\mathcal A$.
Summarizing, we have shown $\|U^0Q'U^0\|_{\hs,\infty}<\infty$.\\

Next we examine the operators $Q^2$, $Q'Q$ and $QZQ$.
All of them have in common that $Q$ or its derivatives appear twice,
and therefore we have two of such factors $[E(p)+E(q)]^{-1}$ in the
kernel of these operators. We shall see that these factors give enough decay
to ensure the finiteness of the Hilbert-Schmidt norms of these operators.

\paragraph{Cases $Q^2$ and $Q'Q$:}
We denote the $n$th derivative with respect to time $t$ by a superscript $(n)$.
For $n=0,1$ we estimate
\begin{align}
  &\sup_{t\in\R}\|Q^{(n)}(t)Q(t)\|_2
\nonumber\\
  &\leq \sup_{t\in\R}\sum_{\mu,\nu=0}^3\sum_\pm\left\|\int_{\R^3}dk\;
|P_\pm(p)\alpha^\mu P_\mp(k) \alpha^\nu P_\pm(q)|
\frac{|\widehat \sa^{(n)}_\mu(t,p-k)\widehat \sa_\nu(t,k-q)|}
{[E(p)+E(k)][E(k)+E(q)]}\right\|_{2,(p,q)}
\nonumber\\
  &\leq \xconstr{c:zwoelf}\sum_{\mu,\nu=0}^3\sup_{t\in\R}\left\|\int_{\R^3}dk\;
 \frac{|\widehat \sa^{(n)}_\mu(t,p-k)\widehat \sa_\nu(t,k-q)|}
{[E(p)+E(k)][E(k)+E(q)]}\right\|_{2,(p,q)}
\label{bound QnQ}
\end{align}
with the finite constant
\begin{align}
  \xconstl{c:zwoelf} :=  \sum_{\mu,\nu=0}^3\sum_\pm \sup_{p,k,q\in\R^3} |P_\pm(p)\alpha^\mu P_\mp(k) \alpha^\nu P_\pm(q)|.
\end{align}

Lemma \pref{lem:integral estimates}(\ref{eqn:convolution_integrals_4}) (Integral Estimates)
provides the upper bound $\xconstr{c:neun}\|\widehat \sa_\mu^{(n)}(t)\|_1\|\widehat \sa_\nu(t)\|_2$
for the norm of the integral on the right hand side of (\pref{bound QnQ}).
Thus, the right hand side of (\pref{bound QnQ}) is bounded by:
\begin{align}
  \ldots &\leq \xconstr{c:neun}\xconstr{c:zwoelf}\sum_{\mu,\nu=0}^3 \sup_{t\in\R} \|\widehat \sa_\mu^{(n)}(t)\|_1\|\widehat \sa_\nu(t)\|_2,
\end{align}
which is finite because of $\sa\in \mathcal A$ and inequality (\pref{eqn:sup_norm}).
Hence, we have shown $\|Q^2\|_{\hs,\infty}<\infty$ and
$\|Q'Q\|_{\hs,\infty}<\infty$.
\paragraph{Case $QZQ$:} In this case we find
\begin{align}
\nonumber
  &\left\|Q(t)ZQ(t)\right\|_2\\
  &=\left\|\sum_{\sigma,\tau\in\{-,+\}}\int_{\R^3}dk\int_{\R^3}dj\; Q_{-\sigma,\sigma}(t,p,k)Z_{\sigma,\tau}(t,k,j) Q_{\tau,-\tau}(t,j,q)\right\|_{2,(p,q)}
\nonumber\\
  & \leq 4\sup_{\sigma,\tau\in\{-,+\}} \sum_{\lambda,\mu,\nu=0}^3 \bigg\|\int_{\R^3}dk\int_{\R^3}dj\; |P_{-\sigma}(p)\alpha^\lambda P_{\sigma}(k)\alpha^\mu P_{\tau}(j)\alpha^\nu P_{-\tau}(q)| \times
\nonumber\\
  &\hspace{4cm}\times\frac{|\widehat \sa_\lambda(t,p-k)\widehat \sa_\mu(t,k-j)\widehat \sa_\nu(t,j-q)|}{[E(p)+E(k)][E(j)+E(q)]}\bigg\|_{2,(p,q)}
\nonumber\\
  &\leq \xconstr{c:sieben} \left\|\int_{\R^3}dk\int_{\R^3}dj\;\frac{|\widehat \sa_\lambda(t,p-k)\widehat \sa_\mu(t,k-j)\widehat \sa_\nu(t,j-q)|}{[E(p)+E(k)][E(j)+E(q)]}\right\|_{2,(p,q)}
\label{bound QZQ}
\end{align}
with the finite constant
\begin{align}
  \xconstl{c:sieben} := 4\sup_{\sigma,\tau\in\{-,+\}} \sum_{\lambda,\mu,\nu=0}^3 \sup_{p,k,j,q\in\R^3} |P_{-\sigma}(p)\alpha^\lambda P_{\sigma}(k)\alpha^\mu P_{\tau}(j)\alpha^\nu P_{-\tau}(q)|.
\end{align}
By Lemma \pref{lem:integral estimates}(\ref{eqn:convolution_integrals_5}) (Integral Estimates) we find
the following bound for the right hand side in (\pref{bound QZQ}):
\begin{align}
  \ldots\leq \xconstr{c:sieben}\xconstr{c:neun} \sum_{\lambda,\mu,\nu=0}^3 \sup_{t\in\R} \|\sa_\lambda(t)\|_1\|\sa_\mu(t)\|_2\|\sa_\nu(t)\|_2
\end{align}
which is finite because $\sa\in\mathcal A$ and inequality (\pref{eqn:sup_norm}).
This proves the claim $\|QZQ\|_{\hs,\infty}<\infty$.

Finally, we prove the claims (\pref{bound U0FU0 and G})
and (\pref{bound F and R}).
As a consequence of $\sa\in{\mathcal A}$ and the bound (\pref{eqn:sup_norm}),
using the definition of the operators
$Z(t)$, $Q(t)$, and $Q'(t)$ by their integral kernels given in
the equations
(\pref{eqn:def Z}), (\pref{eqn:operator_Q}), and (\pref{eqn: def Q'}),
we observe the following operator norm bounds:
\begin{align}
\|L\|_1<\infty \quad\mbox{and}\quad\|L\|_\infty<\infty
\quad\mbox{for} \quad L\in\{Z,Z_\ev,Q,Q'\};
\label{opnorms}
\end{align}
recall the definition
(\pref{def norms}) of the norms used here.
Furthermore, we know $\|U\|_\infty=1$, since the one-particle
Dirac time evolution $U$ consists of unitary operators.
Combining these facts proves the claim (\pref{bound F and R}).
To prove the first claim in (\pref{bound U0FU0 and G}), we calculate:
\begin{equation}
U^0F_\odd U^0=U^0Z_\ev Q U^0-U^0QZ_\ev U^0-U^0Q'U^0-U^0QZ_\odd QU^0;
\end{equation}
see also equation (\pref{eqn: Fodd})
Using the bounds in the Hilbert-Schmidt norm proven before,
this implies the claim $\|U^0F_\odd U^0\|_{\hs,\infty}<\infty$.
Finally, using $\|U^0\|_\infty=1$, $\|Q^2\|_{\hs,\infty}<\infty$,
$\|U\|_\infty=1$, and the bounds (\pref{opnorms}), the second claim
$\|G\|_{\hs,\infty}<\infty$ in (\pref{bound U0FU0 and G}) follows also.
This finishes the proof of the lemma.
\end{proof}
We now state and prove the integral estimates that were used in the previous
proof. Recall that the function $E:\R^3\to \R$ is defined by
$E(p)=\sqrt{|p|^2+m^2}$.
\begin{our_lemma}[Integral Estimates]\label{lem:integral estimates}
For $\xconstl{c:neun}:=\|E^{-2}\|_2<\infty$,
the following bounds hold
for all  $A_1,A_3\in L_1(\R^3,\C)$ and $A_2\in L_2(\R^3,\C)$:
  \begin{align*}
    \left\|\frac{A_2(p-q)}{[E(p)+E(q)]^2}\right\|_{2,(p,q)} &\leq \xconstr{c:neun} \|A_2\|_2\tag{i}\label{eqn:convolution_integrals_1}\\
    \left\|\int_{\R^3}dk\; \frac{A_1(p-k)A_2(k-q)}{[E(p)+E(q)][E(k)+E(q)]} \right\|_{2,(p,q)}
    &\leq \xconstr{c:neun} \|A_1\|_1 \|A_2\|_2\tag{ii}\label{eqn:convolution_integrals_2}\\
    \left\|\int_{\R^3}dk\; \frac{A_1(p-k)A_2(k-q)}{[E(p)+E(k)][E(k)+E(q)]}\right\|_{2,(p,q)} &\leq \xconstr{c:neun} \|A_1\|_1 \|A_2\|_2\tag{iii}\label{eqn:convolution_integrals_4}\\
    \left\|\int_{\R^3}dk\;\int_{\R^3}dj\; \frac{A_1(p-j)A_2(j-k)A_3(k-q)}{[E(p)+E(j)][E(k)+E(q)]}\right\|_{2,(p,q)}
    &\leq \xconstr{c:neun} \|A_1\|_1 \|A_2\|_2\|A_3\|_1&\tag{iv}\label{eqn:convolution_integrals_5}
  \end{align*}
\end{our_lemma}

\begin{proof}
{\it Inequality (\ref{eqn:convolution_integrals_1}):}
Substituting $r:=p-q$ and and using $E(p)+E(q)\ge E(p)$, one finds
  \begin{align}
    \left\|\frac{A_2(p-q)} {[E(p)+E(q)]^2} \right\|_{2,(p,q)}
    &\leq \left\|\frac{A_2(r)}{E(p)^2}\right\|_{2,(p,r)} =
\left\|E^{-2}\right\|_2 \|A_2\|_2.
  \end{align}

{\it  Inequality (\ref{eqn:convolution_integrals_2}):}
Let $B=\{\chi\in L_2(\R^3\times\R^3,\C)\;|\;\|\chi\|_2\le 1\}$
denote the unit ball
in  $L_2(\R^3\times\R^3,\C)$.
Using a dual representation of the norm $\|\cdot\|_2$, we get
  \begin{align}
    &\left\|\int_{\R^3}dk\; \frac{A_1(p-k)A_2(k-q)}{[E(p)+E(q)][E(k)+E(q)]} \right\|_{2,(p,q)} \leq\left\|\int_{\R^3}dk\; \frac{|A_1(p-k)A_2(k-q)|}{E(q)^2}\right\|_{2,(p,q)} \nonumber\\
    &\leq\sup_{\chi\in B}
\int_{\R^3}dp\;\int_{\R^3}dq\;\int_{\R^3}dk \; \left|\frac{A_1(p-k)A_2(k-q)}{E(q)^2}
\chi(p,q)\right|.\label{eqn:convolution_integrals_2_first_estimate}
  \end{align}
  Substituting $j:=p-k$, we bound the right hand side in
(\ref{eqn:convolution_integrals_2_first_estimate}) as follows:
  \begin{align}
    \ldots&=\sup_{\chi\in B} \int_{\R^3}dp\;\int_{\R^3}dq\;\int_{\R^3}dj \; \left|\frac{A_1(j)A_2(p-j-q)}{E(q)^2} \chi(p,q)\right|
  \nonumber\\&
 \leq\|A_1\|_1 \sup_{\chi\in B} \sup_{j\in\R^3}\int_{\R^3}dp\;\int_{\R^3}dq\; \left|\frac{A_2(p-j-q)}{E(q)^2} \chi(p,q)\right|.
  \end{align}
  Substituting $r:=p-j-q$ and changing the order of integration
  turns this into
  \begin{align}
    \ldots=\|A_1\|_1 \sup_{\chi\in B} \sup_{j\in\R^3} \int_{\R^3}dq\;\int_{\R^3}dr\; \left|\frac{A_2(r)}{E(q)^2} \chi(r+q+j,q)\right|.
  \end{align}
  By the Cauchy-Schwarz inequality, we bound the last expression as follows:
  \begin{align}
\nonumber
    \ldots &\leq\|A_1\|_1 \sup_{\chi\in B} \sup_{j\in\R^3} \left\|\frac{A_2(r)}{E(q)^2]}\right\|_{2,(q,r)} \left\|\chi(r+q+j,q)\right\|_{2,(q,r)}\\
    &= \left\|E^{-2}\right\|_2 \|A_1\|_1\|A_2\|_2.
  \end{align}

  {\it Inequality (\ref{eqn:convolution_integrals_4}):}
Similarly, we estimate
  \begin{align}
\nonumber
    &\left\|\int_{\R^3}dk\; \frac{A_1(p-k)A_2(k-q)}{[E(p)+E(k)][E(k)+E(q)]}\right\|_{2,(p,q)} \leq \left\|\int_{\R^3}dk\; \frac{|A_1(p-k)A_2(k-q)|}{E(k)^2}\right\|_{2,(p,q)}\\
    &\leq\sup_{\chi\in B} \int_{\R^3}dp\;\int_{\R^3}dq\;\int_{\R^3}dk \; \left|\frac{A_1(p-k)A_2(k-q)}{E(k)^2} \chi(p,q)\right|.
\label{eq1}
  \end{align}
  Although these terms looks similar to
(\pref{eqn:convolution_integrals_2_first_estimate}),
there seems to be no substitution which enables us to use the result (\pref{eqn:convolution_integrals_2_first_estimate}) directly.

Interchanging the order of integration and
substituting first $j:=p-k$ and then $r=k-q$,
the right hand side in (\pref{eq1}) equals
  \begin{align}
    \ldots &=\sup_{\chi\in B} \int_{\R^3}dq\;\int_{\R^3}dk \;\int_{\R^3}dj\; \left|\frac{A_1(j)A_2(k-q)}{E(k)^2} \chi(j+k,q)\right|
\nonumber\\
  &\leq\|A_1\|_1\sup_{\chi\in B} \sup_{j\in\R} \int_{\R^3}dq\;\int_{\R^3}dk \; \left|\frac{A_2(k-q)}{E(k)^2} \chi(j+k,q)\right|
  \nonumber\\
&=\|A_1\|_1\sup_{\chi\in B} \sup_{j\in\R} \int_{\R^3}dk \;\int_{\R^3}dr\; \left|\frac{A_2(r)}{E(k)^2} \chi(j+k,k-r)\right|
  \nonumber\\
&\leq\|A_1\|_1 \sup_{\chi\in B} \sup_{j\in\R^3} \left\|\frac{A_2(r)}{E(k)^2]}\right\|_{2,(r,k)} \left\|\chi(j+k,k-r)\right\|_{2,(r,k)}
\nonumber\\
    &= \left\|E^{-2}\right\|_2 \|A_1\|_1\|A_2\|_2.
  \end{align}

{\it Inequality  (\ref{eqn:convolution_integrals_5}):}
Again, we get
  \begin{align}
\nonumber
    &\left\|\int_{\R^3}dk\;\int_{\R^3}dj\; \frac{A_1(p-j)A_2(j-k)A_3(k-q)}{[E(p)+E(j)][E(k)+E(q)]}\right\|_{2,(p,q)}\\
\nonumber
    &\leq \left\|\int_{\R^3}dk\;\int_{\R^3}dj\; \frac{|A_1(p-j)A_2(j-k)A_3(k-q)|}{E(j)E(k)}\right\|_{2,(p,q)}\\
    &=\sup_{\chi\in B} \int_{\R^3}dp\;\int_{\R^3}dq\;\int_{\R^3}dk\;\int_{\R^3}dj\; \left|\frac{A_1(p-j)A_2(j-k)A_3(k-q)}{E(j)E(k)}\chi(p,q)\right|.
  \end{align}
  Interchanging the integration and substituting $r:=p-j$ and $s:=k-q$,
this equals
  \begin{align}\label{eq 2}
    \ldots &= \sup_{\chi\in B} \int_{\R^3}dk\;\int_{\R^3}dj\;\int_{\R^3}dr\;\int_{\R^3}ds\; \left|\frac{A_1(r)A_2(j-k)A_3(s)}{E(j)E(k)}\chi(r+j,k-s)\right|
.
  \end{align}
  We apply H\"older's inequality twice to bound
(\pref{eq 2}) as follows:
  \begin{align}
    \ldots \leq \|A_1\|_1\|A_3\|_1\sup_{\chi\in B} \sup_{r,s\in\R^3} \int_{\R^3}dk\;\int_{\R^3}dj\; \left|\frac{A_2(j-k)}{E(j)E(k)}\chi(r+j,k-s)\right|
  \end{align}
  Using the Cauchy-Schwarz inequality and then the substitution $u:=j-k$,
this term is bounded from above by
  \begin{align}
    \ldots &\leq \|A_1\|_1\|A_3\|_1 \left\|\frac{A_2(j-k)}{E(j)E(k)}\right\|_{2,(j,k)}
\nonumber\\&
    \leq \|A_1\|_1\|A_3\|_1 \left\|\frac{A_2(u)}{E(u+k)E(k)}\right\|_{2,(u,k)}
\nonumber\\
\nonumber    &\leq \|A_1\|_1\|A_2\|_2\|A_3\|_1\sup_{u\in\R^3}
\left\|\frac{1}{E(u+k)E(k)}\right\|_{2,k}\\
    &\leq \|A_1\|_1\|A_2\|_2\|A_3\|_1 \left\|E^{-2}\right\|_2.
  \end{align}
In the last step, we have once more used the Cauchy-Schwarz inequality.
\end{proof}

\subsection{Identification of Polarization Classes}\label{sec:Ident Pol classes}

In this Subsection we show that there is a one-to-one correspondence of the magnetic components $\vec A$ of the four-vector fields $A=(A_\mu)_{\mu=0,1,2,3}=(A^0,-\vec A)$ to the physically relevant polarization classes $C(A)=\{e^{Q^{A}}V\;|\;V\in C(0)=[\ch_{-}]_{\approx_{0}}\}$, introduced in Definition \pref{def:indpolclasses} (Induced Polarization Classes).
\begin{our_theorem}[Identification of the Polarization Classes]
\label{ident polclass}
For $A,A'\in {\mathcal C}^\infty_c(\R^3,\R^4)$,
the following are
equivalent:
\begin{enumerate}
\item[(a)]
$C(A)=C(A')$
\item[(b)]
$\vec{A}=\vec{A'}$
\end{enumerate}
\end{our_theorem}
On this ground the following notation makes sense:
\begin{our_definition}[Physical Polarization Classes]\label{def:physical pol classes}
  For $A=(A_\mu)_{\mu=0,1,2,3}=(A_0,-\vec A)$ in ${\mathcal C}_c^\infty(\R^3,\R^4)$, we define
  \begin{align*}
    C(\vec{A}):=C(A).
  \end{align*}
\end{our_definition}
For this Subsection it is convenient to use the four-vector notation of special relativity. To avoid confusion, in this section,
three-vectors are labeled with an arrow.
Define the Lorentz metric $
(g_{\mu\nu})_{\mu,\nu=0,1,2,3}=\operatorname{diag}(1,-1,-1,-1)$.
Raising and lowering of Lorentz indices is performed with respect to $g_{\mu\nu}$.
The inner product of two four-vectors
$a=(a^\mu)_{\mu=0,1,2,3}=(a_0,\vec a)$ and
$b=(b^\nu)_{\nu=0,1,2,3}=(b_0,\vec b)$ is given by
\begin{align}
  a\cdot b := a_\mu b^\mu =
\sum_{\mu,\nu=0}^3 a^\mu g_{\mu\nu} b^\nu = a_0b_0-\vec{a}\cdot\vec{b}
\end{align}
where the $\cdot$ on the right hand side
above is the euclidean scalar product on $\R^3$. Within this four-vector notation it is more convenient to write the Dirac $\C^{4\times 4}$ matrices (\pref{eqn:dirac matrices}) as
\begin{align}
  (\gamma^\mu)_{\mu=0,1,2,3}=\beta \alpha^\mu
\end{align}
which then fulfill
\begin{align}
  \{\gamma^\mu,\gamma^\nu\} = 2g^{\mu \nu}.
\end{align}
Recall that the Fourier transform $\widehat A$ of a vector potential
$A=(A_\mu)_{\mu=0,1,2,3}=(A_0,-\vec A)\in {\mathcal C}^\infty_c(\R^3,\R^4)$
was introduced in equation
(\pref{Fourier transform}).
Using Feynman's dagger ${\slashed A}=\gamma^\mu A_\mu$,
the integral kernel $Z=Z^A$, introduced in equation (\pref{eqn:def Z}), reads
\begin{align}
Z(\vec p,\vec q)
=-ie\gamma^0\widehat{\slashed A}(\vec p-\vec q),
\quad
\vec p,\vec q\in\R^3.
\end{align}
Abbreviating again $E(\vec p)=\sqrt{|\vec p|^2+m^2}$,
we define two momentum four-vectors $p_+, p_-$ for $\vec p\in\R^3$ by
\begin{align}
p_+&=(p_{+\mu})_{\mu=0,1,2,3}=(E(\vec p),-\vec p),\\
p_-&=(p_{-\mu})_{\mu=0,1,2,3}=(-E(\vec p),-\vec p)
\end{align}
such that the corresponding projection operators introduced in (\pref{eqn:projector_formula}) then read
\begin{align}
\label{gamma form of P+-}
P_\pm(\vec p)&=\frac{1}{2p_{\pm 0}}(\slashed p_\pm+m)\gamma^0.
\end{align}
\begin{proof}[Proof of Theorem \pref{ident polclass} (Identification of the Polarization Classes)]
Recall that $e^{Q^A}$ and $e^{Q^{A'}}$ are unitary maps on $\ch$,
because $Q^A$ and $Q^{A'}$ are skew-adjoint.

Let $V=e^{Q^A}\ch_-$ and $W=e^{Q^{A'}}\ch_-$. By definition,
$V\in C(A)$ and $W\in C(A')$ hold. We need to show that $V\approx_0 W$ holds
if and only if $\vec A=\vec A'$.

Now, $P_V=e^{Q^{A}}P_-e^{-Q^{A}}$ holds.
Just as in (\pref{first order eQ}), we know that
$e^{\pm Q^A}-(\id_{\ch}\pm Q^A)$ are Hilbert-Schmidt operators.
As a consequence, $P_V$ differs from
$(\id_{\ch}+ Q^A)P_-(\id_{\ch}- Q^A)$ only by a Hilbert-Schmidt
operator. Using that $Q^A$ is odd, we know $Q^AP_-Q^A=[(Q^A)^2]_{++}$.
Because $(Q^A)^2 $ is a Hilbert-Schmidt operator by Lemma \pref{Lemma HS} ($\hs$ Estimates),
it follows that $Q^AP_-Q^A\in\hs(\ch)$. We conclude that
$P_V-\id_\ch-Q^AP_-+P_-Q^A\in\hs(\ch)$.
The same argument, applied to $A'$, shows
that
$P_W-\id_\ch-Q^{A'}P_-+P_-Q^{A'}\in\hs(\ch)$.
Taking the difference, this implies
\begin{align}
P_V-P_W&\in (Q^A-Q^{A'})P_--P_-(Q^A-Q^{A'})+\hs(\ch)
\nonumber\\&
=Q^{A-A'}P_--P_-Q^{A-A'}+\hs(\ch)
=Q^{A-A'}_{+-}-Q^{A-A'}_{-+}+\hs(\ch);
\end{align}
recall that $Q^A$ is linear in the argument $A$.
Using once more that $Q^{A-A'}$ is odd,
this yields the following equivalences:
\begin{align}
V\approx W \Leftrightarrow P_V-P_W\in\hs(\ch)
\Leftrightarrow Q^{A-A'}_{+-}-Q^{A-A'}_{-+}\in\hs(\ch)
\Leftrightarrow Q^{A-A'}\in\hs(\ch)
\end{align}
Now Lemma \pref{lem:Q in I2 iff A zero} (Hilbert-Schmidt Condition for $Q$)
below, applied to $A-A'$, states that $Q^{A-A'}\in\hs(\ch)$
is equivalent to $\vec A=\vec A'$.
Summarizing, we have shown that $V\approx W$ holds if and only if
$\vec A=\vec A'$.

In order to show that in this case $V \approx_0 W$
holds also, it remains to show $\charge(V,W)=0$. 
Now, because $e^{Q^A}|_{\ch_-\to V}$ and $e^{-Q^{A'}}|_{W\to \ch_-}$
are unitary maps, we get
\begin{align}
&\charge(V,W)=\ind(P_W|_{V\to W})=
\ind\left(\left.e^{-Q^{A'}}P_We^{Q^A}\right|_{\ch_-\to\ch_-}\right)
\nonumber\\&
=
\ind\left(\left.P_-e^{-Q^{A'}}e^{Q^A}\right|_{\ch_-\to\ch_-}\right)
=
\ind\left(\left.(e^{-Q^{A'}}e^{Q^A})_{--}\right|_{\ch_-\to\ch_-}\right).
\end{align}
Because $Q^A$ is skew-adjoint and its square $(Q^A)^2$ is a Hilbert-Schmidt
operator, $e^{Q^A}$ is a compact perturbation of the identity $\id_{\ch}$.
The same argument shows that $e^{-Q^{A'}}$ is also a compact perturbation
of the identity. Hence, $(e^{-Q^{A'}}e^{Q^A})_{--}|_{\ch_-\to\ch_-}$ is a
compact perturbation of $\id_{\ch_-}$ and thus has Fredholm index 0.
This shows that $\charge(V,W)=0$ and finishes the proof.
\end{proof}
The lemma used in the proof of Theorem \pref{ident polclass} (Identification of the Polarization Classes) is:
\begin{our_lemma}[Hilbert-Schmidt Condition for $Q$]\label{lem:Q in I2 iff A zero}
For $A=(A_\mu)_{\mu=0,1,2,3}=(A_0,-\vec A)$ in ${\mathcal C}_c^\infty(\R^3,\R^4)$,
the following are equivalent:
\begin{enumerate}
\item
$Q^A\in \hs(\ch)$,
\item
$\vec A=0$.
\end{enumerate}
\end{our_lemma}
\begin{proof}
We calculate the squared Hilbert-Schmidt norm $\|Q^A\|_{\hs}^2$ of $Q^A$.
Using the abbreviations $Q^A_{+-}(\vec p,\vec q)
= P_+(\vec p) Q^A(\vec p,\vec q)P_-(\vec q)$
and $Q^A_{-+}(\vec p,\vec q)
= P_-(\vec p) Q^A(\vec p,\vec q)P_+(\vec q)$, we get
\begin{align}
\|Q^A\|_{\hs}^2&=\int_{\R^3}\int_{\R^3}
\trace[Q^A(\vec p,\vec q)Q^A(\vec p,\vec q)^*]\,dp\,dq
\nonumber\\&
=
\int_{\R^3}\int_{\R^3}
\left(
\trace[Q^A_{+-}(\vec p,\vec q)Q^A_{+-}(\vec p,\vec q)^*]
+\trace[Q^A_{-+}(\vec p,\vec q)Q^A_{-+}(\vec p,\vec q)^*]
\right)
\,dp\,dq.
\label{i2norm}
\end{align}
Inserting the Definition (\pref{eqn:operator_Q}) of $Q^A$,
using that
$[\gamma^0\slashed{A}(\vec p-\vec q)]^*=\gamma^0\slashed{A}(\vec q-\vec p)$
and that $P_+(p)$ and $P_-(q)$ are orthogonal projections
having the representation (\pref{gamma form of P+-}),
we express the first summand as follows:
\begin{align}
&\trace[Q^A_{+-}(\vec p,\vec q)Q^A_{+-}(\vec p,\vec q)^*]
\nonumber
\\&=\frac{e^2}{4p_{+0}q_{-0}(p_{+0}-q_{-0})^2}
\trace\bigg(
[(\slashed p_++m)\gamma^0][\gamma^0 \widehat{\slashed A}(\vec p-\vec q)]
[(\slashed q_-+m)\gamma^0]
\nonumber\\
&\hspace{1cm}\cdot
[(\slashed q_-+m)\gamma^0]^*
[\gamma^0 \widehat{\slashed A}(\vec p-\vec q)]^*[(\slashed p_++m)\gamma^0]^*
\bigg)
\nonumber
\\&=\frac{e^2}{4p_{+0}q_{-0}(p_{+0}-q_{-0})^2}
\trace\left((\slashed p_++m)\widehat{\slashed A}(\vec p-\vec q)(\slashed q_-+m)
\widehat{\slashed A}(\vec q-\vec p)\right)
\end{align}
Now we use the following formulas for traces of products of
$\gamma$-matrices:
\begin{align}
\trace(\gamma^\mu\gamma^\nu)&=4g^{\mu\nu},\\
\trace(\gamma^\mu\gamma^\nu\gamma^\kappa)&=0,\\
\trace(\gamma^\mu\gamma^\nu\gamma^\kappa\gamma^\lambda)&=
4(g^{\mu\nu}g^{\kappa\lambda}+g^{\mu\lambda}g^{\kappa\nu}
-g^{\mu\kappa}g^{\nu\lambda}).
\end{align}
We obtain
\begin{align}
0\le &\trace[Q^A_{+-}(\vec p,\vec q)Q^A_{+-}(\vec p,\vec q)^*]
\nonumber\\&
=
\frac{e^2}{4p_{+0}q_{-0}(p_{+0}-q_{-0})^2}
\trace\left((\slashed p_++m)\widehat{\slashed A}(\vec p-\vec q)(\slashed q_-+m)
\widehat{\slashed A}(\vec q-\vec p)\right)
\nonumber\\&
=
\frac{e^2}{p_{+0}q_{-0}(p_{+0}-q_{-0})^2}
\Big(
(m^2-p_+\cdot q_-) \widehat A(\vec p-\vec q)\cdot \widehat A(\vec q-\vec p)
\nonumber\\&
\quad
+ (p_+\cdot \widehat A(\vec p-\vec q))(q_-\cdot \widehat A(\vec q-\vec p))
+ (p_+\cdot \widehat A(\vec q-\vec p))(q_-\cdot \widehat A(\vec p-\vec q))
\Big)
.
\label{trace1}
\end{align}
The second summand on the right hand side in (\pref{i2norm})
can be calculated in a similar way by exchanging the indices ``$+$''
and ``$-$'':
\begin{align}
0\le &\trace[Q^A_{-+}(\vec p,\vec q)Q^A_{-+}(\vec p,\vec q)^*]
\nonumber\\&
=
\frac{e^2}{p_{-0}q_{+0}(p_{-0}-q_{+0})^2}
\Big(
(m^2-p_-\cdot q_+) \widehat A(\vec p-\vec q)\cdot \widehat A(\vec q-\vec p)
\nonumber\\&
\quad
+ (p_-\cdot \widehat A(\vec p-\vec q))(q_+\cdot \widehat A(\vec q-\vec p))
+ (p_-\cdot \widehat A(\vec q-\vec p))(q_+\cdot \widehat A(\vec p-\vec q))
\Big)
\nonumber\\&
=
\trace[Q^A_{+-}(\vec q,\vec p)Q^A_{+-}(\vec q,\vec p)^*]
.
\label{i2norm2}
\end{align}
Thus, the two summands in (\pref{i2norm}) are the same up to exchanging
$\vec p$ and $\vec q$.
In particular, this yields
\begin{align}
\|Q^A\|_{\hs}^2&=
2\int_{\R^3}\int_{\R^3}
\trace[Q^A_{+-}(\vec p,\vec q)Q^A_{+-}(\vec p,\vec q)^*]
\,dp\,dq
\nonumber\\&
=
\int_{\R^3}\int_{\R^3} \frac{2e^2}{p_{+0}q_{-0}(p_{+0}-q_{-0})^2}
\Big(
(m^2-p_+\cdot q_-) \widehat A(\vec p-\vec q)\cdot \widehat A(\vec q-\vec p)
\nonumber\\&
\quad
+ (p_+\cdot \widehat A(\vec p-\vec q))(q_-\cdot \widehat A(\vec q-\vec p))
+ (p_+\cdot \widehat A(\vec q-\vec p))(q_-\cdot \widehat A(\vec p-\vec q))
\Big)
\,dp\,dq.
\label{i2norm3}
\end{align}
Let us now use this to prove that  $\vec A=0$ implies $Q^A\in \hs(H)$.
In the case $\vec A=0$, formula (\pref{i2norm3})
boils down
to
\begin{align}
\|Q^A\|_{\hs}^2&=
2e^2\int_{\R^3}\int_{\R^3}
\frac{E(\vec p)E(\vec q)-\vec p\cdot\vec q-m^2}
{E(\vec p)E(\vec q)(E(\vec p)+E(\vec q))^2}|\widehat A_0(\vec p-\vec q)|^2
\,dp\,dq
\nonumber\\&=
2e^2\int_{\R^3}\int_{\R^3}
\frac{E(\vec p)E(\vec p-\vec k)-\vec p\cdot(\vec p-\vec k)-m^2}
{E(\vec p)E(\vec p-\vec k)(E(\vec p)+E(\vec p-\vec k))^2}
|\widehat A_0(\vec k)|^2
\,dp\,dk
\nonumber\\&=
2e^2\int_{\R^3}\int_{\R^3}
\frac{E(\vec p)(E(\vec p-\vec k)-E(\vec p))+\vec p\cdot\vec k}
{E(\vec p)E(\vec p-\vec k)(E(\vec p)+E(\vec p-\vec k))^2}
|\widehat A_0(\vec k)|^2
\,dp\,dk
\nonumber\\&\le
2e^2\int_{\R^3}\int_{\R^3}
\frac{E(\vec p-\vec k)-E(\vec p)+\vec p\cdot\vec k/E(\vec p)}
{E(\vec p-\vec k)E(\vec p)^2}
|\widehat A_0(\vec k)|^2
\,dp\,dk
,
\label{i2norm4}
\end{align}
where we have used $E(\vec p)^2-|\vec p|^2=m^2$.
We expand $E(\vec p-\vec k)$ around $\vec k=0$:
For $t\in\R$, one has
\begin{align}
\frac{\partial}{\partial t}E(\vec p-t\vec k)
&=-\frac{\vec k\cdot(\vec p-t\vec k)}{E(\vec p-t\vec k)},\\
\frac{\partial^2}{\partial t^2}E(\vec p-t\vec k)
&=\frac{|\vec k|^2}{E(\vec p-t\vec k)}-
\frac{[\vec k\cdot (\vec p-t\vec k)]^2}{E(\vec p-t\vec k)^3}.
\end{align}
Using
\begin{align}
0\le [\vec k\cdot (\vec p-t\vec k)]^2\le |\vec k|^2 |\vec p-t\vec k|^2
\le |\vec k|^2 E(\vec p-t\vec k)^2
\end{align}
we conclude
\begin{align}
0\le \frac{\partial^2}{\partial t^2}E(\vec p-t\vec k)
\le \frac{|\vec k|^2}{E(\vec p-t\vec k)}
.
\end{align}
By Taylor's formula, we get for some $t_{\vec p,\vec k}\in[0,1]$:
\begin{align}
\label{f1}
0\le E(\vec p-\vec k)-E(\vec p)+\frac{\vec p \cdot \vec k}{E(\vec p)}
\le \frac{|\vec k|^2}{2E(\vec p-t_{\vec p,\vec k} \vec k)}
.
\end{align}
Now, using the variable $\vec q_t:=\vec p-t \vec k$
with  $0\le t\le 1$, we
estimate
\begin{align}
E(\vec p)^2&=E(\vec q_t+t\vec k)^2
=|\vec q_t+t\vec k|^2+m^2
\le 2|\vec q_t|^2+2t^2 |\vec k|^2+m^2
\nonumber\\&
\le \frac{2}{m^2}(|\vec q_t|^2+m^2)(t^2 |\vec k|^2+m^2)\le \frac{2}{m^2}
E(\vec q_t)^2E(\vec k)^2.
\end{align}
This yields for $0\le t\le 1$:
\begin{align}
\frac{1}{E(\vec p-t\vec k)}\le  \frac{\sqrt{2}}{m}\frac{E(\vec k)}{E(\vec p)}.
\label{f2}
\end{align}
Substituting the bounds (\pref{f1}) and (\pref{f2})
for $t=1$ and for $t=t_{\vec p,\vec k}$ in (\pref{i2norm4}),
we conclude
\begin{align}
\|Q^A\|_{\hs}^2&
\le
\frac{2e^2}{m^2}
\int_{\R^3}\frac{dp}{E(\vec p)^4}
\int_{\R^3}
|\vec k|^2E(\vec k)^2|\widehat A_0(\vec k)|^2
\,dk<\infty
.
\label{i2norm5}
\end{align}
Thus $\vec A=0$ implies $\|Q^A\|_{\hs}<\infty$.

We now prove that $\vec A\neq 0$ implies $\|Q^A\|_{\hs}=\infty$.
We split $A=(A_\mu)_{\mu=0,1,2,3}=(A_0,-\vec A)$ into $A=(A_0,\vec 0)+(0,-\vec A)$.
Abbreviating $Q^{A_0}=Q^{(A_0,\vec 0)}$ and $Q^{\vec A}:=Q^{(0,-\vec A)}$,
we conclude
\begin{equation}
Q^A=Q^{A_0}+Q^{\vec A}
.
\end{equation}
The part (b)$\Rightarrow$(a) implies that
the first summand $Q^{A_0}$ is a Hilbert-Schmidt operator.
Hence,
$Q^A$ is  a Hilbert-Schmidt operator
if and only if $Q^{\vec A}$ is a Hilbert-Schmidt operator.
Thus it remains to show that $\vec A\neq 0$ implies
$\|Q^{\vec A}\|_{\hs}=\infty$.

Equation (\pref{i2norm3}) in the special case of a vanishing $0$-component
of the vector potential can be rewritten as follows:
\begin{align}
&\|Q^{\vec A}\|_{\hs}^2=
\int_{\R^3}\int_{\R^3}
\frac{2e^2}{E(\vec p)E(\vec q)(E(\vec p)+E(\vec q))^2}
\Big(
(m^2+E(\vec p) E(\vec q)+ \vec p\cdot \vec q)
|\widehat{\vec A}(\vec p-\vec q)|^2
\nonumber\\&
\quad
- (\vec p\cdot \widehat{\vec A}(\vec p-\vec q))(\vec q\cdot
\widehat{\vec A}(\vec q-\vec p))
- (\vec p\cdot \widehat{\vec A}(\vec q-\vec p))(\vec q\cdot
\widehat{\vec A}(\vec p-\vec q))
\Big)
\,dp\,dq
\end{align}
Using (\pref{i2norm2}), we see that the integrand in this integral
is non-negative. We substitute $\vec k:=\vec p-\vec q$.
For any measurable set $S\subseteq \R^3\times \R^3$, we get a lower bound
by restricting the integration to $S$:
\begin{align}
\|Q^{\vec A}\|_{\hs}^2&\ge
\int_S \frac{2e^2}{E(\vec p)E(\vec p-\vec k)(E(\vec p)+E(\vec p-\vec k))^2}
\Big(
(m^2+E(\vec p) E(\vec p-\vec k)
\nonumber\\&
\quad\quad
+ \vec p\cdot (\vec p-\vec k))
|\widehat{\vec A}(\vec k)|^2- (\vec p\cdot \widehat{\vec A}(\vec k))((\vec p-\vec k)\cdot
\widehat{\vec A}(-\vec k))
\nonumber\\
&\quad\quad
- (\vec p\cdot \widehat{\vec A}(-\vec k))((\vec p-\vec k)\cdot
\widehat{\vec A}(\vec k))
\Big)
\,dp\,dk.
\label{i2norm6}
\end{align}
The following considerations serve to find an appropriate choice of
the set $S$.
By the assumption $\vec A\neq 0$ we can take $\vec l\in\R^3$
such that $\widehat{\vec A}(\vec l)\neq 0$.
For every $\vec a\in\C^3\setminus\{0\}$, there exists
a unit vector
$\vec b\in\R^3$, $|\vec b|=1$,
such that $|\vec b\cdot\vec a|\le|\vec a|/\sqrt{2}$.
One can see this as follows.
We define $\vec c=\vec a$ if
$|\operatorname{Re} \vec a|\ge |\operatorname{Im} \vec a|$,
and $\vec c=i\vec a$ otherwise.
In particular, $|\vec c|=|\vec a|$ and
\begin{equation}
\label{Im c}
2|\operatorname{Im} \vec c|^2\le
|\operatorname{Re} \vec a|^2+|\operatorname{Im} \vec a|^2=|\vec a|^2
.
\end{equation}
Take any unit vector $\vec b\in\R^3$
orthogonal to $\operatorname{Re}\vec c$.
Using (\pref{Im c}), we get
\begin{equation}
|\vec b\cdot\vec a|=|\vec b\cdot\vec c|=|\vec b\cdot\operatorname{Im}\vec c|
\le |\vec b||\operatorname{Im}\vec c|=|\operatorname{Im}\vec c|\le
\frac{|\vec a|}{\sqrt{2}}
.
\end{equation}
We apply this to $\vec a= \widehat{\vec A}(\vec l)$,
taking a unit vector $\vec b\in\R^3$
with $|\vec{b}\cdot  \widehat{\vec A}(\vec l)|
\le |\widehat{\vec A}(\vec l)|/\sqrt{2}$.
Take any fixed number $\xconstl{c:eins}$ such that
$1/\sqrt{2}<\xconstr{c:eins}<1$; then $|\vec{b}\cdot  \widehat{\vec A}(\vec l)|<\xconstr{c:eins}
|\vec b||\widehat{\vec A}(\vec l)|$ holds because of $|\vec b|=1$ and
$|\widehat{\vec A}(\vec l)|>0$.
Now
$\widehat{\vec A}$ is a {\it continuous} function.
Therefore, there is a compact ball
$\bar B_r(\vec l)=\{\vec k\in\R^3 \;|\; |\vec k-\vec l|\le r\}$, centered at $\vec l$
with some radius $r>0$, such that
\begin{equation}
\label{lower bound hat A}
\xconstl{c:fuenf}:=\inf_{\vec k\in \bar B_r(\vec l)} |\widehat{\vec A}(\vec k)|>0
\end{equation}
is true and
$|\vec{b}\cdot  \widehat{\vec A}(\vec k)|<\xconstr{c:eins}
|\vec b||\widehat{\vec A}(\vec k)|$ holds for all $\vec k\in\bar B_r(\vec l)$.
By compactness of the ball, using continuity of the function
$\R^3\times\R^3\ni (\vec p,\vec k)\mapsto |\vec{p}\cdot
\widehat{\vec A}(\vec k)|-\xconstr{c:eins}
|\vec p||\widehat{\vec A}(\vec k)|$,
the set
\begin{equation}
S_1:=
\{\vec p\in\R^3\;|\;\mbox{for all}\; \vec k\in \bar B_r(\vec l)\;\mbox{holds}\;
|\vec p\cdot\widehat{\vec A}(\vec k)|
<\xconstr{c:eins}|\vec p||\widehat{\vec A}(\vec k)|\}
\label{def s1}
\end{equation}
is an open subset of $\R^3$. The set $S_1$ is nonempty because of
$\vec b\in S_1$.
Furthermore, $S_1$ is a homogeneous set in the following sense:
For all $\vec p\in\R^3$ and all $\lambda\in\R\setminus \{0\}$,
$\vec p\in S_1$ is equivalent to $\lambda\vec p\in S_1$.
Note that $|\vec p\cdot\widehat{\vec A}(\vec k)|=
|\vec p\cdot\widehat{\vec A}(-\vec k)|$
holds, as $\widehat{\vec A}(-\vec k)$ and $\widehat{\vec A}(\vec k)$
are complex conjugate to each other.

We set $S=S_1\times  \bar B_r(\vec l)$.
For the following considerations, note that $|E(\vec p-\vec k)-E(\vec{p})|\le |\vec k|$, $|\vec p|\le E(\vec p)$, and
$ (\vec p\cdot \widehat{\vec A}(\vec k))(\vec p\cdot \widehat{\vec A}(-\vec k))=
|\vec p\cdot \widehat{\vec A}(\vec k)|^2$  hold
for all $\vec p,\vec k\in\R^3$, and that $\widehat{\vec A}$ is bounded on the
ball $\bar B_r(\vec l)$. Using this, one sees that
there is a constant $\xconstl{c:zwei}>0$, depending only on the potential
$\widehat{\vec A}$ and on the choice of the compact ball $\bar B_r(\vec l)$,
such that for all $\vec p\in\R^3$ and all $\vec k\in \bar B_r(\vec l)$, one has
\begin{align}
\Big|\Big[
&(m^2+E(\vec p) E(\vec p-\vec k)+ \vec p\cdot (\vec p-\vec k))
|\widehat{\vec A}(\vec k)|^2
\nonumber\\&\quad
- (\vec p\cdot \widehat{\vec A}(\vec k))((\vec p-\vec k)\cdot
\widehat{\vec A}(-\vec k))
- (\vec p\cdot \widehat{\vec A}(-\vec k))((\vec p-\vec k)\cdot
\widehat{\vec A}(\vec k))
\Big]
\nonumber\\&
-\Big[
2E(\vec p)^2
|\widehat{\vec A}(\vec k)|^2
- 2|\vec p\cdot \widehat{\vec A}(\vec k)|^2
\Big]\Big|\le \xconstr{c:zwei} E(\vec p).
\label{b1}
\end{align}
Furthermore, there is another constant
$\xconstl{c:drei}>0$, depending only on the choice of the
compact ball $\bar B_r(\vec l)$,
such that for all $\vec p\in\R^3$ and all $\vec k\in \bar B_r(\vec l)$, one has
\begin{equation}
E(\vec p-\vec k)(E(\vec p)+E(\vec p-\vec k))^2\le \xconstr{c:drei} E(\vec p)^3
.
\label{b2}
\end{equation}
Substituting the bounds (\pref{b1}), (\pref{b2}), the choice
(\pref{def s1}) of $S_1$, and the lower bound (\pref{lower bound hat A})
of $|\widehat{\vec A}|$ on $\bar B_r(\vec l)$
in the lower bound (\pref{i2norm6}) of $\|Q^{\vec A}\|_{\hs}^2$,
we obtain
\begin{align}
\|Q^{\vec A}\|_{\hs}^2&
\ge
\int_S \frac{2e^2}{\xconstr{c:drei} E(\vec p)^4}
\Big(
2E(\vec p)^2
|\widehat{\vec A}(\vec k)|^2
- 2|\vec p\cdot \widehat{\vec A}(\vec k)|^2-\xconstr{c:zwei} E(\vec p)
\Big)
\,dp\,dk
\nonumber\\&\ge
\int_S \frac{2e^2}{\xconstr{c:drei} E(\vec p)^4}
\Big(2(1-\xconstr{c:eins})E(\vec p)^2
|\widehat{\vec A}(\vec k)|^2-\xconstr{c:zwei} E(\vec p)
\Big)
\,dp\,dk
\nonumber\\&\ge
|\bar B_r(\vec l)|\int_{S_1} \frac{2e^2}{\xconstr{c:drei} E(\vec p)^4}
\Big(2(1-\xconstr{c:eins})\xconstr{c:fuenf}^2E(\vec p)^2
-\xconstr{c:zwei} E(\vec p)
\Big)
\,dp
=\infty.
\label{i2norm7}
\end{align}
We have used that $1-\xconstr{c:eins}>0$, and that $S_1$ is a nonempty,
open homogeneous subset of $\R^3$.
Thus the lemma is proven.
\end{proof}

\subsection{Gauge Transformations}\label{sec:gauge}

As an addendum we briefly discuss gauge transformations.
Let $\vec A\in {\mathcal C}^\infty_c(\R^3,\R^3)$ be a vector potential and
$\vec A^\sim=\vec A+\nabla Y$ be a gauge transform of it with
$Y\in {\mathcal C}^\infty_c(\R^3,\R)$.
Let $e^{iY}: \ch\to\ch$ the multiplication operator with $e^{iY}$.
We prove:
\begin{our_theorem}[Gauge Transformations]\label{thm:gauge}
The gauge transformation $e^{iY}$ fulfills:
\begin{equation}
e^{iY}\in\ur^0\big(\ch,C(\vec A);\ch,C(\vec A^\sim)\big)
\end{equation}
\end{our_theorem}
Although the statement of this theorem does not involve time,
we prove it using the time-evolution from Subsection \pref{time-evolution}.
 A ``direct'' proof, avoiding time-evolution and using similar techniques as in Subsection \pref{time-evolution}, is possible. However, the approach presented here
avoids additional analytical considerations.
\begin{proof}
We switch the gauge transformation on  between the times 0 and 1,
using a smooth function $f:\R\to[0,1]$
with $f(t)=0$ and $f(t)=1$ for $t$ in a neighborhood of $0$ and $1$,
respectively.
We define $\mathsf{Y} :\R^4\ni(t,\vec x)\mapsto f(t)Y(\vec x)\in\R$.
Take the static vector potential
$\sa:\R^4\ni(t,\vec x)\mapsto(0,-\vec A(\vec x))\in\R^4$ and
its gauge-transformed version $\sa^\sim=(\sa^\sim_\mu)_{\mu=0,1,2,3}=(\sa_\mu-\partial_\mu \mathsf{Y})_{\mu=0,1,2,3}=(A_0-\partial_0 \mathsf{Y},-\vec A-\nabla\mathsf{Y})$.
In other words,
\begin{align}
\sa^\sim(t,\vec x)=(-f'(t)Y(\vec x),-\vec A(\vec x)-f(t)\nabla Y(\vec x)).
\end{align}
(It is no problem that the vector potentials used here do
in general not have compact support
in time, because we use only times $t\in[0,1]$.)
Note that at time $t=0$, the gauge transformation is turned off:
$\sa(0)=\sa^\sim(0)=(0,-\vec A)$,
and at time $t=1$ it is completely turned on:
$\sa(1)=(0,-\vec A)$ and
$\sa^\sim(1)=(0,-\vec A^\sim)$.
The one-particle Dirac time-evolutions
$U^\sa$ and $U^{\sa'}$ are also related by a gauge transformation as follows:
\begin{equation}
e^{i\mathsf{Y}(t_1)}U^\sa(t_1,t_0)=U^{\sa^\sim}(t_1,t_0)e^{i\mathsf{Y}(t_0)},
\quad t_1,t_0\in[0,1].
\end{equation}
In particular, this includes
$e^{i Y}U^{\sa}(1,0)=U^{\sa^\sim}(1,0)$.
By Theorem
\pref{Hilbert-Schmidt} (Dirac Time-Evolution with External Field), we have
the following:
\begin{align}
  U^\sa(0,1)&\in \ur^0\big(\ch,C(\vec A);\ch,C(\vec A)\big)\\
  U^{\sa^\sim}(1,0)&\in \ur^0\big(\ch,C(\vec A);\ch,C(\vec A^\sim)\big)
\end{align}
This implies the following:
\begin{align}
  e^{i Y}=U^{\sa^\sim}(1,0)U^{\sa}(0,1)
\in \ur^0\big(\ch,C(\vec A);\ch,C(\vec A^\sim)\big)
\end{align}
Thus the claim is proven.
\end{proof}
We infer that in general the gauge transformation $e^{iY}$ changes
the polarization class.
Using varying wedge spaces, it can be second quantized as follows.
Let $\cs\in\ocean(C(\vec A))$ and $\cs^\sim\in\ocean(C(\vec A^\sim))$.
By Theorem \pref{unser abstrakter Shale Stinespring} (Lift Condition),
there exists $R\in \cu(\ell)$ such that we have the following
second-quantized gauge transformation from $\cf_\cs$ to $\cf_{\cs^\sim}$:
\begin{align*}
  \xymatrix{
    \cf_\cs \ar@/_/[rd]_{\rop R \lop{e^{iY}}}
\ar[r]^{\lop{e^{iY}}} & \cf_{(e^{iY}\cs)}
    \ar[d]^{\rop R}\\
    & \cf_{\cs^\sim}}
\end{align*}

\section{Summary: the Second Quantized Time-Evolution}\label{sec:second quantized time-evolution}

Combining
Theorem \pref{ident polclass} (Identification of the Polarization Classes),
Theorem \pref{Hilbert-Schmidt}
(Dirac Time-Evolution with External Field),
Theorem \pref{unser abstrakter Shale Stinespring} (Lift Condition),
Corollary \pref{col:uniqueness up to a phase}
(Uniqueness of the Lift up to a Phase) and \pref{thm:gauge} (Gauge Transformations) we have proven the following:

\begin{our_mainresult}[Second Quantized Dirac Time-Evolution]\label{thm:second quant dirac}
  Let $\sa=(\sa_\mu)_{\mu=0,1,2,3}=(\sa_0,-\vec{\sa})\in{\mathcal C}^\infty_c(\R^4,\R^4)$
be an external vector potential. Let $\big(U^{\sa}(t_1,t_0):\ch\to\ch\big)_{t_0,t_1\in\R}$ be the corresponding one-particle Dirac time-evolution.\\
\mbox{}\\
We have constructed the following:
\begin{enumerate}[1.]
 \item natural polarization classes $C(\vec{\sa}(t))\in\pol(\ch)/{\approx}_0$, $t\in\R$, introduced in Definition \pref{def:physical pol classes}, depending only on the three-vector potential $\vec\sa(t)$ at time $t\in\R$, but neither on the history $\sa(s)$, $s\neq t$, nor on the electric potential $\sa^0(t)$, and
 \item for any choice of a vacuum $\Phi$ a natural family of Fock spaces $\cf_{t}$, $t\in\R$, as follows: Given a  (separable, infinite dimensional) Hilbert space $\ell$, for any $\Phi\in\ocean_\ell(C(0))$ one has the family of equivalence classes
\begin{align}
  \cs(t):=[e^{Q^{\sa(t)}}\Phi]_\sim
\in\ocean_\ell(C(\vec\sa(t)))/{\sim},\; t\in\R,
\end{align}
and, hence, the corresponding family of Fock spaces $\cf_{t}=\cf_{\cs(t)}$, $t\in\R$.
\end{enumerate}
We have obtained the following:
\begin{enumerate}[1.]
\addtocounter{enumi}{2}
\item For $t_0,t_1\in\R$ it holds that
\begin{equation}
 U^\sa(t_1,t_0)\in\ur^0\big(\ch,C(\vec{\sa}(t_0));\ch,C(\vec{\sa}(t_1))\big).
\end{equation}
This yields on the above family of Fock spaces a natural second-quantized Dirac time-evolution 
\begin{align*}
 \widetilde U^\sa(t_{1},t_{0}):\cf_{t_{0}}\to\cf_{t_{1}},
\end{align*}
completely determined up to a phase. It is given as follows: abbreviating $U^\sa:=U^\sa(t_1,t_0)$, there  is $R\in \mathrm{U}(\ell)$ such that
\begin{equation}
\widetilde U^\sa(t_{1},t_{0})=\rop {R} \lop{U^\sa}:\cf_{t_0}\to \cf_{t_1}
\end{equation}
is a unitary map between the Fock spaces $\cf_{t_0}$ and
$\cf_{t_1}$. $\widetilde U^\sa(t_{1},t_{0})$ is unique up to a phase
in the following sense: for any two such choices $R_1, R_2\in\mathrm{U}(\ell)$ with $\rop {R_1} \lop{U^\sa},\rop {R_2} \lop {U^\sa}
:\cf_{t_0}\to \cf_{t_1}$, the operator $R_1^{-1}R_2$ has a determinant $\det (R_1^{-1}R_2)=e^{i\varphi}$ for some $\varphi\in\R$, and it holds
\begin{equation}
\rop {R_2} \lop {U^\sa}=e^{i\varphi}\rop {R_1} \lop {U^\sa}.
\end{equation}
 \item The formalism is gauge invariant although gauge transformations may change the polarization classes, and therefore the induced second-quantized gauge transformations may act between varying Fock spaces.
\end{enumerate}
\end{our_mainresult}
We emphasize that the family of Fock spaces is defined in terms of the equivalence classes $\cs(t)$, $t\in\R$. Hence, the dependence on the choice of $\Phi$ is weak. Up to a natural isomorphism, being unique up to a phase, $\cf_{t}$ depends only on the polarization class $[\range e^{Q^{\sa(t)}}\Phi]_{\approx_{0}}$; cf. Lemma \pref{lem:connection sim approx} (Connection between $\sim$ and $\approx_0$). Changing the choice of $\Phi$ within the same equivalence class can be viewed as a Bogolyubov transformation.

An application of this theorem is the computation of transition amplitudes. Consider given $\Wedge \Psi^{\inn}\in\cf_{\cs(t_0)}$ and $\Wedge \Psi^{\out}\in\cf_{\cs(t_1)}$, which represent ``in'' and ``out'' states at times $t_0$ and $t_1$, respectively. The transition amplitude is according to the above theorem given by
\begin{align*}
  \left|\sk{\Wedge \Psi^{\out},\rop {R_1} \lop U\Wedge\Psi^{\inn}}\right|^2&=|e^{i\varphi}|^2\left|\sk{\Wedge \Psi^{\out},\rop {R_2} \lop U\Wedge\Psi^{\inn}}\right|^2\\
&=\left|\sk{\Wedge \Psi^{\out},\rop {R_2} \lop U\Wedge\Psi^{\inn}}\right|^2
\end{align*}
which is therefore independent on our specific choice of the matrix $R_{1}$ or $R_{2}$, and also gauge invariant.


\section*{Acknowledgments}
The authors would like to thank Torben Kr\"uger and S\"oren Petrat for valuable 
comments. Furthermore, D.-A. Deckert gratefully acknowledges financial support 
within the scope of the \emph{BayEFG} of the \emph{Freistaat Bayern} and the 
\emph{Universi\"at Bayern e.V.}.

\bibliographystyle{alpha}
\bibliography{qed}

\begin{thebibliography}{GGK90}

\bibitem[And33]{Anderson:36}
C.~D. Anderson.
\newblock The {P}ositive {E}lectron.
\newblock {\em The Physical Review}, 43(6):491--494, 1933.

\bibitem[Bel75]{Bellisard:75}
J.~Bellisard.
\newblock Quantized {F}ields in {I}nteraction with {E}xternal {F}ields {I}.
  {E}xact {S}olutions and {P}erturbative {E}xpansions.
\newblock {\em Commun. math. Phys.}, pages 235--266, 1975.

\bibitem[Bel76]{Bellisard:76}
J.~Bellisard.
\newblock Quantized {F}ields in {I}nteraction with {E}xternal {F}ields {II}.
  {E}xistence {T}heorems.
\newblock {\em Commun. math. Phys.}, pages 53--74, 1976.

\bibitem[Dir34a]{dirac_discussion_1934}
P.~A.~M. Dirac.
\newblock Discussion of the infinite distribution of electrons in the theory of
  the positron.
\newblock {\em Mathematical Proceedings of the Cambridge Philosophical
  Society}, 30(02):150--163, 1934.

\bibitem[Dir34b]{Dirac:34}
P.A.M. Dirac.
\newblock Theorie du {P}ositron.
\newblock {\em Selected Papers on Quantum Electrodynamics Edited by J.
  Schwinger, Dover Publications, Inc., New York}, 1934.

\bibitem[DP07]{PicklDuerr:08-1}
D.~D\"urr and P.~Pickl.
\newblock On {A}diabatic {P}air {C}reation.
\newblock {\em Commun. math. Phys.}, 282(1):161--198, 2007.

\bibitem[DP08]{PicklDuerr:08-2}
D.~D\"urr and P.~Pickl.
\newblock Adiabatic {P}air {C}reation in {H}eavy {I}on and {L}aser {F}ields.
\newblock {\em EPL.}, 81, February 2008.

\bibitem[Fin06]{finster_principle_2006}
Felix Finster.
\newblock {\em The principle of the fermionic projector}, volume~35 of {\em
  {AMS/IP} Studies in Advanced Mathematics}.
\newblock American Mathematical Society, Providence, {RI}, 2006.

\bibitem[Fin08]{finster_regularized_2008}
Felix Finster.
\newblock On the regularized fermionic projector of the vacuum.
\newblock {\em Journal of Mathematical Physics}, 49(3):032304, 60, 2008.

\bibitem[Fin09a]{finster_action_2009}
Felix Finster.
\newblock An action principle for an interacting fermion system and its
  analysis in the continuum limit.
\newblock {\em http://arxiv.org/abs/0908.1542}, August 2009.

\bibitem[Fin09b]{finster_formulation_2009}
Felix Finster.
\newblock A formulation of quantum field theory realizing a sea of interacting
  dirac particles.
\newblock {\em http://arxiv.org/abs/0911.2102}, November 2009.

\bibitem[FS79]{fierz-scharf-79}
H.~Fierz and G.~Scharf.
\newblock Particle interpretation for external field problems in {QED}.
\newblock {\em Helv. Phys. Acta}, 52(4):437--453 (1980), 1979.

\bibitem[GGK90]{gohberg-90}
I.~Gohberg, S.~Goldberg, and M.~A. Kaashoek.
\newblock {\em Classes of linear operators. {V}ol. {I}}, volume~49 of {\em
  Operator Theory: Advances and Applications}.
\newblock Birkh\"auser Verlag, Basel, 1990.

\bibitem[HLS05]{hainzl:05}
Christian Hainzl, Mathieu Lewin, and {\'E}ric S{\'e}r{\'e}.
\newblock Existence of a stable polarized vacuum in the
  {B}ogoliubov-{D}irac-{F}ock approximation.
\newblock {\em Comm. Math. Phys.}, 257(3):515--562, 2005.

\bibitem[LM96]{langmann:96}
E.~Langmann and J.~Mickelsson.
\newblock Scattering matrix in external field problems.
\newblock {\em J. Math. Phys.}, 37(8):3933--3953, 1996.

\bibitem[Mic98]{mickelsson:98}
J.~Mickelsson.
\newblock Vacuum polarization and the geometric phase: gauge invariance.
\newblock {\em J. Math. Phys.}, 39(2):831--837, 1998.

\bibitem[PS86]{pressley:86}
A.~Pressley and G.~Segal.
\newblock {\em Loop groups}.
\newblock Oxford Mathematical Monographs. The Clarendon Press Oxford University
  Press, New York, 1986.
\newblock Oxford Science Publications.

\bibitem[Rui77a]{Ruijsenaars:77-1}
S.~N.~M. Ruijsenaars.
\newblock Charged particles in external fields. {I}. {C}lassical theory.
\newblock {\em J. Mathematical Phys.}, 18(4):720--737, 1977.

\bibitem[Rui77b]{Ruijsenaars:77-2}
S.~N.~M. Ruijsenaars.
\newblock Charged particles in external fields. {II}. {T}he quantized {D}irac
  and {K}lein-{G}ordon theories.
\newblock {\em Comm. Math. Phys.}, 52(3):267--294, 1977.

\bibitem[Sch95]{scharf:95}
G.~Scharf.
\newblock {\em Finite {Q}uantum {E}lectrodynamics}.
\newblock Texts and Monographs in Physics. Springer-Verlag, Berlin, second
  edition, 1995.
\newblock The causal approach.

\bibitem[Sim05]{Simon:05}
B.~Simon.
\newblock {\em Trace ideals and their applications}, volume 120 of {\em
  Mathematical Surveys and Monographs}.
\newblock American Mathematical Society, Providence, RI, second edition, 2005.

\bibitem[SS65]{shale:65}
D.~Shale and W.~F. Stinespring.
\newblock Spinor representations of infinite orthogonal groups.
\newblock {\em J. Math. Mech.}, 14:315--322, 1965.

\bibitem[SW86]{scharf:86}
G.~Scharf and W.~F. Wreszinski.
\newblock The causal phase in quantum electrodynamics.
\newblock {\em Nuovo Cimento A (11)}, 93(1):1--27, 1986.

\bibitem[Tha92]{thaller:92}
B.~Thaller.
\newblock {\em The {D}irac Equation}.
\newblock Texts and Monographs in Physics. Springer-Verlag, Berlin, 1992.

\end{thebibliography}
\immediate\closeout\xrf %

\end{document}